# RUSSIAN ACADEMY OF SCIENCES
## NATIONAL GEOPHYSICAL COMMITTEE

# РОССИЙСКАЯ АКАДЕМИЯ НАУК
## НАЦИОНАЛЬНЫЙ ГЕОФИЗИЧЕСКИЙ КОМИТЕТ

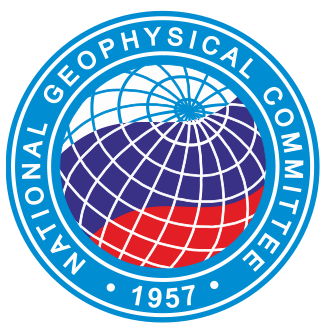

# NATIONAL REPORT

for the
International Association of Geodesy
of the
International Union of Geodesy and Geophysics
2011–2014

# НАЦИОНАЛЬНЫЙ ОТЧЕТ

для
Международной ассоциации геодезии
Международного
геодезического и геофизического союза
2011–2014

Москва   2015   Moscow

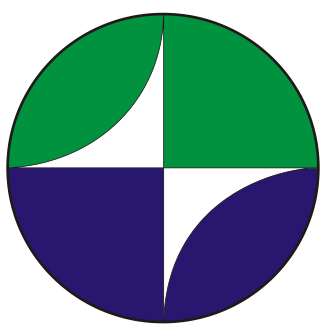

Presented to the XXVI General Assembly
of the
International Union of Geodesy and Geophysics

К XXVI Генеральной ассамблее
Международного геодезического и геофизического
союза

# RUSSIAN ACADEMY OF SCIENCES

National Geophysical Committee

# NATIONAL REPORT

for the
International Association of Geodesy
of the
International Union of Geodesy and Geophysics
2011–2014

Presented to the XXVI General Assembly
of the
IUGG

2015
Moscow

In this National Report are given major results of researches conducted by Russian geodesists in 2011–2014 on the topics of the International Association of Geodesy (IAG) of the International Union of Geodesy and Geophysics (IUGG). This report is prepared by the Section of Geodesy of the National Geophysical Committee of Russia. In the report prepared for the XXVI General Assembly of IUGG (Czhech Republic, Prague, 22 June – 2 July 2015), the results of principal researches in geodesy, geodynamics, gravimetry, in the studies of geodetic reference frame creation and development, Earth's shape and gravity field, Earth's rotation, geodetic theory, its application and some other directions are briefly described. For some objective reasons not all results obtained by Russian scientists on the problems of geodesy are included in the report.

В данном Национальном отчете представлены основные результаты исследований, проводимых российскими геодезистами в 2011—2014 гг., по темам, соответствующим направлениям деятельности Международной ассоциации геодезии (МАГ) Международного геодезического и геофизического союза (МГГС). Данный отчет подготовлен Секцией геодезии Национального геофизического комитета Российской академии наук. В данном отчете, подготовленном к XXVI Генеральной ассамблее МГГС (Чехия, Прага, 22 июня — 2 июля 2015 г.), представлены основные результаты исследований в области геодезии, геодинамики, гравиметрии, создания геодезических систем отсчета, формы и гравитационного поля Земли, вращения Земли, теории геодезии и ее приложений. По понятным причинам, в отчет были включены не все результаты, полученные российскими учеными в области геодезии.

# Contents

# Executive Summary

This review, submitted to the International Association of Geodesy (IAG) of the International Union of Geodesy and Geophysics (IUGG), contains the results obtained by Russian geodesists in 2011-2014. This review was prepared for the XXVI General Assembly of IUGG (Czech Republic, Prague, 22 June – 2 July 2015). It briefly describes the results of principal research in geodesy, geodynamics, gravimetry, in the studies of geodetic reference frame creation and development, the Earth's shape and gravity field, the Earth's rotation, geodetic theory, its application and some other areas of research.

The review is organized as a sequence of abstracts of principal publications and presentations for symposia, conferences, workshops etc. Each of the review paragraphs includes a list of scientific papers published in 2011–2014 including those prepared in cooperation with Russian scientists and their colleagues from other countries. Some interesting international and national scientific events are also mentioned in the review.

For some objective reasons not all the results obtained by Russian scientists on the problems of geodesy are included in the review.

**The more principal studies** are listed below.

The investigation of the impact of the Galactic aberration on the CRF, TRF, end EOP. For accurate modelling of this effect, the current best estimate of the Galactic aberration constant was obtained as $A = 5.0 \pm 0.3$ μas/yr.

Study of systematic errors of the ICRF and various aspects of combination procedures. In particular, the analysis has shown that using the full correlation matrices leads to substantial change in the orientation parameters between the compared catalogues.

New algorithms of Molodensky theory application for determining the Earth's shape.

Construction and development of new absolute gravity meters.

Propagation of the international gravity reference system to the national gravity reference frame.

Study of coseismic gravity changes in relation with the May 24, 2013 Okhotsk deep-focus earthquake.

Geodynamic study of the West Pacific, Near Baltic, Caucasus and Lake Baikal regions.

First geodetic observations of coseismic crustal displacements caused by a deep-focus (611 km) earthquake

Complex geophysical research of theTohoku earthquake phenomenon.

'Paradox' resolution of high and low strain velocities under the concept that the anomalous recent geodynamics is caused by parametric excitation of deformation processes in fault zones in conditions of a quasistatic regime of loading.

Research on relation between core nutation and geomagnetic activity.

It was found that the amplitude and phase of the Free Core Nutation (FCN) variations derived from VLBI observations are correlated with geomagnetic jerks.

The comparison of the epochs of the changes in the FCN amplitude and phase with the epochs of the GMJs indicated that the observed extremes in the FCN amplitude and phase variations were closely related to the GMJ epochs.

The detailed investigtion of the structure of the Chandler Wobble (CW) revealed that it consists of six principal components with the periods from 11 to 75 years. It was also found that the CW variations may be connected with the Sun activity, Markovits's waves, and the Kp and Ap geomagnetic indices.

The algorithm of calculation of plan rectangular coordinates, declinations and scale of Gauss projection in 6º zone by geodetic coordinates.

The weighted modifications of correlation coefficient and Allan variance.

The implementation of the Finsler geometry in geodesy to take a new look at its traditional tasks and to contribute to the construction of new approaches to problem areas of space geodesy and astrometry.

*Section of geodesy*
*Dr. V.P. Savinykh, Chairman, Corr. Member of RAS*
*Dr. V.I. Kaftan, Vice-chairman*

# Reference Frames

**Kaftan V.[1], Malkin Z.[2], Pobedinsky G.[3], Stoliarov I.A.[3]**

[1]Geophysical Center of the Russian Academy of Sciences, Moscow, Russia
[2]Pulkovo Observatory, Saint Petersburg, Russia
[3]Federal Scientific-Technical Center of Geodesy, Cartography and Spatial Data Infrastructure, Moscow, Russia

The latest research is devoted to the problems of International Celestial Reference Frame (ICRF) development.

In recent years much attention has been paid to the astrometric implications of the galactic aberration in proper motions (GA). This effect causes systematic errors in ICRF at a µas level already substantial for results of the VLBI observations used for simultaneous determinations of CRF, TRF and EOP [Malkin, 2011b, 2012c, 2014b; Liu et al., 2012]. Therefore, this correction must be taken into account during highly accurate astrometric and geodetic data processing. Its accuracy depends, in the first place, on accuracy of the Galactic rotation parameters. It was found from analysis of the all available determinations of the Galactic rotation parameters $R_0$ and $\Omega_0$ made during last 10 years that the most probable value of the Galactic aberration constant A = 5.0 ± 0.3 µas/yr [Malkin, 2012d; 2013d, 2013t, 2013f, 2014c].

Systematic errors of the ICRF are discussed in more detail in several papers. As follows from the many-decades experience of classical astronomy, the most accurate catalog of celestial objects forming the ICRF can be obtained from a combination. Various aspects of combination procedures are discussed in [Sokolova, Malkin, 2012, 2013a, 2013b, 2014]. In particular, the analysis has shown that using the full correlation matrices leads to substantial change in the orientation parameters between the compared catalogues.

Correct estimate of the random errors of the catalogs is important for many tasks, such as catalog comparison, computation of the weights of the catalogs during combination etc. Formal uncertainties of the source positions provided in the catalog are generally substantially smaller than the real position accuracy. These estimates can be improved if the correlation between catalogs is accounted for. In [Malkin, 2013a; 2013b, 2013g, 2014a] one of possible approaches to solve this task using a modified and generalized "3-cornered hat" method is considered.

An international project of a use of a radio-telescope in Sierra Negra for VLBI method realization is presented in [Krilov et al., 2014].

Compact satellite laser ranging (SLR) meters "Sazhen-TM" produced by Open Joint-stock Company "Research-and-Production Corporation "Precision Systems and Instruments" are installed at Quasar VLBI network observatories [Finkelstein A. et al, 2013]. The measurement system allows determining distances of 400-6000 km (day) and 400-23000 km (night) with the accuracy of 1 cm. Local ties between GNSS and SLR markers are determined with the precession of 1-3 mm [Finkelstein A. et al., 2012].

VLBI antenna rotation centers were connected with Fundamental Astro-Geodetic Network (FAGN) points with the accuracy of 1-5 mm for plan and 1-10 mm for height components. Loop misclosures were determined as sums of local tie vectors and baseline VLBI and GNSS vectors between points of VLBI and FAGN. Root mean square errors of vector components were received as $m_x$=15 mm, $m_y$=11 mm и $m_z$=14 mm for distances of several thousand kilometers.

The latest International Terrestrial Reference Frame (ITRF) realizations are derived from four space geodesy techniques: VLBI, GPS, SLR, and DORIS, whereas the International Celestial Reference Frame (ICRF) is a result of global VLBI solution. The latter is tied to the ITRF datum using an arbitrary set of reference stations. VLBI also shares responsibility with SLR for ITRF scale. All the techniques contribute to positions and velocities of ITRF stations. As a consequence, we faced with systematic errors and mutual impact of CRF and TRF realizations, which cannot be fixed by datum correction during current combination. These problems have been discussed in [Boehm, 2012a 2012b; Malkin, 2012a, 2012b; Malkin et al., 2012].

The International Terrestrial Reference Frame considers the position at a reference epoch plus a linear velocity term for station coordinates. However, the actual station movement also includes several tidal and non-tidal corrections (e.g., solid Earth tides, ocean and atmosphere loading) recommended by the IERS Conventions as well as unmodelled non-linear displacements. The increasing accuracy of Very Long Baseline Interferometry (VLBI) observations and the growing time span of available data allow the determination of seasonal signals in station positions which still remain unmodelled in the conventional analysis approach. It was shown that neglecting the seasonal station motion leads to UT1 systematic errors at the μs level [Malkin, 2013g]. It was also found that the seasonal station movements do not yield any significant systematic effect on the CRF but can cause a significant change in position of radio sources with small number of sessions non-evenly distributed over the year fraction [Krasna et al., 2013, 2014]

DORIS analysis center operates at the Institute of Astronomy of the Russian Academy of Science (INASAN). It produces SINEX weekly free network solutions, geocenter motion, EOP and STCD series. The latest Gipsy software version (GIPSY 6.0) is used for time series production. Dynamic Regression Modeling is proposed and used for geocenter motion prediction at 6-25 weeks intervals [Kuzin S., Tatevian S., 2011].

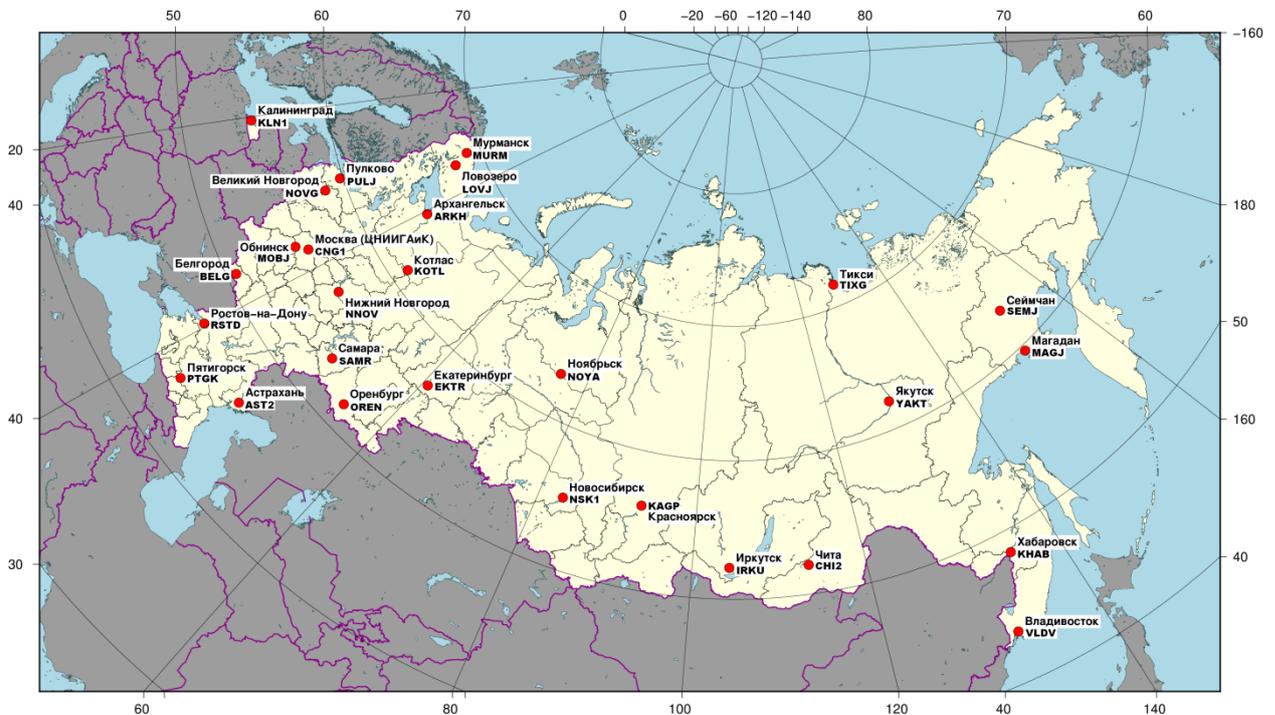

Fig.1 Russian Fundamental Astro-Geodetic Network (FAGN)

Global space reference frame realized as precise GLONASS ephemeris was developed by the Federal State Budgetary Establishment "Federal Scientific-Technical Center of Geodesy, Cartography and Spatial Data Infrastructure". The special web-site (http://rgs-centre.ru) is under construction and works in a test mode. Permanent observations of the Russian Fundamental Astro-Geodetic Network (FAGN) are used for precise ephemeris production (see Fig.1). Figure 1 shows that the FAGN fragment is in the possession of the Federal State Budgetary Establishment "Federal Scientific-Technical Center of Geodesy, Cartography and Spatial Data Infrastructure". FAGN fragments of other ownerships are reflected in [Savinykh V.P. et al., 2014].

New state geocentric coordinate reference system (reference frame) GSK-2011 is developed and officially adopted [Demianov G.V. et al., 2011c, Gorobets V.P. et al., 2012, 2013a, b, c, Kaftan V.I., 2011]. Coordinate transformation parameters between the national coordinate system SK-95 and GSK-2011 are determined [Gorobets V.P., 2013]. SK-95 operates in remote

Russian areas. The special study of the SK-95 state in the Russian Arctic is emphasized in [Khodakov P., Basmanov A.]. The national coordinate system SK-95 is realized by FAGN and two more dense satellite geodetic networks: Precise Geodetic Network (PGN) and 1$^{st}$ order Satellite Geodetic Network (SGN-1) [Demianov G.V. et al., 2011].

GLONASS system has a space realization of the global coordinate reference frame PZ-90.11. Its main development results, role and place in the national coordinate infrastructure are studied and described in [Vdovin V.S., 2013].

The results of a first experiment on the use of full constellation of the GLONASS system for the precise positioning are described. To compare the positioning accuracy estimated by the use of GLONASS and GPS, measurements obtained at 15 sites of the Russian FAGN were analyzed. The outcome of the performed computations shows that sites of the Russian geodetic network were determined with the precision (rms) 3-10 mm in spite of short period of measurements. The differences between coordinates of these sites, estimated by only GPS or GLONASS measurements, are in the same limits. It is considered that the models used for data processing with GLONASS should be more studied and developed [Tatevian S.K., Kuzin S.P., Demjanov G.V., 2013].

A combined use of GPS/GLONASS techniques for the development of the Russian geodetic reference network is studied and described in [Tatevian S., Kuzin S., 2011].

The Geodesy Section [http://geodesy-ngc.gcras.ru/en/] of the National Geophysical Committee [http://ngc.gcras.ru/index_eng.html] of the Russian Academy of Sciences announced an initiative on unifying the observation networks affiliated to different national and departmental organizations into a single regional cluster. A number of meetings of the Geodesy Section held in 2012-2103 were devoted to solving the organizational problem. During this period, the texts of the Statute of the International Commission on the Regional Terrestrial Reference Frame for North East Eurasia (NEEREF) and the Agreement between the institutions and organizations were elaborated, and the signing of the Agreement started. The NEEREF structure should be established by approximate analogy with EUREF. The entire research is based on the unified observation network which provides the initial observation data obtained not only by GNSS but also by other satellite and terrestrial observation techniques. The observation data will be transmitted to the data centers and/or analysis centers for primary treatment and/or final solution determination. The measurement data and processing results should be available to a wide range of users [Savinykh V.P. et al., 2014].

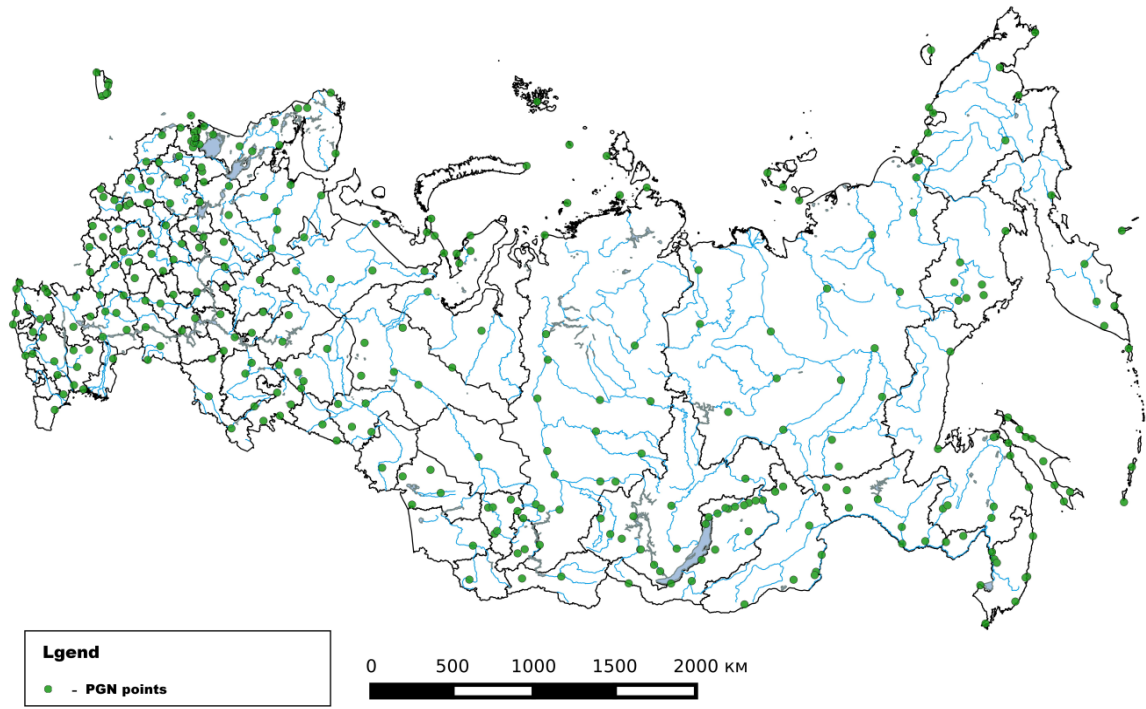

Fig.2 Russian Precise Geodetic Network

One of NEEREF objectives is to conduct research not only in the frame of conventional geodetic problems but also the related geophysical ones. For example, the interaction with geomagnetic observation networks, which is significant for better understanding of the interrelationship of the terrestrial and natural external processes, is proposed [Kaftan V.I., Krasnoperov R.I., 2015].

The first kinematic coordinate reference frame of Russia is a common work result of the above mentioned Russian institutions. The coordinate solutions were obtained based on the ITRF08 catalogue. The coordinate accuracy values of daily Bernese solutions were 0.8 and 1.7 mm for the horizontal and vertical components, accordingly. The velocity vector values of the sites of the Russian Fundamental Astro-Geodetic Network (FAGN) are derived from the data of continuous GPS observations conducted in 01/2010-12/2011. The velocities are determined from the time series analysis of two-year observations. The accuracy of determination of the displacement rates obtained from the time series of daily coordinate solutions attained 0.2-0.3 and 0.4 mm/yr for the horizontal and vertical components, respectively [Gorobets et al., 2012].

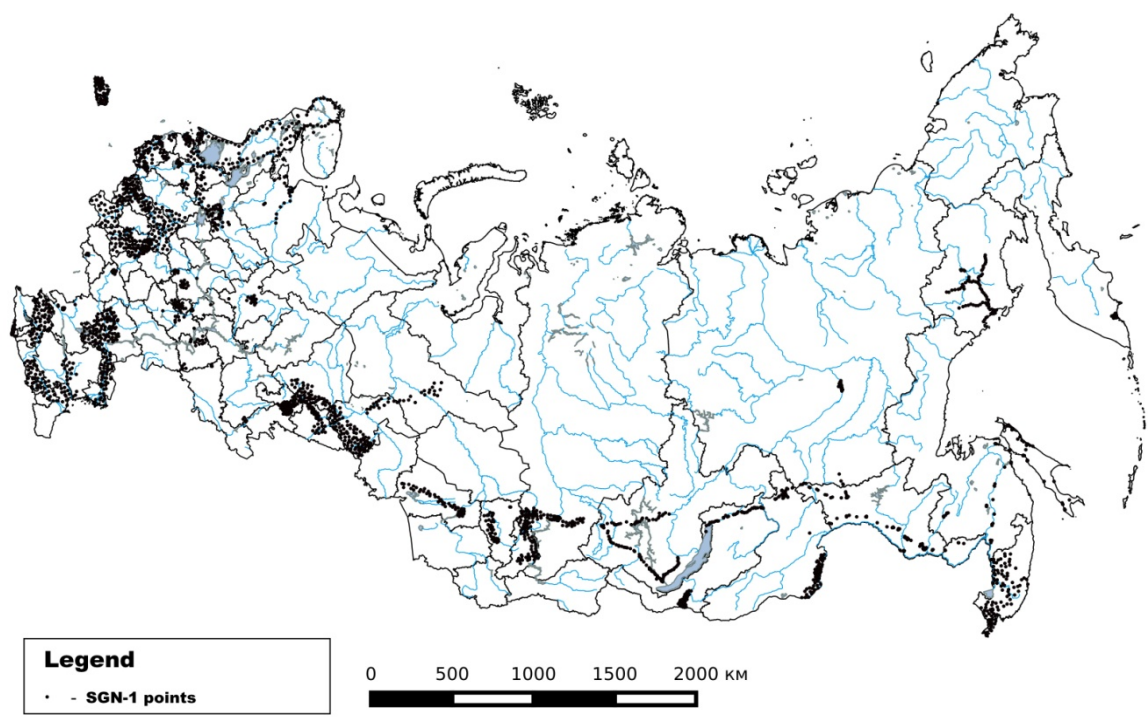

Fig.3 Russian 1st order Satellite Geodetic Network

National vertical precise reference frame (the Main height frame) is developing permanently. The modern state of the Russian vertical reference frame is presented in Fig.4. The Russian vertical reference frame was recently connected with national leveling networks of Belorussia, Finland and Norway. The works on the 1st and 2nd order leveling modernization are initiated at the Crimea territory. The Russian vertical reference frame realizes the Baltic normal height system of 1977.

The history of creation and development of the Russian leveling network is briefly described in [Basmanov A.V., 2013]. The oldest Russian leveling city network of Saint-Petersburg is inspected and reconstructed [Bogdanov A.S. et al., 2013].

The research on registration and identification of measurement gross errors on results of leveling network adjustment by parametric method is described in [Stoliarov I.A., 2013].

The development of state gravity network in Russia (USSR) has begun in the fourth decade of the 20th century. All gravity measurements of that epoch were performed at the four reference stations in Moscow, Pulkovo, Poltava and Kazan, that were directly connected to Potsdam station. The further advancements of gravity measurements resulted in creating the State Gravimetric Reference Frame of the First order during the period from 1965 to 1970.

The modern State gravimetric network was established using differential pendulum method from 1979 to 1994. More than 1000 stations were installed and observed during that time, creating the foundation for further densification of the network and gravimetric surveys. The First order network consisted from 11 fundamental stations where measurements were made using Russian ballistic gravimeters with high precision. The gravimetric network for epoch 1995 was created by combining measurements from fundamental and First order stations.

The special attention is given to the development of Fundamental gravimetric network in the last decade (see Fig.5). Newly created stations are included in complex sites of the FAGN and PGN and joined with First order stations of the previous period if possible. Thus repeated gravity measurements are being made over all national territory more than 20 years later that allowed the accuracy of the previous generation network to be checked and the crustal motion together with repeated leveling to be studied.

All measurements at the stations of the Fundamental Network are made using absolute ballistic gravimeters. New generation of gravity meters of GBL-M series were produced since 2009 in IAiE RAS. TSNIIGAiK has three instruments of the series that are used during regular field measurements. Some measurements are also made with FG-5, GBL-P, GABL-E and other gravity meters. The root mean square errors of the latest gravity measurement vary from 0.8 to 3.3 μGal at the stationary fundamental gravity points. The extension of the gravimetric network on the Russian part of Arctic has started in 2012. There are plans to develop a fundamental gravimetric network in Antarctica from 2014 to 2017 at the sites of active Russian Antarctic stations. National and international comparisons of absolute gravity meters are being made at Russian gravimetric stations.

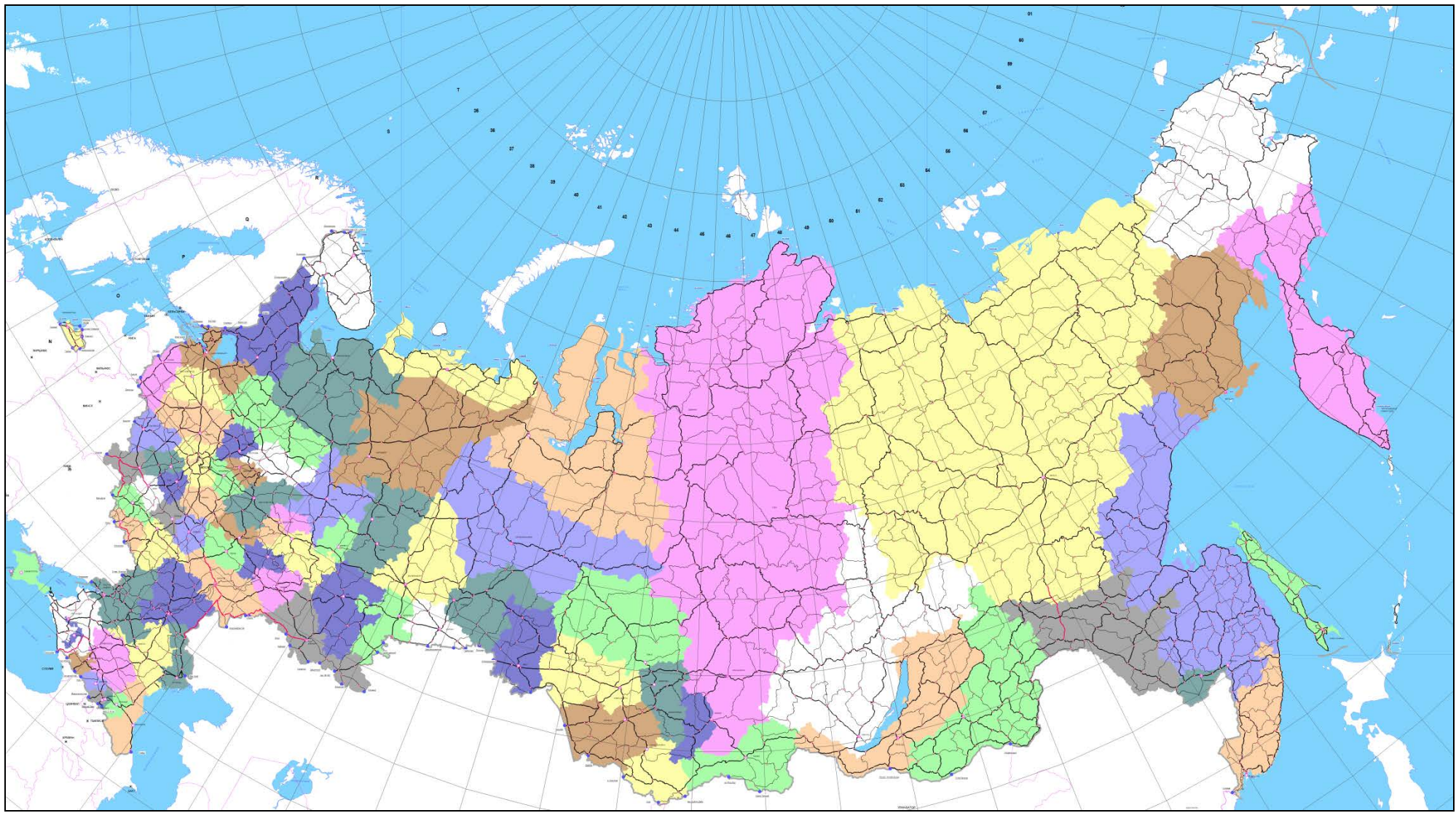

Fig.4 Main Russian vertical reference frame (1st and 2nd order precise levelings). Black solid lines are the 1st order leveling. Thin black lines are the 2nd order leveling. State border of Russia is indicated by gray line. Blue dots – connection points between national networks. Red lines – resent relevelings.

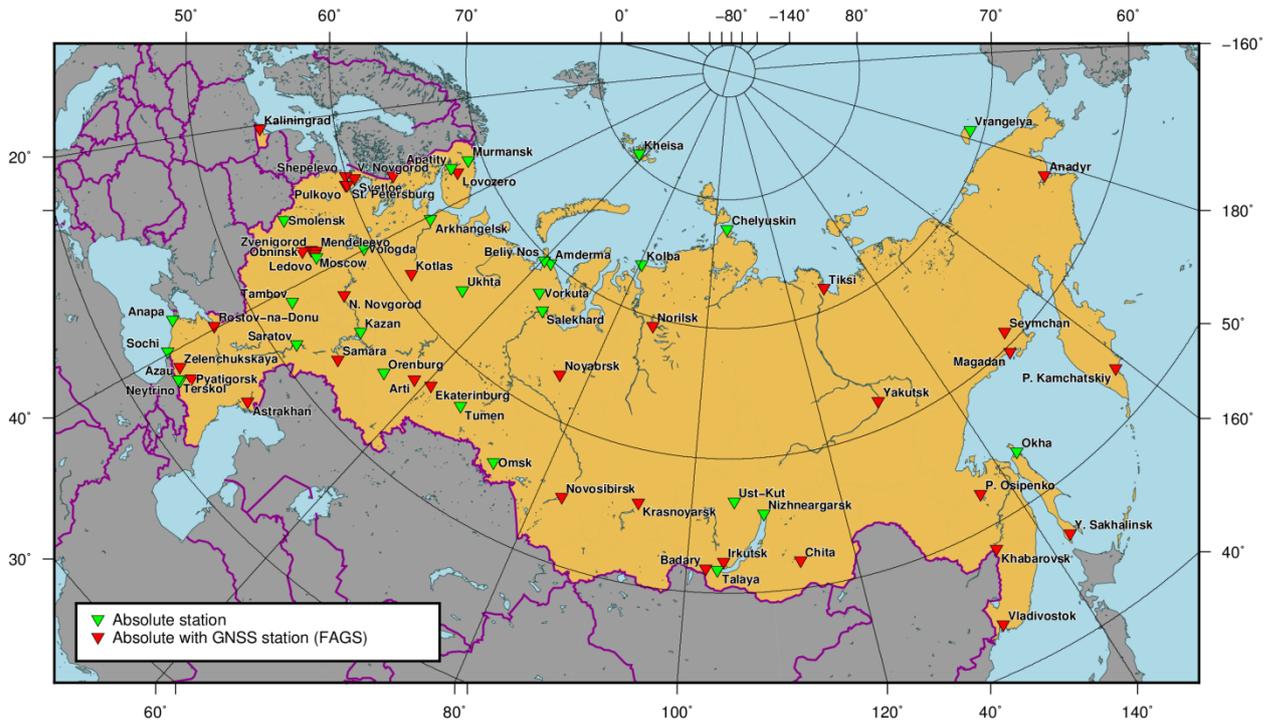

Fig.5 Russian Fundamental Gravity Network

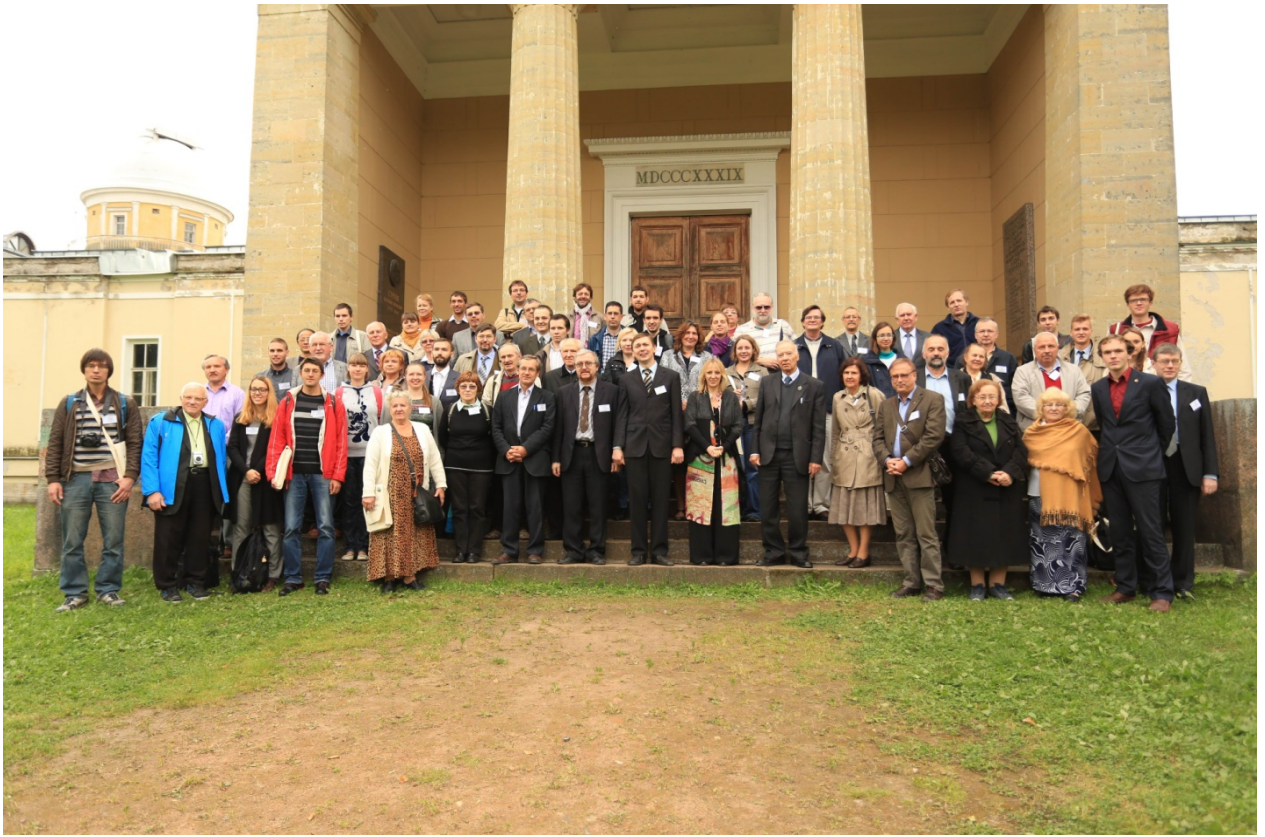

Fig.6. The Journées 2014 participants at Pulkovo Observatory 22 - 24 September 2014

The Journées 2014 "Systèmes de référence spatio-temporels", with the sub-title "Recent developments and prospects in ground-based and space astrometry" were organized at Pulkovo Observatory from 22 to 24 September 2014. The main purpose of the meeting is to provide an international forum for advanced discussion in the fields of space and time reference systems, Earth rotation, astrometry and time. These Journées are included in the program of celebrating of the 175th anniversary of the Pulkovo observatory. Common photo is presented in Fig.6.

The information on gravity study is presented in the Gravity Field section.

**References**

Abdrukhmanov R.Z., Demianov G.V., Kaftan V.I., Pobedinsky G.G. (2013) Methodical questions of creation of global and regional geodetic networks. Абдрахманов Р.З., Демьянов Г.В., Кафтан В.И., Побединский Г.Г. Методические вопросы построения глобальных и региональных геодезических сетей // Автоматизированные технологии изысканий и проектирования.- 2013.- №1(48).- с. 80-85.

Abdrukhmanov R.Z., Demianov G.V., Kaftan V.I., Pobedinsky G.G. (2013) Methodical questions of creation of global and regional geodetic networks. Абдрахманов Р.З., Демьянов Г.В., Кафтан В.И., Побединский Г.Г. Методические вопросы построения глобальных и региональных геодезических сетей // Автоматизированные технологии изысканий и проектирования.- 2013.- №2(49).- с. 67-70.

Basmanov A.V. (2013) History of creation and development of the Russian leveling network. Басманов А.В. История создания и развития нивелирной сети России / Физическая геодезия. Научно-технический сборник ЦНИИГАиК. – М.: Научный мир, 2013. – с.151-163

Boehm J., Malkin Z., Lambert S., Ma C. (2012a) Challenges and perspectives for celestial and terrestrial reference frame determination. 7th IVS General Meeting: Launching the Next-Generation IVS Network, Madrid, Spain, March 4-9 2012, Abstract's Book, 55. http://www.oan.es/gm2012/gm2012AbstractsFinal.pdf

Boehm J., Malkin Z., Lambert S., Ma C. (2012b) Challenges and Perspectives for TRF and CRF Determination. In: IVS 2012 General Meeting Proc., ed. D. Behrend, K.D. Baver, NASA/CP-2012-217504, 2012, 309-313. http://ivscc.gsfc.nasa.gov/publications/gm2012/boehm.pdf

Boehm J., Jacobs C., Arias F., Boboltz D., Bolotin S., Bourda G., Charlot P., A. de Witt, Fey A., Gaume R., Gordon D., Heinkelmmann R., Lambert S., Ma C., Nothnagel A., Malkin Z., Seitz M., Skurikhina E., Souchay J., Titov O. (2014) ICRF-3: Current Status and Interaction with the Terrestrial Reference Frame. Geophysical Research Abstracts, 2014, v. 16, EGU2014-3355.

Bogdanov A.S., Kaptüg V.B., Angelov K.K., Bryn M.Ya. (2013) On the activities to preserve markers of the St.Petersburg's first vertical reference frame. Богданов А.С., Капцюг В.Б., Ангелов К.К., Брынь М.Я. О работах по сохранению знаков первой высотной основы Санкт-Петербурга. Изыскательский вестник, 2013, № 2 (17), 3-15. http://www.spbogik.ru/vestnik/410--17-2-2013.html

Demianov G.V. (2013) Geodesy and GLONASS. Демьянов Г.В. Геодезия и ГЛОНАСС/ Физическая геодезия. Научно-технический сборник ЦНИИГАиК. – М.: Научный мир, 2013. – с.13-24

Demianov G.V., Kaftan V.I., Mazurova E.M., Tatevian S.K. (2011) Reference frames / National Report for the International Association of Geodesy of the International Union of Geodesy and Geophysics 2007-2010. Ed. by V.P.Savinikh and V.I.Kaftan // Международный научно-технический и производственный электронный журнал «Науки о Земле» (International scientific, technical and industrial electronic journal «Geo Science»).- 2011.- №1.- p. 7-13. http://geo-science.ru/wp-content/uploads/GeoScience-01-2011-p-05-36.pdf

Demianov G.V., Mayorov A.N., Pobedinskiy G.G. (2011a) Geodetic coordinate systems and its development on the base of global navigation satellite system usage. Демьянов Г.В., Майоров А.Н., Побединский Г.Г. Системы геодезических координат и их развитие на основе применения глобальных навигационных спутниковых систем // Геодезия и картография-2011.- №6.- с. 7-11. http://elibrary.ru/item.asp?id=21943050

Demianov G.V., Mayorov A.N., Pobedinskiy G.G. (2011b) Problems of permanent development of the state geodetic network and geocentric coordinate system of Russia. Демьянов Г.В., Майоров А.Н., Побединский Г.Г. Проблемы непрерывного совершенствования ГГС и геоцентрической системы координат России // Геопрофи.-2011.- №2.- с. 11-13 http://www.geoprofi.ru/default.aspx?mode=binary&id=1381

Demianov G.V., Mayorov A.N., Pobedinskiy G.G. (2011c) Problems of permanent development of the state geodetic network and geocentric coordinate system of Russia. Демьянов Г.В., Майоров А.Н., Побединский Г.Г. Проблемы непрерывного совершенствования ГГС и геоцентрической системы координат России (продолжение) // Геопрофи.-2011.- №3.- с. 21-27 http://www.geoprofi.ru/default.aspx?mode=binary&id=1405.

Demianov G.V., Mayorov A.N., Pobedinskiy G.G. (2011d) Problems of permanent development of the state geodetic network and geocentric coordinate system of Russia. Демьянов Г.В., Майоров А.Н., Побединский Г.Г. Проблемы непрерывного совершенствования ГГС и геоцентрической системы координат России (окончание) // Геопрофи.-2011.- №4.- с. 49-55 http://www.geoprofi.ru/default.aspx?mode=binary&id=1421

Demianov G.V., Mayorov A.N., Sermiagin R.A. (2011) The joint height system establishment tasks // Международный научно-технический и производственный электронный журнал «Науки о Земле» (International scientific, technical and industrial electronic journal «Geo Science»).-2011.- №1.- p. 37-39. http://issuu.com/geo-science/docs/01-2011

Dokukin P.A., Poddubskiy A.A. (2011) Analysis of satellite observation of etalon baseline. Докукин П.А., Поддубский А.А. Анализ спутниковых

наблюдений эталонного базиса // Землеустройство, кадастр и мониторинг земель. – 2011. – №1.- с.093-099. http://elibrary.ru/item.asp?id=18413163

Duev D.A. (2011) VLBI observations of GLONASS satellites. Proceedings of the International youth science forum «LOMONOSOV-2011» ed. A.I. Andreev, A.V. Andriyanov, E.A. Antipov, M.V. Chistyakova. — Moscow, MAKS Press, 2011.

Finkelstein AM.., Gayazov I.S., Ipatov A.V., Smolentsev S.G., Shargorodsky V.D., Mitriaev V.A. (2012) Laser ranging system Sadgen-TM installation at VLBI observatories. Финкельштейн А.М., Гаязов И.С., Ипатов А.В.,Смоленцев С.Г., Шаргородский В.Д., Митряев В.А.Оснащение обсерваторий РСДБ-комплекса «Квазар-КВО» квантово-оптическими системами «Сажень-ТМ» // Труды ИПА РАН, вып. 23, 2012. С.78-83. http://elibrary.ru/item.asp?id=20193540

Finkelstein A., Gayazov I., Shargorodsky V., Smolentsev S., Mitryaev V. (2012) Installing SLR systems at the "Quasar" VLBI network observatories // Proceedings of the 17th International Workshop on Laser Ranging. Frankfurt am Main, 2012. P.358-362.

Finkelstein A.M., Ipatov A.V., Bezrukov I.A., Gayazov I.S., Kaydanovsky M.N., Kurdubov S.L., Mishin V.Yu., Mikhailov A.G., Salnikov A.I., Surkis I.F., Skurikhina E.A., Yakovlev V.A. (2012) GLONASS rapid providing by universal time data at complex QUAZAR in the mode of e-VLBI. Финкельштейн А.М., Ипатов А.В., Безруков И.А., Гаязов И.С., Кайдановский М.Н., Курдубов С.Л,, Мишин В.Ю., Михайлов А.Г., Сальников А.И.,Суркис И.Ф., Скурихина Е.А., Яковлев В.А. Оперативное обеспечение системы ГЛОНАСС данными о всемирном времени в режиме e-РСДБ на радиоинтерферометрическом комплексе «Квазар-КВО» // Труды ИПА РАН, вып. 23, 2012. С.89-98. http://elibrary.ru/item.asp?id=20193542

Finkelstein A., Ipatov A., Gayazov I., Shargorodsky V., Smolentsev S., Mitryaev V., Diyakov A., Olifirov V., Rahimov I. (2012) Co-location of Space Geodetic Instruments at the "Quasar" VLBI Network Observatories. IVS 2012 General Meeting Proceedings, p.157-160.
http://ivscc.gsfc.nasa.gov/publications/gm2012/fnkelstein1.pdf

Finkelstein A.M., Ipatov A.V., Skurikhina E.A., Surkis I.F., Smolentsev S.G., Fedotov L.V. (2012) Geodinamic observations on the QUASAR VLBI network in 2009-2011. Astronomy Letters. 2012. V.38. No 6. P.349-398

Gayazov I., Mitryaev V., Smolentsev S., Rahimov I., Diyakov A., Shpilevski V., Pshenkin V., Rets Ya. (2012) SLR Observations at the "Quasar" network stations // International Scientific-Technical Conference WPLTN-2012. Book of abstracts. 2012. P. 41-42.

Gayazov I.S., Goubanov V.S., Bratseva O.A., Kourdoubov S.L. (2012) Software design for join processing of different observation tips. Гаязов И.С., Губанов В.С., Братцева О.А., Курдубов С.Л. Разработка программных средств совместной обработки различных типов наблюдений // Труды ИПА РАН, вып. 23, 2012. С. 136-141. http://elibrary.ru/item.asp?id=20193549

Gayazov I.S., Souvorkin V.V. (2012) Determination of transformation parameters between State geocentric coordinate system and International terrestrial reference frame. Гаязов И.С., Суворкин В.В. Определение параметров связи Государственной геоцентрической системы координат и Международной земной системы координат // Труды ИПА РАН, вып. 23, 2012. С.47-54. http://elibrary.ru/item.asp?id=20193536

Gorobets V.P. (2013) Transformation determination between geocentric coordinate system and SK-95. Горобец В.П. Определение связи между геоцентрической системой координат и СК-95 / Физическая геодезия. Научно-технический сборник ЦНИИГАиК. – М.: Научный мир, 2013. – с.95-98

Gorobets V.P., Demianov G.V., Mayorov A.N., Pobedinskiy G.G. (2012) Results of creation of the state geocentric coordinate system of Russian Federation in the frame of the Federal special purpose program GLONASS. Горобец В.П., Демьянов Г.В., Майоров А.Н., Побединский Г.Г. Результаты построения государственной геоцентрической системы координат Российской Федерации в рамках Федеральной целевой программы «ГЛОНАСС». Геодезия и картография, 2012, №2, с. 53-57 http://elibrary.ru/item.asp?id=21760750

Gorobets V.P., Demyanov G.V., Mayorov A.N., Pobedinsky G.G. (2013) Recent state and development direction of the geodetic provision of Russian Federation. Горобец В.П., Демьянов Г.В., Майоров А.Н., Побединский Г.Г. Современное состояние и направления развития геодезического обеспечения РФ. Системы координат (начало) // Геопрофи. - 2013. - № 6. – с. 4 – 9. http://www.geoprofi.ru/default.aspx?mode=binary&id=1721

Gorobets V.P., Demyanov G.V., Mayorov A.N., Pobedinsky G.G. (2014) Recent state and development direction of the geodetic provision of Russian Federation. Горобец В.П., Демьянов Г.В., Майоров А.Н., Побединский Г.Г. Современное состояние и направления развития геодезического обеспечения РФ. Высотное и гравиметрическое обеспечение (окончание) // Геопрофи. - 2014. - № 1. – с. 5 – 11. http://www.geoprofi.ru/default.aspx?mode=binary&id=1739

Gorobets V.P., Demianov G.V., Mayorov A.N., Pobedinsky G.G. (2013) State geocentric coordinate system. Problems of support and development in the frame of the new Federal program "GLONASS". Горобец В. П., Демьянов Г. В., Майоров А. Н., Побединский Г. Г.. Государственная геоцентрическая система координат. Проблемы поддержания и развития в рамках новой ФЦП «ГЛОНАСС» / 14-й Международный научно-промышленный форум «Великие реки'2012». Труды конгресса. Том 1. Нижний Новгород, ННГАСУ, 2013, с. 385 - 389. http://www.nngasu.ru/cooperation/2012-tom1.pdf

Gorobets V.P., Demianov G.V., Pobedinsky G.G., Yablonsky L.I. (2013) State geocentric coordinate system of Russian Federation. The proceedings include presentations presented at plenary sessions of VIII International scientific congress «Interexpo GEO-Siberia-2013». Горобец В.П., Демьянов Г.В.,

Побединский Г.Г., Яблонский Л.И. Государственная геоцентрическая система координат Российской Федерации» / Интерэкспо ГЕО-Сибирь-2013. IX Междунар. науч. конгр., 15–26 апреля 2013 г., Новосибирск: Пленарное заседание: сб. материалов в 2 т. Т. 2. – Новосибирск: СГГА, 2013. с. 76 – 94. http://www.nngasu.ru/cooperation/2012-tom1.pdf

Ipatov A., Gayazov I., Smolentsev S. (2012) "Quasar" VLBI network observatories as co-location sites // International Scientific-Technical Conference WPLTN-2012. Book of abstracts. 2012. P.39-40.

Ipatov A., Ivanov D., Ilin G., Fedotov L., Gayazov I., Kaidanovsky M., Mardyshkin V., Salnikov A., Smolentsev S., Surkis I. (2012) The Russian VLBI-network of new generation // International Scientific-Technical Conference WPLTN-2012. Book of abstracts. 2012. P.38.

Ipatov A.V., Finkelstein A.M., Gayazov I.S., Mardishkin V.V., Mikhailov A.G., Surkis I.F., Ilyin G.N., Ivanov G.V., Kaydanovsky M.N., Salnikov A.I., Fedotov L.V. (2012) VLBI systems for GLONASS support. Ипатов А.В., Финкельштейн А.М., Гаязов И.С., Мардышкин В.В., Михайлов А.Г., Суркис И.Ф., Ильин Г.Н., Иванов Д.В., Кайдановский М.Н., Сальников А.И., Федотов Л.В.. РСДБ системы для поддержки глобальной навигационной системы ГЛОНАСС // Труды ИПА РАН, вып. 24, 2012. С. 12-23. http://elibrary.ru/item.asp?id=20329788

Jacobs, C. S., Arias, F., Boboltz, D., Boehm, J., Bolotin, S., Bourda, G., Charlot, P., de Witt, A., Fey, A., Gaume, R., Gordon, D., Heinkelmann, R., Lambert, S., Ma, C., Malkin, Z., Nothnagel, A., Seitz, M., Skurikhina, E., Souchay, J., Titov, O. (2014) ICRF-3: Roadmap to the Next Generation ICRF. In: Proc. Journees 2013 Systemes de Reference Spatio-temporels, Paris, France, 16-18 Sep 2013, Ed. N.Capitaine, Paris, 2014, 51-56

Kaftan V.I. (2011) The necessity of creation of the state geocentric coordinate reference system. Кафтан В.И. Необходимость установления государственной геоцентрической системы отсчета// Кадастр недвижимости .- 2011.- №3.-(24).-с.87-91 http://www.roscadastre.ru/?id=720

Kaftan V.I., Krasnoperov R.I. (2015) Geodetic observations at geomagnetic observatories. Geomagnetism and Aeronomy. 2015. Vol. 55. No 1. PP. 118-123.

Karpik A.P., Gienko E.G., Kosarev N.S. (2014) Error source analyses of coordinate transformation of points of satellite geodetic networks. Карпик А.П., Гиенко Е.Г., Косарев Н.С. Анализ источников погрешностей преобразования координат пунктов спутниковых геодезических сетей. Известия высших учебных заведений. Геодезия и аэрофотосъемка. 2014. № S4. С. 55-62. http://www.miigaik.ru/journal.miigaik.ru/arhiv_zhurnalov/vypuski_za_2014_/20150120172913-1664.pdf

Khodakov P., Basmanov A. (2014) On the state of geodetic points in Russian Arctic (on example of Laptev and East Siberian seas) Ходаков П. А., Басманов А.В. О состоянии геодезических пунктов в секторе Российской Арктики (на примере морей Лаптевых и Восточно-Сибирского) Известия высших учебных заведений. Геодезия и аэрофотосъемка. 2014. № 4. С. 21-25.

http://www.miigaik.ru/journal.miigaik.ru/arhiv_zhurnalov/vypuski_za_2014_/20150120172913-4566.pdf

Krasna H., Malkin Z., Boehm J. (2013) Impact of non-linear station motions on the ICRF. IAG Scientific Assembly 2013, Potsdam, Germany, 1-6 Sep 2013, Book of Abstracts, p. 300. http://www.iag2013.org/IAG_2013/Welcome_files/Abstracts_iag_2013.pdf

Krasna H., Malkin Z., Boehm, J. (2014) Impact of seasonal station displacement models on radio source positions. In: Proc. Journees 2013 Systemes de Reference Spatio-temporels, Paris, France, 16-18 Sep 2013, Ed. N.Capitaine, Paris, 65-68.

Krilov V.I., Kokina T.N., Mendosa A.D. (2014) Project of a use of a radio-telescope in Sierra Negra for VLBI method realization. Крылов В.И., Кокина Т.Н., Мендоза А.Д. Проект использования радиотелескопа в Сьерра Негра для реализации метода РСДБ. // Сборник статей по итогам научно-технических конференций: Приложение к журналу «Известия вузов. Геодезия и аэрофотосъёмка», № 6. – вып.7, в двух частях. Часть первая. 2014. С. 10-13

Kuzin S., Tatevian S. (2011) INASAN analysis center status report, IDS AWG meeting 23-29 May, Paris, 2011. http://ids-doris.org/report/meeting-presentations/ids-awg-05-2011.html

Kuzin S., Tatevian S. (2013) INA AC processing status and plans for ITRF2013, IDS AWG meeting, 4-5, April, Toulouse, France, 2013. http://ids-doris.org/images/documents/report/AWG201304/IDSAWG1304-Kuzin-INASANprocessingStatus.pdf

Liu J.-C., N. Capitaine, S.B. Lambert, Z. Malkin, Z. Zhu. (2012) Systematic effect of the Galactic aberration on the ICRS realization and the Earth orientation parameters. Astron. Astrophys., 2012, v. 548, A50. DOI: 10.1051/0004-6361/201219421. http://www.aanda.org/index.php?option=com_article&access=standard&Itemid=129&url=/articles/aa/abs/2012/12/aa19421-12/aa19421-12.html

Malkin, Z. (2011a) Pulkovo IVS Analysis Center (PUL) 2010 Annual Report. In: IVS 2010 Annual Report, Eds. D. Behrend, K. D. Baver, NASA/TP-2011-215880, 2011, 247-249. ftp://ivscc.gsfc.nasa.gov/pub/annual-report/2010/pdf/acpul.pdf

Malkin Z.M. (2011b) The Influence of Galactic Aberration on Precession Parameters Determined from VLBI Observations. Astronomy Reports, 2011, Vol. 55, No. 9, 810-815. DOI: 10.1134/S1063772911090058

Malkin Z. (2012a) Connecting terrestrial to celestial reference frames. In: IAU XXVIII General Assembly, 2012, Abstract Book, 918-919. http://www.referencesystems.info/iau-joint-discussion-7.html

Malkin Z. (2012b) Connecting terrestrial to celestial reference frames. Proc. IAU, 2012, Vol. 10, Issue H16, 223-224.

Malkin Z. (2012c) On the impact of the Galactic aberration on VLBI-derived precession model. In: Schuh H., Boehm S., Nilsson T., Capitaine N. (Eds.) Proc.

Journees 2011 Systemes de Reference Spatio-temporels, Vienna, Austria, 19-21 Sep 2011, Vienna: Vienna University of Technology, 2012, 168-169.

Malkin Z. (2012d) The current best estimate of the Galactocentric distance of the Sun based on comparison of different statistical techniques. arXiv:1202.6128, 2012.

Malkin, Z. (2012e) Pulkovo IVS Analysis Center (PUL) 2011 Annual Report. In: IVS 2011 Annual Report, Eds. D. Behrend, K. D. Baver, NASA/TP-2012-217505, 2012, 256-258. ftp://ivscc.gsfc.nasa.gov/pub/annual-report/2011/pdf/acpul.pdf

Malkin Z. (2013a) A new approach to the assessment of stochastic errors of radio source position catalogues. Astron. Astrophys., 2013, v. 558, A29. DOI: 10.1051/0004-6361/201322334

Malkin Z. (2013b) On Application of the 3-Cornered Hat Technique to Radio Source Position Catalogs. Proc. 21st Meeting of the EVGA, Eds. N. Zubko, M. Poutanen, In: Rep. Finn. Geod. Inst., 2013, 2013:1, 175-177. ISBN: 978-951-711-296-3 http://evga.fgi.fi/sites/default/files/u3/Proceedings_EVGA2013.pdf

Malkin Z. (2013c) Using modified Allan variance for time series analysis. In: Reference Frames for Applications in Geosciences, Z. Altamimi, X. Collilieux (eds.), IAG Symposia, 2013, v. 138, 271-276. DOI: 10.1007/978-3-642-32998-2_39 http://link.springer.com/chapter/10.1007/978-3-642-32998-2_39 (ГАО: G1041)

Malkin Z.M. (2013d) Analysis of Determinations of the Distance between the Sun and the Galactic Center, Astronomy Reports, 2013, v. 57, No. 2, 128-133. DOI: 10.1134/S1063772913020078

Malkin Z. (2013t) Statistical analysis of the determinations of the Sun's Galactocentric distance. In: Advancing the Physics of Cosmic Distances, Proc. IAU Symp. 289, R. de Grijs (Ed.), 2013, 406-409. DOI: 10.1017/S1743921312021825

Malkin Z.M. (2013f) Some results of the statistical analysis of the Sun galaxy-centric distance determination. Малкин З.М. Некоторые результаты статистического анализа определений галактоцетрического расстояния Солнца. Тр. Всероссийской астрометрической конф. "Пулково-2012", Изв. ГАО, 2013, No. 220, 401-406.

Malkin Z.M. (2013g) Random error determination of radio-sources catalogue coordinates. Малкин З.М. Об определении случайных ошибок каталогов координат радиоисточников. Тр. Всероссийской астрометрической конф. "Пулково-2012", Изв. ГАО, 2013, No. 220, 59-64.

Malkin Z. (2014a) Errors of radio source position catalogs. In: Proc. Journees 2013 Systemes de Reference Spatio-temporels, Paris, France, 16-18 Sep 2013, Ed. N.Capitaine, Paris, 2014, 69-71.

Malkin Z. (2014b) The implications of the Galactic aberration in proper motions for the Celestial Reference Frame. MNRAS, 2014, 445(1), 845-849. DOI: 10.1093/mnras/stu1796.

Malkin Z. (2014c) The Galactic aberration constant. In: Proc. Journees 2013 Systemes de Reference Spatio-temporels, Paris, France, 16-18 Sep 2013, Ed. N.Capitaine, Paris, 44-45.

Malkin Z., Jacobs C., and IAU ICRF3 Working Group. (2014) The ICRF-3: Status, plans, and progress on the next generation International Celestial Reference Frame. In: Journees 2014 Systemes de Reference Spatio-temporels, St. Petersburg, Russia, 22-24 Sep 2014, Book of Abstracts, 3.

Malkin Z., Schuh H., Ma C., Lambert S. (2012) Interaction between celestial and terrestrial reference frames and some considerations for the next VLBI-based ICRF. In: Schuh H., Boehm S., Nilsson T., Capitaine N. (Eds.) Proc. Journees 2011: Earth rotation, reference systems and celestial mechanics: Synergies of geodesy and astronomy, Vienna, Austria, Sep 19-21, Vienna: Vienna University of Technology, 2012, 66-69. http://syrte.obspm.fr/jsr/journees2011/malkin1.pdf

Malkin Z., Sokolova Ju. (2012) Assessment of stochastic errors of radio source position catalogues. In: IAU XXVIII General Assembly Abstract Book, 948. http://www.referencesystems.info/iau-joint-discussion-7.html

Malkin Z., Sokolova Ju. (2013) Pulkovo IVS Analysis Center (PUL) 2012 Annual Report. In: IVS 2012 Annual Report, Eds. K.D. Baver, D. Behrend, K.L. Armstrong, NASA/TP-2013-217511, 2013, 305-308. ftp://ivscc.gsfc.nasa.gov/pub/annual-report/2012/pdf/acpul.pdf

Malkin Z., Sokolova Yu. (2014) Pulkovo IVS Analysis Center (PUL) 2013 Annual Report. In: IVS 2013 Annual Report, Eds. K.D. Baver, D. Behrend, K.L. Armstrong, NASA/TP-2014-217522, 2014, 312-315.

Malkin Z., Sun J., Boehm J., Boehm S., Krasna H. (2013a) Searching for an Optimal Strategy to Intensify Observations of the Southern ICRF sources in the framework of the regular IVS observing programs. In: Proc. 21st Meeting of the EVGA, Eds. N. Zubko, M. Poutanen, Rep. Finn. Geod. Inst., 2013, 2013:1, 199-203. ISBN: 978-951-711-296-3 http://evga.fgi.fi/sites/default/files/u3/Proceedings_EVGA2013.pdf

Malkin Z., Sun J., Boehm J., Boehm S., Krasna H. (2013b) Searching for optimal strategy to intensify observations of the Southern ICRF sources in the framework of the regular IVS observing programs. In: 21st Meeting of the European VLBI Group for Geodesy and Astrometry, Espoo, Finland, March 5-8, 2013, Book of abstracts, 17. http://evga.fgi.fi/sites/default/files/Abstract_book.pdf

Mazurov B.T. (2014) Theoretical foundations of a cable bridge dynamics from geodetic observation. Мазуров Б.Т. Теоретические основы моделирования динамики вантовых мостов по геодезическим наблюдениям. Интерэкспо Гео-Сибирь. 2014. Т. 1. № 1. С. 170-175.

Mazurova E., A. Karpik. (2014) The recent progress of the Russian terrestrial reference frame, IAG Commission 1 Symposium: Reference Frames for Applications in Geodetic Science, 13-17 October, 2014, Luxembourg. http://geophy.uni.lu/users/tonie.vandam/REFAG2014/SESS_IV_Reg_Ref_Frames/Mazurova.pdf

Mazurova E., A. Mikhaylov. (2013) Algorithm for transforming the coordinates of lunar objects while changing from various coordinate systems into the selenocentric one, Geophysical Research Abstracts, Vol.15, EGU2013-PREVIEW, EGU General Assembly 2013, 07-12 April, Vienna, Austria. http://adsabs.harvard.edu/abs/2013EGUGA..15.2472M

Salnikov P.A. (2011) Development of technique of precise leveling. Сальников П.А. Разработка методики высокоточного геометрического нивелирования. Международный научно-технический и производственный журнал «Науки о Земле» - 2011. - №2 - с.28-34. http://issuu.com/geo-science/docs/02-2011

Savinykh V., Bykov V., Karpik A., Moldobekov B., Pobedinsky G., Demianov G., Kaftan V., Malkin Z., Steblov G. (2013) Organization of the North East Eurasia Reference Frame, Савиных В.П., Быков В.Г., Карпик А.П., Молдобеков Б., Побединский Г.Г., Демьянов Г.В., Кафтан В.И., Малкин З.М., Стеблов Г.М., Татевян С.К. Организация Международной комиссии по региональной земной геодезической основе Северо-Восточной Евразии / «Фундаментальное и прикладное координатно-временное и навигационное обеспечение» (КВНО-2013), 15-19 апреля 2013 г., Санкт-Петербург, Россия. Тезисы докладов. Санкт-Петербург: ИПА РАН, 2013.- с.185-188

Savinykh V.P., Bykov V.G., Krapik A.P., Moldobekov B., Pobedinsky G.G., Demianov G.V., Kaftan V.I., Malkin Z.M., Steblov G.M. (2014) Organization of the North East Eurasia reference frame.- International scientific, technical and industrial electronic journal «Geo Science» 01/2014; №1/2-2014:16-25. http://issuu.com/geo-science/docs/geoscience_1-2-2014

Sokolova Ju., Malkin Z. (2012) New Pulkovo combined catalogues of the radio source positions. In: IAU XXVIII General Assembly, 2012, Abstract Book, 937-938. http://www.referencesystems.info/iau-joint-discussion-7.html

Sokolova Ju.R., Malkin Z.M. (2013a) Impact of the correlation information on the orientation parameters between celestial reference frames. Соколова Ю.Р., Малкин З.М. О влиянии учета корреляционной информации на параметры взаимной ориентации небесных систем отсчета. Вестник СПбГУ, Сер. 1, 2013, Вып. 4, 146-151. http://vestnik.unipress.ru/pdf13/s01/s01v4_13.pdf

Sokolova Y., Malkin Z. (2013b) Impact of the correlation information on the orientation parameters between celestial reference frames. IAG Scientific Assembly 2013, Potsdam, Germany, 1-6 Sep 2013, Book of Abstracts, 303. http://www.iag2013.org/IAG_2013/Welcome_files/Abstracts_iag_2013.pdf

Sokolova Yu. R., Malkin Z. M. (2014) Pulkovo Combined Catalogue of Radio Source Positions PUL 2013. Соколова Ю.Р., Малкин З.М. Пулковский сводный каталог координат радиоисточников PUL 2013. Письма в Астрон. журн., 2014, т. 40, N 5, 306-315. DOI: 10.7868/S0320010814050040

Stoliarov I.A. (2013) On the question of registration and identification of measurement gross errors on results of leveling network adjustment by parametric method. Столяров И.А. К вопросу обнаружения и идентификации грубых ошибок измерений по результатам уравнивания нивелирных сетей параметрическим способом / Физическая геодезия. Научно-технический сборник ЦНИИГАиК. – М.: Научный мир, 2013. – с.122-134.

Tatevian S., Kuzin S. (2011) On the combined use of GPS/GLONASS techniques for the development of the Russian geodetic reference network, Advances in Geosciences, Vol. 26: Solid Earth (2010), Ed. Kenji Satake, World Scientific Publishing Company, 2011, pp. 23-32.

Tatevian S.K., Kuzin S.P., Demjanov G.V. (2013) On the Use of GLONASS for Precise Positioning, Journal of Remote sensing Technology, Vol.1, Iss. 2, pp 31-35 (2013). http://www.bowenpublishing.com/jrst/scopepaper.aspx?ScopeID=2866&researchfield=Highly Accurate Navigation and Position Technique

Tornatore V., Haas R., Duev D., Pogrebenko S., Casey S., Molera Calvès G., Keimpema A. (2011) Single baseline GLONASS observations with VLBI: data processing and first results. Proceedings of the 20th EVGA Meeting and 12th Analysis Workshop, 29-31 March 2011.

Vdovin V.S. (2013) The PZ-90 System. The main development results, its role and place in the national system of coordinate, time and navigation support. Вдовин В.С. Система ПЗ-90. Основные итоги развития. Роль и место в единой системе координатно-временного и навигационного обеспечения страны. // Труды ИПА РАН, вып. 27, 2013. С. 132-142. http://elibrary.ru/item.asp?id=21511714

# Gravity Field

## Kaftan V.[1], Sermiagin R.[2], Zotov L.[3]

[1]Geophysical Center of the Russian Academy of Sciences, Moscow, Russia
[2]Federal Scientific-Technical Center of Geodesy, Cartography and Spatial Data Infrastructure
[3]Sternberg Astronomical Institute, Moscow Univercity, Moscow, Russia

Problems of the Earth's dynamics in relation to General Relativity effects are studied by Kopeikin et al. (2014). A concept of Relativistic Geoid is proposed.

General problems of space geodetic measurements for global changes monitoring are discussed in [Tatevian et al., 2012, 2014a, b].

Studies of the Geocenter dynamics by the analysis of the measurements of the GPS and DORIS satellite systems were performed by Valeev et al. (2011).

The specialities of deformation of continental and ocean lithosphere revealed by geodetic technique are considered as an evidence of the north movement of the Earth's core in [Goncharov et al., 2011].

The problems of modern figure of the Earth theory are discussed in [Pik & Yurkina, 2013]. The Molodensky theory is one of the few precise methods of the Earth shape theory. However, it is unfairly neglected or insufficiently used. Many resent publications disseminate an idea that modern geodesy cannot dispense with Gauss-Listing geoid and Molodensky theory is not reliable enough. As a result, Japan has changed its height system from normal to ortometric.

An example of a departure of right reason and logic is a spreading of special and general relativity theories. This and several other examples of this kind are related to insufficiency of mathematical education in many countries of the world. Computation substitutes mathematic knowledge. The authors [Pik & Yurkina, 2013] give the definition and explanation of a normal height and quazigeoid height. They provide the formulation of disturbing gravity potential using refined gravity anomalies and develop formulas of deflection.

The representation of gravity potential coefficients through gravity anomaly coefficients is presented in [Brovar, 2013].

Modern geodetic GNSS technologies presuppose the necessity of the knowledge of the quasigeoid height having an accuracy of about $5*10^{-5}$. The theoretic assumption of V.V. Brovar was checked with the special purpose in view. A numerical experiment approved the accuracy of V.V. Brovar method not less than $5*10^{-5}$ [Brovar & Stolarov, 2013]. It is equivalent to 1 mm for the Caucasus test region.

A spherical approximation is the basis of a majority of formulae in physical geodesy. However modern accuracy of the disturbing potential definition demands an ellipsoidal approximation. The purpose of the work [Mazurova & Yurkina, 2011] is to construct the Green's function for an ellipsoidal Earth. The Green's function depends only on surface geometry with given boundary values. Thus, it can be calculated irrespective to gravimetric data completeness. Any changes in gravitational data are not reflected in the Green's function and if it is already known, the changes can be just considered. Therefore the solution can become useful for the definition of the disturbing potential of an ellipsoidal Earth.

The outcomes of a research related to the development of a methodology for assessing the quality of models of the Earth gravity field used in geodesy and adjacent areas are presented. Requirements for such models were analyzed. Questions relating to the classification of gravity models by various characteristics were considered. It is shown that the quality of the models includes the quality of their design and implementation. The authors [Nepoklonov et al., 2014] have established connection between the quality of implementation of the models and their main functional and performance features. The general scheme of quality evaluation of modern gravity models is proposed. The authors propose a technique for estimation of accuracy of gravity models as one of the main characteristics defining their quality.

Classical methods of the definition of anomaly height demand knowledge of continuous faultless values of a gravity anomaly on the total surface area of Planet Earth. In fact, the M.S. Molodensky's combined method is used in practice. According to the method, the surface of the Earth is divided into some "near" and farfield zones.

As a rule, a detailed gravimetric surveying with the subsequent definition of the transforms of the gravitational field is performed by numerical integration in the "near" zone. The influence of farfield zone is considered by decomposition of a gravity anomaly in a series of the spherical functions. The transforms of the gravitational field are very difficult to calculate with the classical methods of numerical integration—even with accuracy of zero approximation and extremely with accuracy of the first and the subsequent approximations. Now wavelet-transformation has wide popularity at digital information processing. The algorithms of calculation of the height anomaly with accuracy of the first approximation of the M.S. Molodensky's theory are executed on the basis of wavelet-transformation. The results of calculation transforms of the gravitational field are presented for Central Alps area [Mazurova & Lapshin, 2011].

A method of discrete linear transformations is used effectively to calculate deflections of the vertical on the basis of discrete values of gravity anomaly. Fourier Transformation algorithms, Short-Time Fourier Transformation, and wavelet-transformation are used for realization of the method. The results of calculation of deflections of the vertical that were executed on the basis of classical Fourier Transform (FT), Short-Time Fourier Transform (STFT), and Wavelet-transformation (WT) are presented as 3D-models which illustrate action of the Heisenberg's uncertainty principle in the specified algorithms [Mazurova et al., 2013].

A new free-fall absolute ballistic gravimeter ABG-VNIIM-1 was fabricated at the D.I. Mendeleyev Research Institute for Metrology (VNIIM). For this gravimeter the authors [Vitushkin & Orlov, 2014a, 2014b] have developed an original mechanical system of ballistic unit, a compact iodine-stabilized in frequency Nd:YVO$_4$/KTP diode-pumped solid-state laser at the wavelength of 532 nm and the laser interferometer. The path of free fall of the test body in a vacuum chamber is about 10 cm. The electronic system for the fast acquisition of the length and time intervals during the free fall is based on the NI PXI platform. A special software GROT was developed to control of all the systems and to evaluate the measured gravity acceleration. A passive vibration isolation of the reference reflector in the laser interferometer is based on the seismometer. The gravimeter ABG-VNIIM-1 was tested at the gravimetric site "Lomonosov-1" at the Lomonosov branch of VNIIM. The estimated total instrumental uncertainty of ABG VNIIM-1 was determined to be $2 \cdot 10^{-8}$ m s$^{-2}$. The typical residuals in the least square evaluation of the trajectory of the test body in a single drop at the "Lomonosov-1" site are from 0.3 to 0.8 nm.

Absolute gravity determinations were determined by the Federal Scientific-Technical Center of Geodesy, Cartography and Spatial Data Infrastructure from 2011 to 2014 at 35 gravity stations of Russia. Especial efforts were done for the North territory of Russia, sea shore and islands of the Arctic Ocean. Part of stations is placed at permafrost territory. Repeated absolute gravity observation was performed at 10 stations of geodynamic test areas and FAGN stations.

The Russian-Finland comparisons of absolute gravity meters were done in June-July 2013 in the frame of international cooperation between the Federal Scientific-Technical Center of Geodesy, Cartography and Spatial Data Infrastructure and Finnish Geodetic Institute. Five absolute meters of four institutions were used in the comparison. It were FG5x-221 (Finnish Geodetic Institute), FG5-110 and GBL-M-002 (TsNIIGAiK), GABL-PM (Institute of Automation and Electrometry, Siberian Branch of the Russian Academy of Sciences), and GAPL-M (Niimorgeofizika-Service.Com.).

The measurements were executed at six pillars of the four fundamental points of FAGN (two of them are IGS points) located in different physical-geographic conditions: Pulkovo, Svetloe, TsNIIGAiK (pillars 110A and 109A), and Zwenigorod (pillars A and B).

The gravity meter FG5x-221 is a primary etalon of Finland. It took part in the International Comparison of Absolute Gravity meters (ICAG2013) at Walferdange (Luxemburg) on November 2013. In such a way the gravity unit transfer from international etalon to the Russian FAGN stations was performed taking into account the offsets of every Russian gravity meters.

The Federal Scientific-Technical Center of Geodesy, Cartography and Spatial Data Infrastructure is developing a new global gravity model. At the first step of the work the update digital relief model was developed using digital topographic maps of the territory of Russia. The main steps of relief model creation are:
- Estimation of data sources and a choice of the best
- Combination of data sorces
- Accuracy estimation of the developed model.

An accuracy estimation was prepared using independent control data – geodetic network points, leveling benchmarks etc.

A new digital elevation model called RDTM2014.0 was constructed and tested. The mean difference between the model and control points was received equal to -1.6 m, mean absolute difference – 1.8 m, and median difference - -0.4 m.

The model is used as a base for quazigeoid model creation.

Wavelet representation is analyzed as the variant of new gravitation and quazigeoid models creation.

Modern state verification schedule for free fall acceleration is criticized in [Staklo et al., 2014]. Autors discussed disadvantages of the unit etalon conception proposed in recent national standard of 2012. Group national etalons are proposed to create on a base of the Federal fundamental astro-geodetic and gravimetric networks.

State and perspectives of modern instrumental gravimetry is recounted, a historical review of foreign works was performed by Soviet and Russian specialists [Basmanov et al., 2011]. Main historical moments of creation of state gravimetric network are presented. The need of taking into account the world experience at carrying out gravimetric works is noted.

Coseismic gravity changes, that mainly occur due to vertical deformation of layer boundaries with density contrast (i.e. surface and Moho) were detected using the Gravity Recovery and Climate Experiment (GRACE) satellites for the 2013 May 24 Okhotsk deep-focus earthquake (Mw8.3) [Tanaka et al., 2015]. This enables to suggest GRACE as a perspective tool to map vertical ground movements of deep earthquakes over both land and ocean.

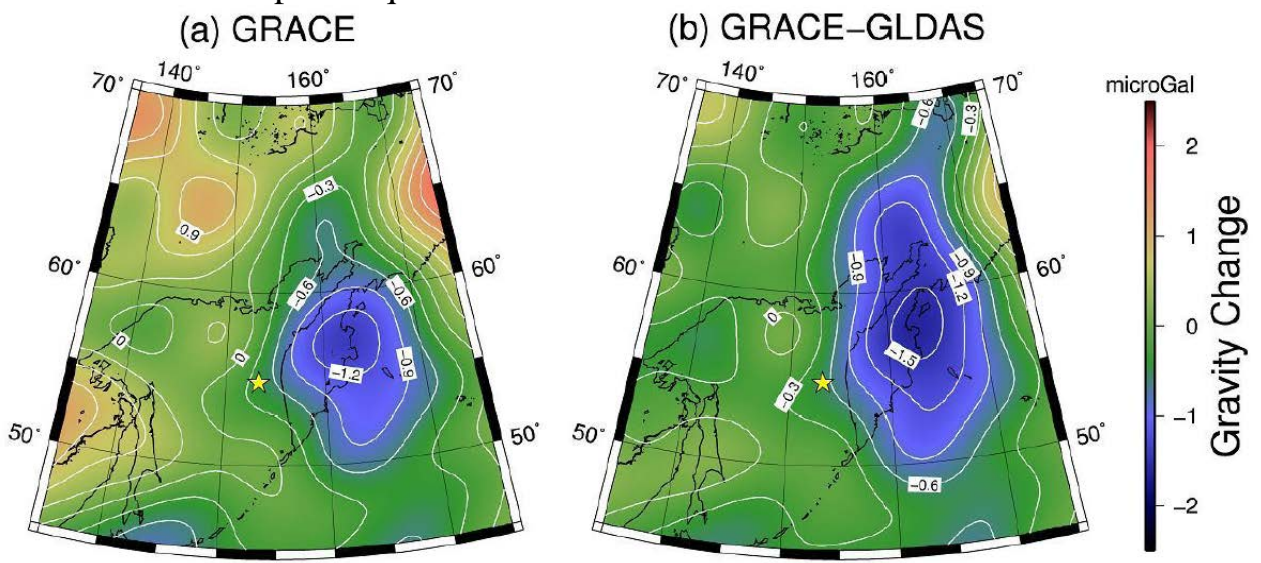

**Figure.** (a) The distribution of the coseismic gravity change caused by the 2013 Okhotsk deep earthquake observed by GRACE. The star shows the epicenter of the earthquake, and the contour interval is 0.3 μGal. (b) Same as (a) but the land hydrological signals have been corrected using the GLDAS model [Tanaka et al., 2015].

Gravity Recovery And Climate Experiment (GRACE) twin satellites have been observing the mass transports of the Earth inferred by the monthly gravity field solutions in terms of spherical harmonic coefficients since 2002. In particular, the GRACE temporal gravity field observations revolutionize the study of basin-scale hydrology, because gravity data reflect mass changes related to ground and surface water redistribution, ice melting, and precipitation accumulation over large scales. However, to use the GRACE data products, de-striping/filtering is required. The researchers [Zotov et al., 2015] applied the multichannel singular spectrum analysis (MSSA) technique to filter GRACE data and separate its principal components (PCs) at different periodicities. Data averaging over the 15 largest river basins of Russia was performed. In spring 2013 the extremely large snow accumulation occurred in Russia, while the autumn 2014 was quite dry. The maxima and minima are evident in GRACE observations, which correspond to Amur River flood in 2013, Volga River dry period in 2010 etc. They can be compared to the hydrological models, such as Global Land Data Assimilation System (GLDAS) or WaterGAP Global Hydrology Model (WGHM), and gage data. Long-periodic climate-related changes were separated into PC 2. Finally, it was observed that there were mass increases in Siberia and decreases around the Caspian Sea [Zotov et al., 2015].

**References**

Baranov V.N., Korolevich V.V. (2011) An example of estimating the accuracy of the model EGM 2008 according to astronomical and geodetic data. Баранов В.Н., Королевич В.В. Пример оценки точности модели EGM 2008 по астрономо-геодезическим данным /Международный научно-технический и производственный журнал «Науки о Земле» - 2011. - №2 - с.39-43. http://geo-science.ru/wp-content/uploads/39-43.pdf

Basmanov A.V., Popadiov V.V., Sermiagin R.A. (2011) Viet Nam state gravimetric network development. Басманов А.В., Попадьев В.В., Сермягин Р.А. Развитие государственной гравиметрической сети Вьетнама // Геодезия и картография-2011.- №5.- с. 16-19. http://elibrary.ru/item.asp?id=21943036

Boyarsky E.A., Mazurova E.M., Vitushkin L.F. (2011) "Gravity Field", National Report for the International Association of Geodesy of the International Union of Geodesy and Geophysics 2007−2010, Moscow, pp. 21-30. http://www.iag-aig.org/index.php?tpl=text&id_c=52&id_t=510

Brovar B.V. (2013) Representation of expanding coefficients of potential through expanding coefficients of gravity acceleration. Бровар Б.В. Представление коэффициентов разложения потенциала через коэффициенты разложения аномалий ускорения силы тяжести. / Физическая геодезия. Научно-технический сборник ЦНИИГАиК. – М.: Научный мир, 2013. – с.69-73.

Brovar B.V., Gusev N.A. (2013) On changes of the content of geodetic and gravimetric data caused by using of satellite technologies. Бровар Б.В., Гусев Н.А. Об изменении состава геодезических и гравиметрических данных, обусловленных применением спутниковых технологий. / Физическая геодезия. Научно-технический сборник ЦНИИГАиК. – М.: Научный мир, 2013. – с.25-43.

Brovar B.V., Stoliarov I.A. (2013) On verification of high precession method of V.Brovar for external perturbing potential determination of the real Earth and quasigeoid heights on test models. Бровар Б.В., Столяров И.А. О проверке высокоточного метода В.Бровара для определения внешнего возмущающего потенциала реальной Земли и высот квазигеоида на тестовых моделях / Физическая геодезия. Научно-технический сборник ЦНИИГАиК. – М.: Научный мир, 2013. – с.74-94.

Brovar V.V. (2013) Role of gravity field in geodesy. Бровар В.В. Роль гравитационного поля в геодезии / Физическая геодезия. Научно-технический сборник ЦНИИГАиК. – М.: Научный мир, 2013. – с.220-225.

Chuykova N.A., Nasonova L. P., Maximova T.G. (2011) Greenland mantle gravity anomalies and its geologic-geophysical interpretation. Чуйкова Н.А., Насонова Л.П., Максимова Т.Г. Мантийные гравитационные аномалии Гренландии и их геолого-геофизическая интерпретация. Материалы XXXVIII сессии Международного семинара «Вопросы теории и практики геологической интерпретации гравитационных, магнитных и электрических полей». Пермь,24-28 января 2011, с.83-85.

Crossley D., Vitushkin L., Wilmes H. (2013) Global systems for the measurement of the gravity field of the Earth: from Potsdam to Global Geodynamics Project and further to International System of Fundamental Absolute Gravity Stations, «Фундаментальное и прикладное координатно-временное и навигационное обеспечение» (КВНО-2013), 15-19 апреля 2013 г., Санкт-Петербург, Россия. Тезисы докладов. Санкт-Петербург: ИПА РАН, 2013. с. 147 - 151.

Dementiev Yu.V., Kalenitskiy A.I., Krapik A.P., Seredovish V.A. (2014) On a full topographic gravity reduction. Дементьев Ю.В., Каленицкий А.И., Карпик А.П., Середович В.А. О полной топографической редукции силы тяжести. Известия высших учебных заведений. Геодезия и аэрофотосъемка. 2014. № 3. С. 13-16. http://www.miigaik.ru/journal.miigaik.ru/arhiv_zhurnalov/vypuski_za_2014_/20150120172913-9427.pdf

Ebauer K., Sorokin N. (2013) High precise methods of numerical integration of artifitical Earth satellites' motion equations for Chebyshev approximation for processing laser observations of a satellite. Эбауэр К.В., Сорокин Н.А. Высокоточные методы численного интегрирования уравнений движения ИСЗ с Чебышевской аппроксимацией для обработки лазерных наблюдений ИСЗ // Известия высших учебных заведений. Геодезия и аэрофотосъемка, 2013, № 3, с.3-8. http://www.miigaik.ru/journal.miigaik.ru/2013/20130830165548-6482.pdf

Jiang Z., Palinkas V., Francis O., Baumann H., Mäkinen J., Vitushkin L., Merlet S., Tisserand L., Jousset P., Rothleitner C, Becker M., Robertsson L., Arias E.F. (2013) On the gravimetric contribution to watt balance experiments, Metrologia, 2013, v.50, n 5, pp 452-471. http://iopscience.iop.org/0026-1394/50/5/452

Mäkinen J., Bilker M., Wilmes G., Falk R., Kaftan V.I., Gusev N.A., Korolev N.N., Yushkin V.D. (2013) Results of international comparisons in 2005. Мяккинен Я., Билкер М., Вилмес Г., Фальк Р., Кафтан В.И., Гусев Н.А., Королев Н.Н., Юшкин В.Д. Результаты международных сравнений в 2005 году //Труды симпозиума международной ассоциации геодезии (IAG) TGSMM2013 «Наземная, морская и аэрогравиметрия: измерения на неподвижных и подвижных основаниях» 17-19 сентября 2013 года.

Mazurov B.T., Nekrasova O.I. (2014) Approximation of gravitation influence of local topography by the use of digital models. Мазуров Б.Т., Некрасова О.И. Аппроксимация гравитационного влияния локального рельефа по его цифровым моделям. Геодезия и картография. 2014. № 7. С. 2-4. http://elibrary.ru/item.asp?id=21831896

Mazurova E. (2011) "Quasigeoid Height Evaluation on the basis of Discrete Linear Transforms", poster, XXV General Assembly IUGG-2011, 28 June- 7 July, Melbourne, Australia. http://www.iugg2011.com/

Mazurova E., Lapshin A. (2013) "On the action of Heisenberg's uncertainty principle in discrete linear methods for calculating the components of the deflection of the vertical", Geophysical Research Abstracts, Vol.15, EGU2013-PREVIEW, EGU General Assembly 2013, 07-12 April, Vienna, Austria. http://www.egu2013.eu

Mazurova E., A. Kozlova (2011) "About calculation of the components of the deflection of the vertical through discrete linear transformations", EGU General Assembly 2011, 03-08 April, Vienna, Austria. http://meetingorganizer.copernicus.org/EGU2011/EGU2011-1801.pdf

Mazurova E.M., Lapshin A.Yu. (2011) Height anomaly computation in a level of the first approximation of the M.S. Molodensky's theory in the "near" zone with wavelet-transformations. Мазурова Е.М., Лапшин А.Ю. Вычисление аномалии высоты с точностью первого приближения теории Молоденского в ближней зоне на основе вейвлет-преобразования. Известия вузов. Геодезия и аэрофотосъёмка. №6, 2011, стр.41-43. http://www.miigaik.ru/journal.miigaik.ru/2011/20120224140213-1560.pdf

Mazurova E.M., Lapshin A.Y., Menshova E.V. (2013) "On the Heisenberg's uncertainty principle in calculating the components of deflection of the vertical". Izvestiya Vuzov. Geodeziya i Aerofotos'yomka (News of Higher schools. Geodesy and air photography), № 2, 2013, pp. 31-35. http://miigaik.ru/journal.miigaik.ru/2013/20130918172221-4297.pdf

Mazurova E.M., M.I. Yurkina (2011a) Use of Green's function for determining the disturbing potential of an ellipsoidal Earth, 2011, Stud. Geophys. Geod., Vol. 55, pp.455-464. http://link.springer.com/article/10.1007/s11200-011-0026-1#page-2

Mazurova E.M.,Yurkina M.I. (2011b) On the problem of Green function determination for the ellipsoidal Earth. Мазурова Е.М., М.И. Юркина К вопросу определения функции Грина для эллипсоидальной Земли. Известия вузов.

Геодезия и аэрофотосъёмка. №5, 2011, стр. 3-10. http://www.miigaik.ru/journal.miigaik.ru/2011/20111117145225-8969.pdf

Nepoklonov V.B., Lidovskaya T.A., Spesivtsev A.A. (2014) Quality estimation of Earth's gravity field models. Непоклонов В.Б., Лидовская Е.А., Спесивцев А.А. Оценка качества моделей гравитационного поля Земли. // Изв. вузов. Геодезия и аэрофотосъемка.-2014. М.:- № 2 .- С.24-32 http://www.miigaik.ru/journal.miigaik.ru/arhiv_zhurnalov/vypuski_za_2014_/20150120172913-5457.pdf

Neuman Yu.M., Sugaipova L.S., Popadiev V.V. (2012) Satellite gradientometry experiments. Нейман Ю.М., Сугаипова Л.С., Попадьёв В.В. Эксперименты со спутниковой градиентометрией, Геодезия и картография.-2012.-№ 12.-с.77-79.

Neyman Yu.M., Sugaipova L.S. (2014a) On global geopotential model adaptation to regional specialties (Part 1). Нейман Ю.М., Сугаипова Л.С. Об адаптации глобальной модели геопотенциала к региональным особенностям (Часть 1) Известия высших учебных заведений. Геодезия и аэрофотосъемка. 2014. № 3. С. 3-12. http://www.miigaik.ru/journal.miigaik.ru/arhiv_zhurnalov/vypuski_za_2014_/20150120172913-9427.pdf

Neyman Yu.M., Sugaipova L.S. (2014b) On global geopotential model adaptation to regional specialties (Part 2). Нейман Ю.М., Сугаипова Л.С. Об адаптации глобальной модели геопотенциала к региональным особенностям (Часть 2) Известия высших учебных заведений. Геодезия и аэрофотосъемка. 2014. № 4. С. 3-7. http://www.miigaik.ru/journal.miigaik.ru/arhiv_zhurnalov/vypuski_za_2014_/20150120172913-4566.pdf

Panteleev V.L., Chesnokova T.S. (2011) Task of deconvolution and inertial gravimetry. Пантелеев В.Л., Чеснокова Т.С. Задача деконволюции и инерциальной гравиметрии. Вестник МГУ, Физ. Астрономия, 2011, № 1, с.75-79.

Pick M., Yurkina M.I. (2013) On modern theory of the figure of the Earth. Пик М., Юркина М.И. О современной теории фигуры Земли / Физическая геодезия. Научно-технический сборник ЦНИИГАиК. – М.: Научный мир, 2013. – с.55-68.

Staklo A.B., Brovar B.V., Gusev N.A., Sermiagin R.A., Oschepkov I.A., Popadiev V.V. Ensuring the uniformity of measurementas in gravimetry. (2014) Стакло А. В., Бровар Б. В., Гусев Н. А., Сермягин Р. А., Ощепков И. А., Попадьёв В. В. Обеспечение единства измерений в гравиметрии // Геофизический вестник. - 2014. - № 2. - С. 15 - 18.

Sugaipova L.S. (2012) Creating regular grid of mean values of the Earth potential second derivatives after GOCE project's outcomes. Сугаипова Л.С. Создание регулярной сетки усредненных значений вторых производных геопотенциала по результатам проекта GOCE, Изв.ВУЗов, Геодезия и

аэрофотосъёмка, № 5, 2012. http://www.miigaik.ru/journal.miigaik.ru/2012/20121108172714-3921.pdf

Sugaipova L.S. (2014) On harmonic analysis of results of satellite gradiometry. Сугаипова Л.С. О гармоническом анализе по результатам спутниковой градиентометрии. Известия высших учебных заведений. Геодезия и аэрофотосъемка. 2014. № 2. С. 19-24. http://www.miigaik.ru/journal.miigaik.ru/arhiv_zhurnalov/vypuski_za_2014_/20150120172913-5457.pdf

Tanaka, Y., K. Heki, K. Matsuo, and N. V. Shestakov (2015), Crustal subsidence observed by GRACE after the 2013 Okhotsk deep-focus earthquake, Geophys. Res. Lett., 42, doi:10.1002/2015GL063838.

Vitushkin L.F., Orlov O.A. (2013) Absolute ballistic gravimeter ABG-VNIIM-1 of D.I.Mendeleyev Institute for Metrology, Abstracts of IAG Symposium on Terrestrial Gravimetry @Static and Mobile Measurements – TGSMM-2013, SRC of RF "Concern ELEKTROPRIBOR", 17-20 September 2013, p 29.

Vitushkin L. F., Orlov O.A. (2014a) Absolute Ballistic Gravimeter ABG-VNIIM-1 by D.I. Mendeleyev Research Institute for Metrology, Gyroscopy and Navigation, 2014, vol.5, n. 4, 283-287.

Vitushkin L.F., Orlov O.A. (2014b) Absolute ballistic gravimeter ABG-VNIIM-1 – development of VNIIM named after D.I.Mendeleev. Витушкин Л.Ф., Орлов О.А. Абсолютный баллистический гравиметр АБГ-ВНИИМ-1 разработки ВНИИМ имени Д.И.Менделеева. Гироскопия и навигация. 2014. № 2 (85). С. 95-101. http://elibrary.ru/item.asp?id=22401426

Vitushkin L.F., Wilmes H. (2013) Absolute ballistic gravimetry: measuring techniques and metrology, «Фундаментальное и прикладное координатно-временное и навигационное обеспечение» (КВНО-2013), 15-19 апреля 2013 г. Санкт-Петербург, Россия. Тезисы докладов. Санкт-Петербург: ИПА РАН, 2013.- с. 78-81.

Yushkin V.D. (2012a) Estimation of environmental influence on gravity acceleration from vertical gradient data. Юшкин В. Д. Оценка влияния окружающей среды на ускорение силы тяжести по данным вертикальных градиентов. Геодезия и картография.-2012, № 3, стр. 3 – 7. http://elibrary.ru/item.asp?id=21760755

Yushkin V.D. (2012b) Gravitation anomaly of the Elbrus and rock density of its cone. Юшкин В.Д. Гравитационная аномалия Эльбруса и плотность пород его конуса. Труды симпозиума международной ассоциации геодезии (IAG) TGSMM2013 «Наземная, морская и аэрогравиметрия: измерения на неподвижных и подвижных основаниях» 17-19 сентября 2013 года».

Yushkin V.D., Sapunov A.N., Stus Yu.F., Kalish E.N., Bunin I.A., Nosov D.E. (2013) Creation of absolute gravity reference test area in permafrost conditions. Юшкин В.Д., Сапунов А.Н., Стусь Ю.Ф., Калиш Е.Н., Бунин И.А., Носов Д.Е. Создание абсолютного опорного полигона в условиях вечной мерзлоты / Физическая геодезия. Научно-технический сборник ЦНИИГАиК. – М.: Научный мир, 2013. – с.135-141

Zotov L.V., C.K. Shum, N.L. Frolova (2015) Gravity changes over Russian rivers basins from GRACE, Chapter 3 in Planetary Exploration and Science: Recent Results and Advances, Springer.

# Geodynamics

**Gorshkov V.[1], Kaftan V.[2], Malkin Z.[1], Shestakov N.[3,4], Steblov G.[5]**

[1]Pulkovo Observatory, Saint Petersburg, Russia
[2]Geophysical Center of the Russian Academy of Sciences, Moscow, Russia
[3]Far Eastern Federal University, Vladivostok, Russia
[4]Institute of Applied Mathematics, FEB RAS, Vladivostok, Russia
[5]Geophysical Survey of the Russian Academy of Sciences, Obninsk, Russia

The problems of Earth's dynamics in relation to General Relativity effects are studied by Kopeikin et al. (2014). A concept of Relativistic Geoid is proposed.

General problems of space geodetic measurements for global changes monitoring are discussed in [Tatevian et al., 2012, 2014a, b].

The studies of the Geocenter dynamics by the analysis of the measurements of the GPS and DORIS satellite systems performed by Valeev et al. (2011).

Especialities of deformation of continental and ocean lithosphere revealed by geodetic technique are considered as an evidence of the north movement of the Earth's core in [Goncharov et al., 2011].

In [Malkin, 2014e] the authors presented the results of a study, which have been performed to investigate the impact of the cut-off elevation angle (CEA) and elevation-dependent weighting (EDW) on the Earth orientation parameters (EOP)and baseline length estimates obtained from astrometric and geodetic VLBI observations. For this test, 2-week continuous CONT05 VLBI observations were processed with different CEA and EDW settings, keeping all other options the same as used during the routine data processing. For the baseline length, the repeatability test was used to investigate the impact of the analysis options under investigation. For EOP, the uncertainties and correlations between estimated parameters have been investigated, as well the differences between VLBI and GPS results obtained during the CONT2005 period. It has been shown that applying a small CEA up to about 8-10 degrees does not have large impact on the results, except a small degradation of the baseline length repeatability, whereas applying EDW allows smaller errors for the baseline length, polar motion and UT1 to be made. No substantial impact was found on the celestial pole offset. Finally, we conclude that an inclusion of the low-elevation observations, properly weighted, improves the baseline length repeatability and EOP results.

Permanent and field GNSS measurements at nearly 40 sites surrounding the Gulf of Finland (South Finland, Estonia and Russia geodetic networks) were used for geodynamic researches of this region [Galaganov et al., 2011; Gorshkov et al., 2012b, 2013b, 2015]. This region is interesting being a transient zone between the Baltic shield and East-European platform or in geology aspect between Archean (3.5 billion years) to Carboniferous (350 million years). The authors used a state-of-art approach to calculation of site positions by GIPSY 6.3 with all modeling corrections including loading ones. It was revealed that low-frequency variations of loading corrections besides nearly seasonal components have also a bias, different for various stations, at that hydrology loading corrections very seldom correspond to real seasonal station variations. The dynamics of the free from the low-frequency components of the station coordinates and base lines between them were also used to estimate the type of distribution of errors [Gorshkov, Shcherbakova, 2012]. These errors have mainly flicker and Gaussian white noise distribution for different stations. So the corrected for loading effects station velocities and its errors were used to assess the strain field of this flexure region by GRID_STRAIN package. This strain field has a weak (up to 3 nanostrain per year) almost meridian compression and possibly a slow counterclockwise rotation of the Baltic shield with respect to the East European platform.

The same approach was used for estimation of strain field of the region of the Gulfs of Finland and Bothnia intersection and Baltic Sea up to Kaliningrad by using observation data of Finnish, Sweden, Estonian and Latvian GPS stations [Assinovskaya et al., 2011, 2013]. Seismic hazards in the Eastern Baltic region are traditionally considered having quite a low frequency and intensity; therefore seismic data alone do not provide sufficient constraints on the geodynamic models of this region. Therefore the analysis of the GPS-based regional crustal motions, strains, and co-seismic deformations was applied to develop geodynamic models for this Baltic Sea region. The GPS results were compared with seismic data because it is known that the strong Osmussaar earthquake (M = 4.6) had occurred in this region in 1976 and the Kaliningrad earthquake (M = 5.0) in 2004. Comparison between the seismic and GPS results permitted to characterize the active regional faults more accurately. These data and earthquake focal mechanisms provided for the Lake Ladoga, Gulf of Finland, and Kaliningrad earthquakes may be useful for the "Seismic Hazard Harmonization in Europe" (SHARE) project, which is devoted to updating seismic hazard models throughout Europe.

Another strong and deep (610 km) earthquake beneath the Sea of Okhotsk (2013, M = 8.3) was used to estimate the limits of GPS kinematic technology (1 sec resolution) to register seismic waves [Gorshkov et al., 2013a]. This earthquake had the distant and large enough macro seismic manifestations over the world. It was found that the closest to earthquake focus GPS-station in Kamchatka (PETS, 550 km) has the same GPS-signal shape as a seismic one.

The pole tide (PT) triggering of seismicity was studied in [Gorshkov, Vorotkov, 2012; Gorshkov, 2014]. PT is generated by the centrifugal effect of polar motion on the chandler ($f_{cw}$ = 0.84 cpy) and annual frequencies. These frequencies, their beat frequency (0.16 cpy) and doubled frequency of chandler wobble (1.66 cpy) were revealed in seismic intensity spectrum of weak (M < 5) earthquake. The failure time for such earthquakes (1 - 10 years) is in a good agreement with the periodicity of stress oscillations excited by PT in the Earth's crust. The NEIC and CMT global seismic databases (1976 – 2013) were used for the search of the PT influence on the intensity of seismic process. The normal and shear stresses excited by PT were calculated for 32.2 thousand seismic events from CMT. The phases of the PT stresses for each earthquake were assessed and subsequently used for statistical estimation of PT triggering of seismicity. The PT stress oscillations excite the weak earthquakes of thrust-slip fault type only on 95% significance level by $\chi^2$ and Schuster's statistical test.

The North Eurasian deformation array (NEDA), established and developed by GS RAS since 1997, is intensively used over the last few years for monitoring active deformation belts around Russian territory. These tectonically active belts comprise the Far East and the North Caucuses, where the local deformation arrays have been also deployed. Various kinds of tectonic processes in these regions are investigated, such as interseismic, coseismic, postseismic activity, as well as the tectonic plates' configuration and interaction. As for the seismic activity during 2011-2014 the following phenomena were revealed and analyzed: the anomalous postseismic motion after the great Simushir doublet in 2006-2007 and the 2013 Okhotsk deep (611 km) earthquake, all the events having the magnitude Mw exceeding 8. Regarding the interseismic deformations, in 2011-2014 the modern motions of the Earth crust were investigated in the Ossetia part of the Great Caucasus. Finally, the kinematics of microplates in the North-East Asia was estimated with the new set of measurements. These topics are outlined below and addressed in more detail.

In 2006–2007, a doublet of great earthquakes ($M$w > 8) struck in the center of the Kuril subduction zone, a thrust event followed by an extensional event. Our observations of the Kuril GPS Array in 2006–2009 outlined a broad zone of postseismic deformation with initial horizontal velocities to 90 mm/a, and postseismic uplift. Prior to the earthquakes, all observation sites of the Kuril network were moving towards the continent due to the subduction deformation of the continental margin. After the events, the direction of displacement had changed to the opposite direction at the stations located on the Matua, Ketoy, and Kharimkotan Islands, which were the nearest to the seismic events, and experienced a significant turn on the Urup Island nearby. We showed that most of the postseismic signal after the great Kuril doublet is caused by the viscoelastic relaxation of shear stresses in the weak asthenosphere with the best-fitting Maxwell viscosity in the range of (5–10) × 1017 Pa s, an order of magnitude smaller than was estimated for several subduction zones. We predict that the postseismic deformation will die out in about a decade after the earthquake doublet [Kogan M.G. et al., 2011, 2013, Vladimirova I.S. et al., 2011].

The researchers analyzed the first ever GPS observations of static surface deformation from a deep earthquake: the 24 May 2013 $Mw$ 8.3 Sea of Okhotsk, 611 km-deep, event. Previous studies of deep earthquake sources relied on seismology and might have missed evidence for slow slip in the rupture. We observed coseismic static offsets on a GPS network of 20 stations over the Sea of Okhotsk region. The offsets were inverted for the best fitting double-couple source model assuming a layered spherical Earth. The seismic moment calculated from static offsets is only 7% larger than the seismological estimate from Global

Centroid Moment Tensor (GCMT). Thus, GPS observations confirm shear faulting as the source model, with no significant slow-slip component. The relative locations of the U.S. Geological Survey hypocenter, GCMT centroid, and the fault from GPS indicate slip extending for tens of kilometers across most of the slab thickness [Steblov G.M., 2014; Shestakov N.V. et al., 2014].

The Ossetian part of the Greater Caucasus, being one of the most tectonically active regions of the Caucasus, until recently was not covered by the high-precision geodetic measurements based on satellite methods. Since 2010 the network of satellite geodetic campaign sites for periodical observations by the mobile GPS equipment was deployed in this region as well as three permanent GPS sites were established. The velocities of horizontal motions were estimated in three reference frames: in the International Terrestrial Reference Frame ITRF2008, in Eurasia-fixed frame and in the local reference frame (defined by the local permanent site ARDN). The obtained results show, first, general submeridional motion of the region caused by the convergence of the Eurasian and Arabian plates, and, second, reveal a number of the tectonic features caused, apparently, by continuing local processes of the tectonic structure formation in this region [Milyukov V. et al., 2012, 2014].

The question concerning the integrity of major tectonic plates is still unclear for several regions covering the plate junction zones. The Northeast Asia is one such region, where there is no common concept of the configuration of plate boundaries. From the classical viewpoint, the dynamics of Northeast Asia is determined by the superposition of the relative rotations of the three major plates (Eurasian, North American and Pacific). According to the alternative viewpoint, the fragments that were split from these plates rotate independently in the form of microplates (Bering, Okhotsk, and Amur). The analysis of kinematics for the GPS stations located in eastern Chukotka, western Alaska, and on the Bering Sea islands suggests the existence of the Bering microplate rotating clockwise relative to the North American plate [Gabsatarov et al., 2013].

The Far East of Russia is the more geodynamic active region. In addition of the national reference frame the special geodynamic observation network is established by the Far Eastern branch of the Russian Academy of Sciences [Shestakov et al., 2012, Sorokin et al., 2013a, 2013b].

The present tectonics of Northeast Asia has been extensively investigated during the last 12 years by using GPS techniques in a frame of international cooperation [Shestakov et al., 2011]. Nevertheless, a crustal velocity field of the southeast of Russia near the northeastern boundaries of the hypothesized Amurian microplate has not been defined yet. The GPS data collected between 1997 February and 2009 April at sites of the regional geodynamic network were used to estimate the recent geodynamic activity of this area. The calculated GPS velocities indicate almost internal (between network sites) and external (with respect to the Eurasian tectonic plate) stability of the investigated region. It has not found clear evidences of any notable present-day tectonic activity of the Central Sikhote-Alin Fault as a whole. This fault is the main tectonic unit that determines the geological structure of the investigated region. The obtained results speak in favor of the existence of a few separate blocks and a more sophisticated structure of the proposed Amurian microplate in comparison with an indivisible plate approach.

The more intensive regional study was devoted to the 2011 Great Tohoku earthquake [Shestakov et al., 2011, 2012, 2013, 2014]. It was clearly reflected in the Earth's surface deformation at continental territory. Small coseismic offsets detectable using GPS techniques were found more than 2300 km away from the Great Tohoku 2011 earthquake epicenter. Area of the most intense far-field co- and postseismic deformations with the maximum offset values exceeding 40 and 18 mm, respectively, extends westward from Honshu Island to the Korean Peninsula, northeastern China and southern Far East Russia. The Sakhalin Island does not exhibit notable displacements caused by the earthquake, in contrast to the adjacent territories. A rectangular fault model with uniform slip was developed based on the GPS-detected far-field coseismic displacements using the spherically layered Earth assumption. Both far- and near-field coseismic deformations are generally well described by a single-segment rupture of $200 \times 96$ km$^2$, characterized by thrust slip with minor strike-slip component of about 33 m and by the seismic moment value of $1.9 \cdot 10^{22}$ N·m (Mw = 8.8), which roughly constrains the major slip area. The resultant compact fault geometry revealed that the main portion of the seismic moment had been realized in a relatively small-sized rupture segment. The sensitivity of far-field GPS data to the major slip area might also be used in the development of a seismically generated giant tsunami warning system [Shestakov et al., 2012].

The far-field coseismic ionospheric disturbances induced by the 2011 Great Tohoku earthquake using different GPS data sources (IGS data, continuously and periodically observed regional geodynamic GNSS networks and other GNSS observations applicable for this study) were determined and analyzed. The total electron content (TEC) data extracted from the original GPS observations were used to study the ionospheric response to this seismic event in the far-field zone (Fig. 1). The TEC disturbances with periods of 5-15 min propagating from the rupture were successfully detected by GPS methods at distances up to more than 2000 km away from the epicenter. Their intensities decreased away from the quake epicenter. It was found some irregularity of the TEC disturbance attenuation in different directions. A comparative analysis was carried out of the distribution of ionosphere disturbances and of the far-field coseismic displacements [Shestakov et al., 2013].

The dynamics of the Baykal rift zone is studied using GNSS observation [Serebriakova et al., 2013]. The observation show that not only transverse extensions take place at main regional faults but oppositely directed displacements of fault flanks also occur.

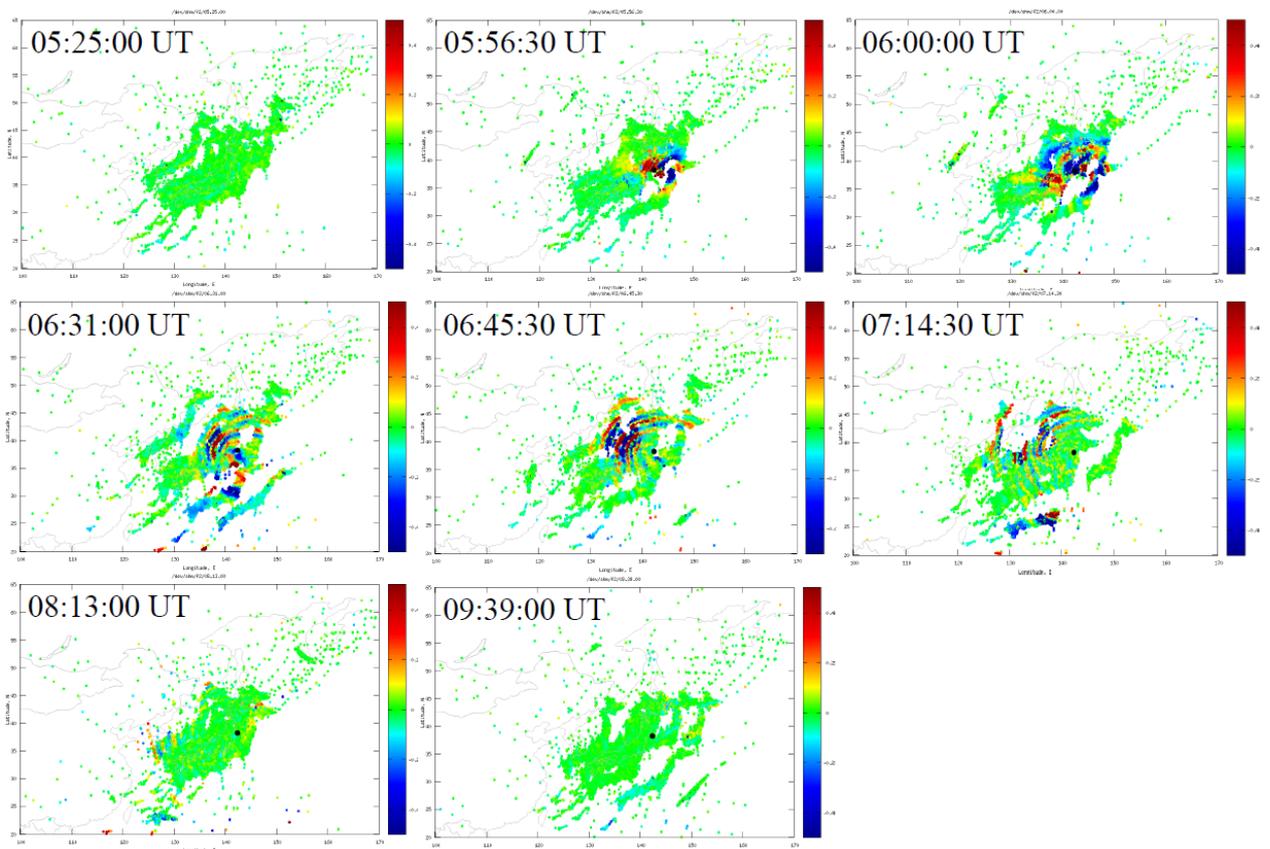

Fig. 1 Ionospheric response to the Great Tohoku earthquake. Legended. Samples of filtered TEC maps showing the propagation of different scale traveling ionospheric disturbances (TIDs). Upper-left plot shows no TEC perturbations before the earthquake. The next samples show large scale coseismic and postseismic TIGs. The last figure renders the TEC conditions after the main TIDs have passed away.

The Baikal rift zone is a giant tectonic structure producing catastrophic earthquakes. Lake Baikal is the biggest reservoir of fresh water all over the world. Geodynamical monitoring of the region is a high important activity of scientific and stakeholder community. Historical seismic records showed that the main deformation of the region is an extension across the axis line of the Baikal rift. Different repeated geodetic measurements were done in the region. Several GNSS field companies are executed in the area from 1994 to 2014. The observation data were processed and deformation characteristics were received. The first epochs of absolute gravity measurements are executed. The received displacement vectors demonstrate the existence of movements of the order of 3 mm/yr in general. The deformations of the territory have the mean level of 10-6. The study shows that the south and north parts of the Lake Baikal are in a state of different deformation tendencies in different time intervals. The line of zero-deformation is close to the continuation of the Obruchev fault zone dividing the Lake Baikal into north and south parts.

Recent deformation processes taking place in real time are analyzed on the basis of data on fault zones which were collected by long-term detailed geodetic survey studies with application of field methods and satellite monitoring [Kuzmin, 2014a, 2014b]. A new category of recent crustal movements is described and termed as parametrically induced tectonic strain in fault zones. It is shown that in the fault zones located in seismically active and aseismic regions, super intensive displacements of the crust (5 to 7 cm per year, i.e. (5 to 7)·$10^{-5}$ per year) occur due to very small external impacts of natural or technogenic / industrial origin. The spatial discreteness of anomalous deformation processes is established along the strike of the regional Rechitsky fault in the Pripyat basin. It is concluded that recent anomalous activity of the fault zones needs to be taken into account in defining regional regularities of geodynamic processes on the basis of real-time measurements.

The results of analyses of data collected by long-term (20 to 50 years) geodetic surveys in highly seismically active regions of Kopetdag, Kamchatka and California are presented. It is evidenced by instrumental geodetic measurements of recent vertical and horizontal displacements in fault zones that deformations are 'paradoxically' deviating from the inherited movements of the past geological periods. In terms of the recent geodynamics, the 'paradoxes' of high and low strain velocities are related to a reliable empirical fact of the presence of extremely high local velocities of deformations in the fault zones (about $10^{-5}$ per year and above), which take place at the background of slow regional deformations, with lower velocities by the order of 2 to 3. Very low average annual velocities of horizontal deformation are recorded in the seismic regions of Kopetdag and Kamchatka and in the San Andreas fault zone; they amount to only 3 to 5 amplitudes of the tidal deformations of the Earth per year. A 'faultblock' dilemma is stated for the recent geodynamics of faults in view of interpretations of monitoring results. The matter is that either a block is an active element generating anomalous recent deformation and a fault is a 'passive' element, or a fault zone itself is a source of anomalous displacements and blocks are passive elements, i.e. host medium. 'Paradoxes' of high and low strain velocities are explainable under the concept that the anomalous recent geodynamics is caused by parametric excitation of deformation processes in fault zones in conditions of a quasistatic regime of loading. Based on empirical data, it is revealed that recent deformation processes migrate in fault zones both in space and time. Two types of waves, 'interfault' and 'intrafault', are described. A phenomenological model of autowave deformation processes is proposed; the model is consistent with monitoring data. A definition of 'pseudowave' is introduced. Arrangements to establish a system for monitoring deformation autowaves are described. When applied to geological deformation monitoring, new measurement technologies are associated with result identification problems, including 'ratios of uncertainty' such as 'anomaly's dimensions – density of monitoring stations' and 'anomaly's duration – details of measurements in time'. It is shown that the InSAR interferometry method does not provide for an unambiguous determination of ground surface displacement vectors [Kuzmin, 2014a, 2014b].

Intensive gathering of permanent GNSS observation data stimulates a development of kinematic visualization of observed deformation processes. It helps to do reconnoiter analysis of the data and propose more plausible mechanisms of observed processes. Developments in kinematic data visualization are described in [Kaftan et al., 2011a, b]. The developed techniques have allowed analyzing 5yr GNSS data observation near Parkfield earthquake epicenter [Kaftan & Krasnoperov, 2012, 2013].

The well known Parkfield earthquake (M 6.0), which occurred on September 28, 2004 was expected by American geophysicists since late 1980-s. In 1990-s in the vicinity of the epicenter of this earthquake a dense permanent GPS-observation network was established. This network provided important information on the Earth's surface changes of this region both before and after the earthquake. Earlier it was proposed to describe the Earth's surface diplacements within seismo-generating strike-slip faults using a simple mechanism of elastic rebound. This model was verified later as far as repeated geodetic observation results were accumulated [Pevnev, 2011, 2013a, b, c, d]. The 2004 Parkfield earthquake enabled the possibility of verification of the elastic rebound mechanism using repeated GPS observations.

Observational data for the Parkfield GPS network used in this research was acquired from the SOPAC internet archive. The research also included development of a special software package for adjustment of baseline vector differences and visual and animated representation of data. As a result spatiotemporal animated models of displacements and deformations were received. The elastic rebound mechanism was demonstrated.

The fault behavior within different segments of the observation network was analyzed in a period of 2.5 years before the earthquake. It was demonstrated that before the earthquake the fault is not completely locked as it was earlier proposed in the abovementioned Reid's mechanism. Nevertheless, certain creep deceleration was registered somewhile before the seismic event. Analysis of separate segments of the observation network made it possible to assume that deceleration of fault flanks migrates and moves towards the future epicenter of an earthquake. Thus it might be assumed that the fault falls under locking stepwise before the earthquake moment while moving towards the epicenter from the North West boarder of the observation network [Kaftan & Krasnoperov, 2012, 2013].

GPS data was also used for analysis of the aftershock sequence caused by postseismic deformation process in the Parkfield earthquake source area [Kaftan & Rodkin, 2011, Rodkin & Kaftan, 2012].

The brief theory of two-dimensional problem of deformation is presented by Gerasimenko. In order to calculate simpler and uniquely the bearings of principal deformation axes formula is proposed [Gerasimenko, 2014].

Local geodynamic test areas are being constructed and explored in deferent parts of Russia. It is especially important in the zones of accumulation of radioactive waste. Initial field GNSS campaigns are described in [Tatarinov, 2014].

The special computer algorithms and software for deformation analysis in GNSS networks are proposed and described in [Yambaev & Markuze, 2014].

A simultaneous development of the continental-scale and regional-scale satellite geodetic networks is a keystone for understanding the kinematics and geodynamics of the interplate deformations. The large-scale observation networks allow, first, to implement the stable reference frame, and, second, to reveal active deformation belts at the plate boundaries and to estimate intensity of the boundary deformation accumulation. At the same time regional densification of the observational networks allows detailed investigation of the local tectonic structures and their features on top of the interplate deformations.

Today the dynamic processes of the Earth as a unit system need to be studied cross-functionally by means of different sciences. Interconnection between different dynamic processes has a special interest for better understanding of the nature as a whole. Some researches are devoted to find out interrelations between different terrestrial and external processes.

An attempt is made to identify the relationship between the free nutation of the Earth's core, expressed by changes in the parameters of the Earth's rotation, and geomagnetic activity [Malkin, 2013].

Research of regularities of Caspian Sea level was carried out continuously with the use of terrestrial observation techniques. The closed relationships between level changes, solar activity and Earth's rotation variation were estimated over the period of the last centuries. The last decades are characterized by the intensive development of satellite and space observation techniques. Nowadays the accuracy and spatial-temporal resolution of sea level and cosmo-geophysical processes observation has considerably increased. Therefore the large current data ensuring the study of cause-and-effect relations between the Caspian Sea level and geophysical processes of global and space scales were collected. The results of the resent precise observation data analysis with high resolution as well as the long Caspian Sea level time-series combining terrestrial and space observation are proposed to the research community. Spectral characteristics of the Caspian Sea level changes, Earth's rotation parameters (LOD), solar activity and other processes are studied. High amplitude oscillation components having close periods are revealed in the spectra of all analyzed processes. Caspian Sea level oscillations are following in an antiphase to the solar activity changes. The results of the analysis provide the new and important information facilitating to reveal the causes of the regional climate changes [Kaftan et al., 2014].

**REFERENCES**

Assinovskaya, B., Shchukin Ju., Gorshkov V., Shcherbakova N. (2011) On recent geodynamics of the Eastern Baltic Sea region. Baltica, 2011, V. 24. No. 2, 61-70.

Assinovskaya, B.A., Gorshkov V.L., Shcherbakova N.V., Panas N.M. (2013) Active faults revealed in the Baltic Sea using geodinamic observation data. Ассиновская Б.А., Горшков В.Л., Щербакова Н.В., Панас Н.М. Активные разломы, выявленные по данным геодинамических наблюдений в Балтийском море. Инженерные изыскания, 2013, N 2, 50-55. http://elibrary.ru/item.asp?id=19028722

Baek Jeongho, Shin Young-Hong, Na Sung-Ho, Shestakov Nikolay V., Park Pil-Ho, Cho Sungki. (2012) Coseismic and postseismic crustal deformations of the Korean Peninsula caused by the 2011 Mw 9.0 Tohoku earthquake, Japan, from global positioning system data // Terra Nova, Vol. 24, Issue 4, 2012, p. 295-300, DOI: 10.1111/j.1365-3121.2012.01062.x.

Barkin Yu.V. (2011) Earth's shape centurial variation in the current epoch. Баркин Ю.В. Вековые вариации фигуры Земли в современную эпоху. Современное состояние наук о Земле (Материалы международной конференции, посвященной памяти Виктора Ефимовича Хаина, г. Москва, 1-4 февраля 2011 г.) М.: Изд-во Геологический факультет МГУ имени М.В.Ломоносова. (CD-ROM). 2011. С. 183-188. http://khain2011.web.ru/khain-2011-theses.pdf

Bondur V.G., Voronova O.S. (2012) Drift long wave emission variation in preparation and occurrence of strong earthquakes on Russia territory in 2008-2009. Бондур В.Г., Воронова О.С. Вариации уходящего длинноволнового излучения при подготовке и протекании сильных землетрясений на территории России в 2008 и 2009 году // Известия ВУЗов. Геодезия и Аэрофотосъемка. – 2012, №1, с. 79–85. http://www.miigaik.ru/journal.miigaik.ru/2012/20120220145822-9603.pdf

Bondur V.G., Zverev A.T., Gaponova E.V., Zima A.L. (2012a) Aerospace research of deformation waves as earthquake precursors manifesting in lineament system dynamics. Бондур В.Г., Зверев А.Т., Гапонова Е.В., Зима А.Л. Исследование из космоса деформационных волн – предвестников землетрясений, проявляющихся в динамике линеаментных систем.// Исследование Земли из космоса – 2012. – №1. с. 3–20.

Bondur V.G., Zverev A.T., Gaponova I.V. Multilevel lineament analysis of space images of West-Siberian oil-gas province. Бондур В.Г., Зверев А.Т., Гапонова Е.В. (2012b) Многоуровневый линеаментный анализ космических изображений Западно-Сибирской нефтегазоносной провинции // в кн. «Аэрокосмический мониторинг объектов нефтегазового комплекса» / под ред. Бондура В.Г. М.: Научный мир, 2012, С. 92–102.

Didenko A.N., Bykov V. G., Shestakov N.V., Bormotov V. A., Gerasimenko M. D., Kolomiets A. G., Vasilenko N. F., Prytkov A. S., Sorokin A.A. (2011). March 11, 2011 Tohoku, Japan Earthquake recorded by FEB RAS network of deformation and seismological observations. Диденко А.Н., Быков В.Г., Шестаков Н.В., Бормотов В.А., Герасименко М.Д., Коломиец А.Г., Василенко Н.Ф., Прытков А.С., Сорокин А.А. Землетрясение Тохоку 11 марта 2011г. Данные сети деформационных и сейсмологических наблюдений ДВО РАН // Вестник ДВО РАН, 2011, № 3, с. 18-24.

Dokukin P.A., Kaftan V.I., Krasnoperov R.I. (2013) On influence of a form of triangles of GNSS network on results of Earth's surface deformation determination. Докукин П.А. Кафтан В.И., Красноперов Р.И. Влияние формы треугольников СРНС сети на результаты определения деформаций земной поверхности / Физическая геодезия. Научно-технический сборник ЦНИИГАиК. – М.: Научный мир, 2013. – с.115-122.

Dokukin P.A., Poddubskiy A.A. (2011) Space geodesy technique usage for geodynamical processes study. Докукин П.А., Поддубский А.А. Применение методов космической геодезии для изучения геодинамических процессов (на примере Чили) // Землеустройство, кадастр и мониторинг земель. – 2011. – №4.

Freymueller J. T., Steblov G. M., Kogan M. G., Titkov N. N., Vasilenko N. F., Prytkov A.S., Frolov D. I. (2013) How Much Surface Deformation Results from Slab Processes Rather than Surface Plate Tectonics? // AGU Fall Meeting, San Francisco, USA, 9-13 December 2013.

Gabsatarov Yu. V., G. M. Steblov, D. I. Frolov. (2013) The new GPS evidence for the region of Bering microplate // Izvestiya, Physics of the Solid Earth. 2013. V. 49, N. 3, P. 411-415. DOI:10.1134/S106935131302002X

Galaganov O.N., Gorshkov V.L., Guseva T.V., Rosenberg N.K., Perederin V.P., Shcherbakova N.V. (2011) Recent crustal motion of Ladoga-Onega Region revealed from satellite and ground measurements. Галаганов О.Н., Горшков В.Л., Гусева Т.В., Розенберг Н.К., Передерин В.П., Щербакова Н.В. Современные движения земной коры Ладого-Онежского региона по данным спутниковых и наземных измерений // Современные проблемы дистанционного зондирования Земли из космоса, т. 8, N 2, 2011, с. 130-136. http://d33.infospace.ru/d33_conf/2011v8n2/130-136.pdf

Gerasimenko M. D., Shestakov N.V., Tereshkina A.A. (2011) Recent vertical movements of Muraviev-Amursky peninsula inferred from geodetic data. (2011). Герасименко М.Д., Шестаков Н.В., Терешкина А.А. Современные вертикальные движения на полуострове Муравьева-Амурского по геодезическим данным //Геологические процессы в обстановках субдукции, коллизии и скольжения литосферных плит. Материалы Всероссийской конференции с международным участием, Владивосток, 20-23 сентября 2011 г., Владивосток, Дальнаука, 2011, с. 401-402. http://conf2011.fegi.ru/docs/05_p397-451.pdf

Gerasimenko M.D. The question of determination of directions of main deformation axis. Герасименко М.Д. К вопросу определения направлений главных осей деформаций // Геодезия и картография. 2014. № 5. С. 28-29. http://elibrary.ru/item.asp?id=21831879

Goncharov M.A., Raznitsin Yu.N., Barkin Yu. V. (2011) Especialities of deformation of continental and ocean litosphere as an evidence of the north movement of the Earth's core. Гончаров М.А., Разницин Ю.Н., Баркин Ю.В. Особенности деформации континентальной и океанской литосферы как свидетельство северного дрейфа ядра Земли. Современное состояние наук о Земле (Материалы международной конференции, посвященной памяти Виктора Ефимовича Хаина, г. Москва, 1-4 февраля 2011 г.) М.: Изд-во Геологический факультет МГУ имени М.В.Ломоносова. (CD-ROM). 2011. С. 461-466. http://khain2011.web.ru/khain-2011-theses.pdf

Gorshkov V.L. (2014) Study of pole tide triggering of seismicity. In: Proc. of X International Conference "Problem of Geocosmos", Editors: V.N. Troyan, N. Yu. Bobrov, A. A. Kosterov, A. A. Samsonov, N. A. Smirnova, and T. B. Yanovskaya. SPb., 2014, P. 163-167.

Gorshkov V.L., Miller N.O., Vorotkov M.V. (2012a) Manifestation of solar and geodynamic activity in the dynamics of the Earth's rotation. Geomagnetism and Aeronomy, 2012, V. 52, No. 7, 944-952.

Gorshkov V.L., Shcherbakova N.V., Assinovskaya B.A. Influence of small and remote earthquakes on GPS-kinematics. Горшков В.Л., Щербакова Н.В., Ассиновская Б.А. (2013a) Влияние слабых или далёких землетрясений на GPS-кинематику. «Изыскательский Вестник», СПб общество геодезии и картографии, 2013, №2(17), с. 34-39.

Gorshkov V.L., Shcherbakova N.V., Assinovskaya B.A. (2013) Results of GNSS observation in the East-Baltic region and their interpretation. Горшков В.Л., Н.В. Щербакова, Б.А. Ассиновская (2013b) Результаты ГНСС-наблюдений в Восточно-Балтийском регионе и их интерпретация. Современные методы обработки и интерпретации сейсмологических данных. Материалы Восьмой Международной сейсмологической школы. Обнинск: ГС РАН, 2013, с. 145 -149.

Gorshkov V.L., Scherbakova N.V., Mohnatkin A.V., Smirnov S.S., Petrov S.D., Trofimov D.A., Guseva T.V., Perederin V.P., Rosenberg N.K. (2015) Deformation of the South-Eastern Baltic Shield from GNSS observations. In: Proc. Journees 2014 Systemes de Reference Spatio-temporels, St. Petersburg, Russia, 22-24 Sep 2014, pp.211-214.

Gorshkov V.L., Smirnov S.S., Scherbakova N.V. (2012) GNSS loading effect in regional geodynamics study. Горшков В.Л., Смирнов С.С., Щербакова Н.В. (2012b) Нагрузочные эффекты в ГНСС-наблюдениях при исследовании региональной геодинамики. Вестник СПбГУ, Серия 1, 2012, вып. 2, 148-156. http://elibrary.ru/item.asp?id=17789287

Gorshkov V., M. Vorotkov. (2012) On the pole tide excitation of seismicity. Proc. of the 9th International Conference "Problems of geocosmos", St.Petersburg, Petrodvorets, October 8-12, 2012, Eds: V. Troyan, V. Semenov, M. Kubyshkina, pp. 142-145.

Gorshkov V., Vorotkov M., Malkin Z., Miller N., Chapanov Ya. On manifestation of solar activity in sea level and Earth's rotation variations. Горшков В., Воротков М., Малкин З., Миллер Н., Чапанов Я. (2012c) О проявлении солнечной активности в вариациях уровня моря и вращения Земли. В кн.: Тр. Всеросс. ежегодной конф. по физике Солнца "Солнечная и солнечно-земная физика - 2012", СПб: ГАО РАН, 2012, 507-510.

Hui H., Wang R., Malkin Z. Application of Titius-Bode law in earthquake study. In: Journees 2014 Systemes de Reference Spatio-temporels, St. Petersburg, Russia, 22-24 Sep 2014, Book of Abstracts, 30.

Izyumov S.F., Kuzmin Yu.O. (2014) Study of resent geodynamic processes in Kopetdag Region. Изюмов С.Ф., Кузьмин Ю.О. Исследование современных геодинамических процессов в Копетдагском регионе. Физика Земли. 2014. № 6. С. 3-16. http://elibrary.ru/download/98031747.pdf

Kaftan V.I. (2012) Geodetic geodynamics in the knowledge system on the Earth. Кафтан В.И. Место геодезической геодинамики в системе знаний о Земле// Кадастр недвижимости .- 2012.- №2.-(27).-с.43-46.

Kaftan V., Komitov B., Lebedev S. Caspian Sea level and cosmo-geophysical processes: satellite and terrestrial data analysis. Space Studies of the Earth's Surface, Meteorology and Climate (A) Scienti_c Exploitation of Operational Missions in Oceanography and Cryosphere, Use of In-situ Data and Assimilation in Models (A2.1). 40th COSPAR Scienti_c Assembly 2014. Electronic abstracts.

Kaftan V., Krasnoperov R. (2013) Earth's surface movements in relation to Parkfield 2004 earthquake: Interpretation of permanent GPS observations. International Association of Geodesy, Scientific Assembly 150th Anniversary of the IAG, Book of Abstracts, Book of Abstracts, September 1-6, 2013, Potsdam, p.187. http://www.iag2013.org/IAG_2013/Publication_files/abstracts_iag_2013_2808.pdf

Kaftan V.I., Krasnoperov R.I. (2012) Elastic rebound mechanism testing using GPS data related to Parkfield 2004 earthquake. Book of abstracts 33rd General Assembly of the European Seismological Commission (GA ESC 2012), 19-24 August 2012, Moscow and Young Seismological Training Course (YSTS 2012), 25-30 August 2012, Obninsk – M., PH "Poligrafiqwik", 2012, p.94.

Kaftan V., R. Krasnoperov, P. Yurovsky. (2011) Elastic rebound mechanism: GPS-observation analysis in relation to the 2004 Parkfield earthquake (M=6.0) // XXV IUGG General Assembly. Earth on the Edge: Science for a Sustainable Planet. 28 June–7 July 2011. Melbourne, Australia. Electronic Storage of Abstracts. http://www.iugg2011.com/abstracts/pdf/abstracts/81106015_KRASNOPEROV01197.pdf

Kaftan V.I., Krasnoperov R.I., Yurovskiy P.P (2011) Geodetic testing of the elastic rebound model in a reference to the Parkfield earthquake (California, USA, Se[t. 28, 2004, M=6). Кафтан В.И., Красноперов Р.И., Юровский П.П. Геодезическая проверка модели упругой отдачи в связи с землетрясением Паркфилд (Калифорния, США, 28.09.2004, М 6) // Проблемы сейсмотектоники: Материалы XVII Международной конференции 20-24 сентября 2011 года / Под ред. акад. А.О. Глико, д.г.-м.н. Е.А. Рогожина, д.г.-м.н. Ю.К.Щукина, к.г.-м.н. Л.И. Надежка. – Москва, 2011. – с.246-250.

Kaftan V.I., Rodkin M.V. (2011) Postseismic relaxation process from geodetic and seismic data/ Кафтан В.И., Родкин М.В. Процесс постсейсмической релаксации по геодезическим и сейсмическим данным // Проблемы сейсмотектоники: Материалы XVII Международной конференции 20-24 сентября 2011 года / Под ред. акад. А.О. Глико, д.г.-м.н. Е.А. Рогожина, д.г.-м.н. Ю.К.Щукина, к.г.-м.н. Л.И. Надежка. – Москва, 2011. – с.250-253.

Kaftan V.I., Steblov G.M., Tatevian S.K., Pevnev A.K. (2011) Geodynamics / National Report for the International Association of Geodesy of the International Union of Geodesy and Geophysics 2007-2010. Ed. by Dr. V.P.Savinikh and Dr.

V.I.Kaftan // Международный научно-технический и производственный электронный журнал «Науки о Земле» (International scientific, technical and industrial electronic journal «Geo Science»).- 2011.- №1.- p. 17-22. http://geo-science.ru/wp-content/uploads/GeoScience-01-2011-p-05-36.pdf

Khanchuk A.I., Safonov D.A., Radziminovich Ya.B., Kovalenko N.S., Konovalov A.V., Shestakov N.V., Bykov V.G., Serov M.A., Sorokin A.A. (2011) The Largest Recent Earthquake in the Upper Amur Region on October 14, 2011: First Results of Multidisciplinary Study. Ханчук А.И., Сафонов Д.А., Радзиминович Я.Б., Коваленко Н.С., Коновалов А.В., Шестаков Н.В., Быков В.Г., Серов М.А., Сорокин А.А. Сильнейшее современное землетрясение в верхнем Приамурье 14 октября 2011 г.: первые результаты комплексного исследования // Доклады академии наук, 2012, том 445, № 3, с. 338-341, DOI: 10.1134/S1028334X12070227.

Kogan M.G., Ekström G., Vasilenko N.F., Prytkov A.S., Frolov D.I., Freymueller J.T., Steblov G.M. (2013) Rapid postseismic relaxation after the great 2006-2007 Kuril earthquakes from GPS observations in 2007-2011 // Journal of Geophysical Research. 2013. Т. 118. № 7. С. 3691-3706. DOI: 10.1002/jgrb.50245 http://onlinelibrary.wiley.com/doi/10.1002/jgrb.50245/abstract

Kogan M.G., Vasilenko N.F., Frolov D.I., Freymueller J.T., Steblov G.M., Levin B.W., Prytkov A.S. (2011) The mechanism of postseismic deformation triggered by the 2006–2007 great Kuril earthquakes // Geophys. Res. Lett., 2011. V.38. N.6. L06304. P.1-5. DOI:10.1029/2011GL046855.

Kogan M.G., Ekström G., Vasilenko N.F., Prytkov A.S., Frolov D.I., Freymueller J.T., Steblov G.M. (2013) Rapid postseismic relaxation after the great 2006-2007 Kuril earthquakes from GPS observations in 2007-2011 // Journal of Geophysical Research. 2013. V. 118. N.7. P. 3691-3706. DOI: 10.1002/jgrb.50245.

Kopeikin S., Mazurova E., Karpik A. (2014) Relativistic Aspects of SLR/LLR Geodesy, 19-th International Workshop on Laser Ranging, 27-31 October, 2014, Annapolis, MD. http://ilrs.gsfc.nasa.gov/ilrw19/docs/2014/Presentations/3144_Kopeikin_presentation.pdf

Krasna H., Malkin Z., Boehm, J. (2014) Impact of seasonal station displacement models on radio source positions. In: Proc. Journees 2013 Systemes de Reference Spatio-temporels, Paris, France, 16-18 Sep 2013, Ed. N.Capitaine, Paris, 2014, 65-68.

Kuzmin Yu.O. (2014a) On the front burner problems of identification of observation results in recent geodynamics. Кузьмин Ю.О. Актуальные проблемы идентификации результатов наблюдений в современной геодинамике. Физика Земли. 2014. № 5. С. 51.

Kuzmin Yu.O. (2014b) Resent geodynamics of fault zones: faulting in a real time scale. Кузьмин Ю.О. Современная геодинамика разломных зон: разломообразование в реальном масштабе времени. Geodynamics &

Tectonophysics. 2014. Т. 5. № 2. С. 401-443. http://gt.crust.irk.ru/images/upload/tblarticle146/magazin146.pdf

Liu J.-C., N. Capitaine, S.B. Lambert, Z. Malkin, Z. Zhu. (2012) Systematic effect of the Galactic aberration on the ICRS realization and the Earth orientation parameters. Astron. Astrophys., 2012, v. 548, A50. DOI: 10.1051/0004-6361/201219421

Malkin Z. (2013) Free core nutation and geomagnetic jerks. J. of Geodynamics, 2013, v. 72, 53-58. DOI: 10.1016/j.jog.2013.06.001

Mazurov B. T. (2012) The analysis of geodetic measurements taking into account dynamics objects of monitoring. Мазуров Б.Т. Анализ геодезических измерений с учетом динамики объектов мониторинга Известия вузов. Геодезия и аэрофотосъёмка. №2/1, 2012, стр.18-21 http://www.miigaik.ru/journal.miigaik.ru/2012/20120726112123-9081.pdf

Mazurov B.T., Nikolaeva O.N., Romashova L.A. (2012a) Experience of use of digital cards for radiatsionno's analysis situations Мазуров Б.Т., Николаева О.Н., Ромашева Л.А. Интегральные экологические карты как инструмент исследования динамики экологической обстановки промышленного центра Известия вузов. Геодезия и аэрофотосъёмка. №2/1, 2012, стр.91-95. http://www.miigaik.ru/journal.miigaik.ru/2012/20120726112123-9081.pdf

Mazurov B.T., Nikolaeva O.N., Romashova L.A. (2012b) Integrated ecological cards as instrument of research loudspeakers ecological situations industrial centre Мазуров Б.Т., Николаева О.Н., Ромашева Л.А. Интегральные экологические карты как инструмент исследования динамики экологической обстановки промышленного центра Известия вузов. Геодезия и аэрофотосъёмка. №2/1, 2012, стр.88-91. http://www.miigaik.ru/journal.miigaik.ru/2012/20120726112123-9081.pdf

Milyukov V.K., Drobishev V.N., Mironov A.P., Steblov G.M., Hubaev H.M. (2014) Ossetian geodetic satellite network: creation and first results of geodinamic monitoring // Vestnik Vladikavkaz scientific center. 2014. V.14. N4. P.2-11.

Milyukov V.K., Kousraev A.V.,Drobishev V.N., Douev D.A., Mironov A.P., Steblov G.M., Torchiniv H.Z., Houbaev H.M. (2012a) Organisation of geodynamic monitoring system in the Ossetia part of the Large Caucasus on the base of mobile GPS/GLONASS measurements. В.К. Милюков, А.В. Кусраев, В.Н. Дробышев, Д.А. Дуев, А.П. Миронов, Г.М.Стеблов, Х.З.Торчинов, Х.М. Хубаев. Организация системы геодинамического мониторинга Осетинской части Большого Кавказа на основе мобильных GPS/ГЛОНАСС измерений // Тезисы докладов III Международной научно-практической конференции «Опасные природные и техногенные геологические процессы на горных и предгорных территориях Северного Кавказа». ВНЦ РАН и СО-А, Владикавказ, 2012.

Milyukov V., A. Mironov, G. Steblov, V. Drobishev, H. Hubaev, A. Kusraev, Kh.-M. Torchinov, V. Shevchenko. (2012b) GPS constrains on modern geodynamical motion in the Ossetia region of the Great Caucasus: preliminary results // "Monitoring and modeling Earth deformation from giant to small scale events" Volume of abstracts 16th General Assembly of WEGENER. University of Strasbourg, Strasburg, France, 2012.

Milyukov V.K., Yushkin V.D., Mironov A.P., Demianov G.V., Sermiagin R.A., Basmanov A.V., Popadiev V.V., Nasretdinov I.F., Zaalishvili V.B., Kanoukov A.S., Dzeranov B.V. (2013a) Geodynamics of the Caucasus Region. Милюков В.К., Юшкин В.Д., Миронов А.П., Демьянов Г.В., Сермягин Р.А., Басманов А.В., Попадьев В.В., Насретдинов И.Ф., Заалишвили В.Б., Кануков А.С., Дзеранов Б.В. Геодинамика Кавказского региона //Измерит. Техника 2013, 10, стр. 3-5.Тезисы докладов.

Milyukov V.K., Yushkin V.D., Mironov A.P., Demianov G.V., Sermiagin R.A., Basmanov A.V., Popadiev V.V., Nasretdinov I.F., Zaalishvili V.B., Kanoukov A.S., Dzeranov B.V. (2013b) Geodynamics of the Caucasus Region. Милюков В.К., Юшкин В.Д., Миронов А.П., Демьянов Г.В., Сермягин Р.А., Басманов А.В., Попадьев В.В., Насретдинов И.Ф., Заалишвили В.Б., А.С. Кануков, Б.В.Дзеранов. Геодинамика Кавказского региона // Труды симпозиума международной ассоциации геодезии (IAG) TGSMM2013 «Наземная, морская и аэрогравиметрия: измерения на неподвижных и подвижных основаниях» 17-19 сентября 2013 года».

Milyukov V.K., Yushkin V.D., Zaalishvili V.B., Kanoukov A.S., Dzeranov B.V. (2013) Gravity change monitoring at reference gravimetric sites of Northern Caucasus by high precise relative gravimeters. Милюков В.К., Юшкин В.Д., Заалишвили В.Б., Кануков А.С., Дзеранов Б.В. Мониторинг приращений силы тяжести на опорных гравиметрических пунктах Северного Кавказа высокоточными относительными гравиметрами// Геология и геофизика Юга России.-№2, 2013, стр. 39-45.

Ohzono, M., H. Takahashi, N. V. Shestacov, and M. D. Grasimenko. (2013) Postseismic deformation after the 2011 Tohoku earthquake around the northeast Asia // 120th meeting of Japan Geodetic Society, P-01, October 29-31, 2013.

Pevnev A.K. (2011) On the place of the geodesic monitoring in the problem of earthquake prediction. Певнев А.К. О месте геодезического мониторинга в проблеме прогноза землетрясений Международный научно-технический и производственный журнал «Науки о Земле» - 2011. - №1 - с.40-49. http://geo-science.ru/wp-content/uploads/GeoScience-01-2011-p-40-49.pdf

Pevnev A.K. (2013a) On possibility of geodetic method in recognition of preparing and crushing earthquake sources by its volume variation. Певнев А.К. О возможностях геодезического метода в обнаружении готовящихся и разрушающихся очагов землетрясений по вариациям объема этих очагов. Труды III Международной научно-практической конференции, приуроченной к 10-летию схода ледника Колка 20 сентября 2002г. Владикавказ. 2013, с.448-453.

Pevnev A.K. (2013b) On the new possibility of geodetic method in earthquake prediction problem solving. Певнев А.К. О новых возможностях геодезического метода в решении проблемы прогноза землетрясений// Геоинжиниринг. 2013. № 1 (17), с.40-43.

Pevnev A.K. (2013c) On the territory of geodetic monitoring in the problem of earthquake prediction. Певнев А.К. О месте геодезического мониторинга в проблеме прогноза землетрясений. Актуальность идей Г.А.Гамбурцева в XXI веке. Янус-К. 2013, с.351-365.

Pevnev A.K. (2013d) Research of source volume variation of expected earthquake – the way to its forecast. Певнев А.К. Исследование вариаций объема очага готовящегося землетрясения – путь к его прогнозу. Международный научно-технический и производственный журнал «Науки о Земле», № 2-3, 2013, с.50-55. http://www.miigaik.ru/journal.miigaik.ru/2012/20120726112123-9081.pdf

Pevnev A.K. (2013e) Usage of a size and a volume of an earthquake source for earthquake prediction. Певнев А.К. Использование вариаций размеров и объема очага землетрясения для прогноза землетрясений//Землеустройство, кадастр и мониторинг земель.2013. №1, с.57-62.

Pevnev A.K., Simonian V.V., Rubtsov I.V. (2013) On possibility of geodetic and tide gauge methods for solving the problem of time earthquake prediction. Певнев А.К., Симонян В.В., Рубцов И.В. О возможностях геодезического и уравнемерного методов в решении проблемы прогноза времени землетрясения// Инженерные изыскания. 2013. № 9, с.29-32.

Rodkin M.V., Kaftan V.I. (2012) Postseismic relaxation from geodetic and seismic data. Book of abstracts 33rd General Assembly of the European Seismological Commission (GA ESC 2012), 19-24 August 2012, Moscow and Young Seismological Training Course (YSTS 2012), 25-30 August 2012, Obninsk – M., РН "Poligrafiqwik", 2012, p.120.

Serebriakova L.I., Gorobets V.P., Sermiagin R.A., Basmanov A.V., Burtovoi V.V., Nasretdinov I.F., Frolov K.E. (2013) High precision results of satellite measurements in the network of the North Baikal test area. Серебрякова Л.И., Горобец В.П., Сермягин Р.А., Басманов А.В., Буртовой В.В., Насретдинов И.Ф., Фролов К.Е. Результаты высокоточных спутниковых измерений в сети Серебайкальского ГДП / Физическая геодезия. Научно-технический сборник ЦНИИГАиК. – М.: Научный мир, 2013. – с.122-134.

Shestakov N. V., M. Ohzono, H. Takahashi, M. D. Gerasimenko, V. G. Bykov, E. I. Gordeev, V. N. Chebrov, N. N. Titkov, S. S. Serovetnikov, N. F. Vasilenko, A. S. Prytkov, A. A. Sorokin, M. A. Serov, M. N. Kondratyev, Pupatenko V. V. (2014) Modeling of Coseismic Crustal Movements Initiated by the May 24, 2013, Mw = 8.3 Okhotsk Deep Focus Earthquake. // ISSN 1028_334X, Doklady Earth Sciences, 2014, Vol. 457, Part 2, pp. 976–981. ( Шестаков Н. В., Ohzono М., Takahashi Н., Герасименко М. Д., Быков В. Г., Гордеев Е. И., Чебров В. Н., Титков Н. Н., Сероветников С. С., Василенко Н. Ф., Прытков А.С., Сорокин А.А., Серов М.А., Кондратьев М.Н., Пупатенко В.В. Моделирование косейсмических движений земной коры, инициированных глубокофокусным Охотоморским землетрясением 24.05.2013 г., Mw= 8.3 // ДОКЛАДЫ АКАДЕМИИ НАУК, 2014, том 457, № 4, с. 1–6. DOI: 10.7868/S086956521422023X).

Shestakov N., Ohzono M., Takahashi H., Gerasimenko M., Nakao S. (2013) Upper mantle rheology of Sea of Japan inferred from postseismic displacements of the Tohoku earthquake // Abstract JpGU2013, 19-24 May 2013, Chiba, Japan, Submission No.:02676.

Shestakov N., Perevalova N., Voeykov S., Ishin A., Yasyukevich Y., Bykov V., Gerasimenko M. (2013) Investigation of coseismic displacements and ionospheric disturbances in the Far East of Russia generated by the Great 2011 Tohoku earthquake // EGU General Assembly 2013, 7 April – 12 May 2014, Vienna, Austria. Geophysical Research Abstracts, Vol. 15, EGU2013-PREVIEW, 2013. http://adsabs.harvard.edu/abs/2013EGUGA..15.3751S

Shestakov N., Takahashi H., Ohzono M., Prytkov A., Bykov V., Gerasimenko M., Luneva M., Gerasimov G., Kolomiets A., Bormotov V., Vasilenko N., Baek J., Park P.-H., Serov M. (2012). Analysis of the far-field crustal displacements caused by the 2011 Great Tohoku earthquake inferred from continuous GPS observations // Tectonophysics, 2012, v. 524-525, p. 76-86. DOI: 10.1016/j.tecto.2011.12.019. http://adsabs.harvard.edu/abs/2012Tectp.524...76S

Shestakov N.V., Baek J., Gerasimenko M. D., Takahashi T., Kolomiets A. G., Gerasimov G.N., Bormotov V. A., Bykov V. G., Park P., Cho J., Tereshkina A.A., Vasilenko N. F. (2011) Large-scale deformations of the Earth crust in the East Asia caused by the Tohoku earthquake of march 11, 2011, Mw=9.0. (2011). Шестаков Н.В., Baek J., Герасименко М.Д., Takahashi T., Коломиец А.Г., Герасимов Г.Н., Бормотов В.А., Быков В.Г., Park P., Cho J., Терешкина А.А., Василенко Н.Ф., Прытков А.С. Крупномасштабные деформации земной коры в Восточной Азии, вызванные японским землетрясением 11 марта 2011 года (mw=9.0), по данным GPS измерений //Геологические процессы в обстановках субдукции, коллизии и скольжения литосферных плит. Материалы Всероссийской конференции с международным участием, Владивосток, 20-23 сентября 2011 г., Владивосток, Дальнаука, 2011, с. 449-451. http://conf2011.fegi.ru/docs/05_p397-451.pdf

Shestakov N.V., Bormotov A.V., Bykov V.G., Pupatenko V.V., Konovalov A.V., Sorokin A.A., Petukhin A.G. (2012) The collocated GNSS/Seismological network effectiveness in studying the giant megathrust earthquake signals //Abstracts 33rd General Assembly of the European Seismological Commission (GA ESC 2012), 19-24 August 2012, Moscow and Young Seismologist Training Course (YSTC 2012),25-30 August 2012, Obninsk, 2012, p. 296.

Shestakov N.V., Bykov V.G., Konovalov A.V., Fleitout L., Trubienko O., Gerasimenko M.D. (2012) Co- and postseismic crustal deformations in the Russian Far East due to the 2011 Great Tohoku earthquake from GPS observations // Abstracts 33rd General Assembly of the European Seismological Commission (GA ESC 2012), 19-24 August 2012, Moscow and Young Seismologist Training Course (YSTC 2012), 25-30 August 2012, Obninsk, 2012, p. 126.

Shestakov N.V., Gerasimenko M. D., Ohzono M. (2011). Crustal displacements and deformations in the Russian Far East caused by the Tohoku earthquake March 11, 2011 and their impact on GNSS observation results.

Шестаков Н.В., Герасименко М. Д., Охзоно М. Движения и деформации земной коры Дальнего Востока Российской Федерации, вызванные землетрясением Тохоку 11.03.2011 г., и их влияние на результаты GNSS-наблюдений // Геодезия и картография, 2011, № 8, с. 35-43. http://elibrary.ru/item.asp?id=21868948

Shestakov N.V., Gerasimenko M.D. (2013) Combination of postseismic crustal displacements by the heterogeneous geodetic nets data. Шестаков Н.В., Герасименко М.Д. Комбинирование косейсмических смещений земной коры по данным разнородных геодезических сетей // Тезисы докладов Четвертой научно-техническй конференция "Проблемы комплексного геофизического мониторинга Дальнего Востока России", 30 сентября - 4 октября 2013 г., г. Петропавловск-Камчатский. - С. 24.

Shestakov N.V., Gerasimenko M.D., Takahashi H., Kasahara M., Bormotov V.A., Bykov V.G., Kolomiets A.G., Gerasimov G.N., Vasilenko N.F., Prytkov A.S., Timofeev V.Yu., Ardyukov D.G., Kato. T. (2011b). Present tectonics of the southeast of Russia as seen from GPS observations // Geophysical Journal International, 184, p. 529-540, doi: 10.1111/j.1365X.2010.04871.x

Shestakov N.V., Takahashi H., Ohzono M., Bykov V.G., Gerasimenko M.D., Prytkov A.S., Bormotov V.A., Luneva M.N., Kolomiets A.G., Gerasimov G.N., Vasilenko N.F., Baek J., Park P.-H., Sorokin A.A., Bakhtiarov V.F., Titkov N.N., Serovetnikov S.S. (2011). Crustal displacements of East Asia caused by the Tohoku earthquake of march 11, 2011, Mw=9.0 // 7th Biennual Workshop on Japan-Kamchatka-Alaska Subduction Processes: Mitigating Risk through International Volcano, Earthquake, and Tsunami Science (JKASP-2011), Russia, Petropavlovsk-Kamchatsky, August 25-30, 2011. Abstracts Volume. Petropavlovsk-Kamchatsky, p. 48-53. http://www.kscnet.ru/ivs/conferences/kasp/tez/index.html

Shestakov Nikolay V., Hiroaki Takahashi, Mako Ohzono, Alexander S. Prytkov, Victor G. Bykov, Mikhail D. Gerasimenko, Margarita N. Luneva, Grigory N. Gerasimov, Andrey G. Kolomiets, Vladimir A. Bormotov, Nikolay F. Vasilenko, Jeongho Baek, Pil-Ho Park, Mikhail A. Serov. (2012c) Analysis of the far-field crustal displacements caused by the 2011 Great Tohoku earthquake inferred from continuous GPS observations // Tectonophysics, Vol. 524–525 (2012), 76–86, DOI: 10.1016/j.tecto.2011.12.019.

Sorokin A.A., Korolev S.P., Shestakov N.V., Konovalov A.V., Girina O.A. (2013a) Information technology software tools for geophysical and video observation in FEB RAS research of dangerous natural hazards in the Far East Russia. Сорокин А.А.Королёв С.П., Шестаков Н.В., Коновалов А.В., Гирина О.А. Организация работы инструментальных сетей наблюдений ДВО РАН для проведения геофизических исследований и мониторинга опасных природных явлений на Дальнем Востоке // Четвертая научно-техническая конференция. Проблемы комплексного геофизического мониторинга Дальнего Востока России. 30 сентября - 4 октября 2013 г., г. Петропавловск-Камчатский. http://www.emsd.ru/conf2013lib/pdf/techn/Sorokin_etc.pdf

Sorokin A.A., Korolev S.P., Urmanov I.P., Verhoturov A.L., Nesterenkova Ja.S., Shestakov N.V., Konovalov A.V. (2013b) Information technology software tools for instrumental observation of natural process and phenomenon in the Far East of Russia. Сорокин А.А., Королёв С.П., Урманов И.П., Верхотуров А.Л. , Нестеренкова Я.С., Шестаков Н.В., Коновалов А.В. Информационное обеспечение работы инструментальных сетей наблюдений для мониторинга природных процессов и явлений на территории Дальнего Востока России // Информационные технологии и высокопроизводительные вычисления. Материалы всероссийской научно-практической конференции. Хабаровск, 25-27 июня 2013 г., Изд. ТОГУ, 2013, с. 322-325.

Steblov G.M. (2012a) Development of geodynamic monitoring in seismic activity areas of the North Eurasia. Стеблов Г.М. Развитие геодинамического мониторинга в сейсмически активных районах Северной Евразии. Вопросы теории и практики. // Седьмая Международная сейсмологическая школа «Современные методы обработки и интерпретации сейсмологических данных» пос. Нарочь, Республика Беларусь, 9 – 14 сентября 2012 г.

Steblov G.M. (2012b) Research of recent lithosphere movements using satellite geodetic data. Стеблов Г.М. Исследование современных движений литосферы по данным спутниковой геодезии // Тез. Десятая всероссийская открытая ежегодная конференция «Современные проблемы дистанционного зондирования Земли из космоса» (Физические основы, методы и технологии мониторинга окружающей среды, природных и антропогенных объектов), Москва, ИКИ РАН, 12-16 ноября 2012 г.

Steblov Grigory M., Göran Ekström, Mikhail G. Kogan, Jeffrey T. Freymueller, Nikolay N. Titkov, Nikolay F. Vasilenko, Meredith Nettles, Yury V. Gabsatarov, Alexandr S. Prytkov, Dmitry I. Frolov and Mikhail N. Kondratyev. (2014) First geodetic observations of a deep earthquake: The 2013 Sea of Okhotsk Mw 8.3, 611 km-deep, event // Geophys. Res. Lett. 2014. V. 41. N. 11. P. 3826–3832. DOI: 10.1002/2014GL060003.

Steblov G. M., Kogan M. G., Freymueller J. T., Titkov N. N., Ekstrom G., Gabsatarov Y. V., Vasilenko N. F., Nettles M., Prytkov A. S., Frolov D. I. (2013) The size and rupture of the great 2013 deep-focus earthquake beneath the Sea of Okhotsk: constraints from GPS // AGU Fall Meeting, San Francisco, USA, 9-13 December 2013.

Steblov G.M., Vladimirova I.S. (2012) Rheological Models of Great Subduction Earthquakes from Simultaneous Inversion of Coseismic and Postseismic GPS Data // AGU Fall Meeting, San Francisco, USA, 3-7 December 2012.

Takahashi H., Ohzono M., Nakao S., Shestakov N., Gerasimenko M., Vasilenko N., Prytkov A., Bykov V., Luneva M., Gordeev E. (2012a) Great subduction earthquakes and plate coupling along Japanese Islands and their impacts to tectonics in northeastern Asia // Program and abstracts, The Seismological Society of Japan, 2012 Fall meeting, Oct. 16-19, Hakodate, Japan, 2012, p. 149.

Takahashi H., Ohzono M., Nakao S., Shestakov N., Gerasimenko M., Vasilenko N., Prytkov A., Bykov V., Luneva M., Gordeev E. (2012b) Evaluation of crustal deformation induced by great subduction earthquakes and plate coupling in NE-Asian continent – In preparation for reevaluation of Amur plate motion // Abstracts of 118th Meeting of the Geodetic Society of Japan, 2012 Fall meeting, Oct. 31- Nov. 2, Japan, 2012, p. 91-92.

Tatarinov V.N., Morozov V.N., Kaftan V.I., Kagan A.Ya. (2014) Geodynamical monitoring as a basis for conservation biosphere at disposal of radioactive waste. Татаринов В.Н., Морозов В.Н., Кафтан В.И., Каган А.Я. Геодинамический мониторинг как основа сохранения биосферы при захоронении радиоактивных отходов // Международный научно-технический и производственный журнал «Науки о Земле» №3-2014.- с. 5-11.

Tatevian S. (2012) On the role of space geodetic measurements for global changes study. Татевян С.К. О роли космической геодезии в изучении глобальных изменений, в Монографии « Современные глобальные изменения природной среды», т.3 «Факторы глобальных изменений», Москва. Научный центр МИР. 2012. с.128-136.

Tatevian S. (2014a) Space Geodesy Application for the Natural Hazards Monitoring of the Russian Far Eastern Territory, Journal of Geodesy and Geomatics Engineering, 1, 2014. С. 38-41.

Tatevian S.K. (2014b) On the use of space Geodesy for Global Geodynamic Studies, Journal of Geodesy and Geomatics Engineering, David ublishing Comp. New-York, USA. Volume 1, Number 1, December 2014 (Serial number 1), 38-41. 2014.

Tatevian S., Kluykov A., Kuzin S. (2012) On the role of space geodetic measurements for global changes monitoring, Russ. J. Earth. Sci., 12, ES3002, doi:10.2205/2012ES000511(2012).
http://elpub.wdcb.ru/journals/rjes/abstract/v12/2012ES000511-abs.html

Tatevian S.K., G.F.Attia et al. (2014) Monitoring of global geodynamic processes using satellite observations. NRIAG Journal of Astronomy and Geophysics,Elsevier, (2014) 3, 46-52 http//dx.doi.org./10.1016.j.nriag .2014.03.00

Tatevian S.K., Kuzin S.P. On the Use of GLONASS for Precise Positioning and Geodynamic Study. Татевян С.К., Кузин С.П. Использование измерений ГЛОНАСС для точного позиционирования и геодинамических исследований, Труды ИПА РАН, том 27, (2013).

Tereshkina A.A., Shestakov N.V. (2013) To the question of using GLONASS observation in psevdokinematic calculation. Терешкина А.А., Шестаков Н.В. К вопросу об использовании наблюдений ГЛОНАСС при псевдокинематической обработке спутниковых измерений // Тезисы докладов Четвертой научно-техническй конференция "Проблемы комплексного геофизического мониторинга Дальнего Востока России", 30 сентября - 4 октября 2013 г., г. Петропавловск-Камчатский. - С. 23.

Timofeev V. Yu., Kazansky A.Yu., Ardyukov D. G., Metelkin D.V., Gornov P.Yu., Shestakov N.V., Boyko E,V., Timofeev A.V., Gilmatova G.Z. (2011).

Rotation Parameters of the Siberian Domain and Its Eastern Surrounding Structures during Different Geological Epochs. Russian Journal of Pacific Geology, 2011, Vol. 5, No. 4, pp. 288-297. Тимофеев В.Ю., Казанский Ф.Ю., Ардюков Д.Г., Метелкин Д.В., Горнов П.Ю, Шестаков Н.В., Бойко Е.В., Тимофеев А.В., Гильманова Г.З. О параметрах вращения сибирского домена и его восточного обрамления в различные геологические эпохи // Тихоокеанская геология, 2011, т. 30, с. 21-31. http://itig.as.khb.ru/POG/30_4R.html

Timofeev V.Yu., Ardyukov D.G., Solov'ev V.M., Shibaev S.V., Petrov A.F., Gornov P.Yu., Shestakov N.V., Boiko E.V., Timofeev A.V. (2012a) Plate boundaries in the Far East region of Russia (from GPS measurement, seismic-prospecting, and seismological data). Тимофеев В.Ю., Ардюков Д.Г., Соловьев В.М., Шибаев С.В., Петров А.Ф., Горнов П.Ю., Шестаков Н.В., Бойко Е.В., Тимофеев А.В. Межплитные границы Дальневосточного региона России по результатам GPS измерений, сейсморазведочных и сейсмологических данных // Геология и геофизика, т. 53, №4, 2012, с. 489-507.

Timofeev V.Yu., Ardyukov D.G., Solov'ev V.M., Shibaev S.V., Petrov A.F., Gornov P.Yu., Shestakov N.V., Boiko E.V., Timofeev A.V. (2012b). Plate boundaries in the Far East region of Russia (from GPS measurement, seismic-prospecting, and seismological data). // Russian Geology and Geophysics, Vol. 53, 2012, p. 321–336. http://www.sciencedirect.com/science/article/pii/S1068797112000429

Valeev S.G., Kluykov A.A., Kuzin S.P., Tatevian S.K., Fashutdinova V.A. (2011) Studies of the Geocenter dynamics by the analysis of the mesurements of the GPS and DORIS satellite systems. Валеев С.Г., Клюйков А.А., Кузин С.П., Татевян С.К., Фасхутдинова В.А. Исследования динамики геоцентра по результатам анализа измерений спутниковых систем DORIS и GPS, Москва, «Геодезия и картография», 2011, №12 стр.32-38. http://elibrary.ru/item.asp?id=21816606

Vityazev V.V., Miller N.O., Prudnikova E. Ja. The use of the Singular Spectrum Analysis for investigating the pole motion. Витязев В.В., Н.О. Миллер, Е.Я. Прудникова. (2012) Использование сингулярного спектрального анализа при исследовании движения полюса. Вестник СПбГУ, Серия 1, 2012, вып. 2, с. 139-147. http://elibrary.ru/item.asp?id=17789285

Vladimirova I. S., G. M. Steblov, D. I. Frolov. (2011) Viscoelastic deformations after the 2006–2007 Simushir earthquakes // Izvestiya, Physics of the Solid Earth. 2011. V. 47, N.11, P. 1020-1025. DOI:10.1134/S1069351311100132.

Yambaev H.K., Markuze Yu.I. (2014) Structure and algorithm of crustal movement analysis from observation results of satellite reference networks. Ямбаев Х.К., Маркузе Ю.И. Структура и алгоритм анализа движений земной коры по результатам наблюдений региональных спутниковых референц-сетей. Интерэкспо Гео-Сибирь. 2014. Т. 1. № 1. С. 223-225.

# Earth's Rotation

## Gorshkov V.[1], Malkin Z.[1], Zotov L.[2]

[1]Pulkovo Observatory, Saint Petersburg, Russia
[2]Sternberg Astronomical Institute, Moscow Univercity, Moscow, Russia

The development of the rotation theory of solid celestial body on the base of Hamilton-Jakoby theory is proposed by Yurkina [2013a]. Leonhard Euler and other scientists have noticed that the direction of the gravity force can be shifted from the centre of mass of the Earth. To obtain a general solution of the task of the Earth's orientation change in space it is necessary to reflect the fact of such shift in the null iteration. The Hamilton-Yakoby equation gives this opportunity.

The history of computation of preseccion and nutation corrections is described in [Yurkina, 2013b]. The author shows an inconvenience of modern aproaches due to not taking into account a point of application of the force of attraction of the Sun and the Moon in the Earth interiour.

The Russian terrestrial segment of VLBI realization consists of three stations of the Quasar VLBI network. The modernization of instrumental and program means allowed an accuracy of Earth rotation parameters observation and a quality of geodynamic study to be improved [Finkelstain A.M. et al, 2012]. Two main observation programs are being realized for Earth rotation parameters (Ru-E) and universal time (Ru-U) determination at diurnal and hour sessions consequently. Root mean square (RMS) residuals of Earth rotation parameters from IERS 08 C04 series are received as 1 mas for terrestrial pole and 0.38 mas for celestial pole coordinates. RMS residuals for universal time determination are equal to 0.034 ms. RMS values for the universal time determination on the Ru-U program are equal to 0.053 ms.

The relationships between different manifestations of solar and geomagnetic activity and the structural peculiarities of the dynamics of the pole wobble and irregularities in the Earth's rotation were studied using singular spectrum analysis [Gorshkov, Miller, 2011; Gorshkov et al., 2011, 2012a, 2012c]. There are two close major peaks and several lower ones in the Chandler wobble (CW) spectrum. Components in the geomagnetic activity were distinguished in the same frequency band (by the Dst and Ap indices). The amplitude of CW and solar activity (SSArea) component was synchronized, i.e. the amplitude of about forty-year variations of each CW component corresponds to a similar variation in the amplitude of the corresponding component of SSA: the CW amplitude increases with solar activity. Five to seven-year oscillations in the Earth's rotation rate with a complex dynamics of amplitude variations are shown in variations in geomagnetic activity. It is revealed that secular (decade) variations in Earth rotation rate on average repeat global variations in the secular trend of the Earth's geomagnetic field with a delay of eight years during the whole observation period. Beginning from the second half of the 20th century, an increase in solar activity generally corresponds to a decrease in Earth rotation rate (vice versa for Dst).

In [Miller, Malkin, 2012b] a joint analysis of the Polar motion (PM) and celestial pole offset (CPO) time series is performed, which is the difference between the actual and modelled precession-nutation angles, and time series of two geomagnetic indices Kp and Dst. The primary goal was to reveal a possible connection between the Earth's magnetic field variations and CW and FCN excitation. This study was based on the extraction of the common principal components in the four analyzed series using the Multi-Channel Singular Spectrum Analysis and their amplitude and phase analysis using the Hilbert transform. Two groups of common principal components (PCs) were found: trends, and quasi-harmonic terms with near-Chandlerian frequencies for PM, Kp and Dst series, and near-FCN frequency for CPO series (both periods are near 430 days. Comparison of the spectra of the investigated series and their amplitude and phase variations showed some interesting common features. However, the obtained results are still not sufficient to quantify the effects of interconnections of the CW, FCN and geomagnetic field.

A detailed investigation of the long-period PM variations was carried out in several studies, the Chandler wobble (CW) in the first place. Among some interesting CW peculiarities, the phase jump of about $180^o$ with simultaneous drastic decreasing of the CW amplitude occurred in the 1920s. It is supposed to be a unique event and a subject of intensive investigation by many authors. However, using a new 170-year PM time series [Miller, Prudnikova, 2011] allowed two more phase jumps in the 1840s and in the 2000s, which were earlier preliminarily detected at the shorter time interval [Malkin, Miller, 2011].

The paper [Miller, 2011] studies the Chandler component of polar motion, obtained from variations in the Pulkovo latitude over 170 years (1840–2009). The author employed different methods of analysis of non-stationary time series, such as wavelet analysis, methods of band-pass filtering, singular spectral analysis, and Fourier and Hilbert transforms. The long observation record and the methods of analysis of non-stationary time series had allowed identifying two similar structures, both well apparent during the periods of 1845–1925 and 1925–2005 in the time variations of phase and amplitude. The presence of this structure indicates that low-frequency regularities may be present in the Chandler polar motion, and one of the manifestations of this may be well known feature in the region of 1925. The superimposed epoch method was used to estimate the period of variations in the amplitude with a simultaneous change of phase of this oscillation, which was found to be 80 years.

The structural features of CW are shown in [Miller, 2013]. The longest time series of the pole coordinates are used for this purpose. Six components in the interval of 1.1-1.3 year were found by the method of singular spectral analysis. The first two components possess repeated structural features. Sum of these components corresponds to two main peaks of the CW spectrum. The spectral analysis of the variations of other components sum amplitude showed the existence of harmonicas with the periods of 11, 20, 27, 40 years which can be connected with Sun activity and Markovits's waves. During the Multidimentional Singular Spectrum Analysis of CW and geomagnetic indexes of Kp, Ap it was found that the behavior of the CW weak component was synchronized with the similar one, found during the decomposition of indexes. There are about 40 variations coordinated with Solar Activity in the field of century variations of CW amplitude. Besides this, such CW parameters as period, quality factor Q, amplitude and phase of the time variations are estimated in this work.

In [Miller, Vorotkov, 2013] the residuals are obtained after an exclusion of the main components from time series C01, C04 IERS. A separation of the components is carried out by the Singular Spectrum Analysis. An analysis of the residuals allows to test quality of allocated components and to investigate a random component of a row. The random component is obtained by a consecutive exclusion of insignificant components with quasi-periodic character. The research of a casual component allows to reveal the dynamics of its parameters variation in time and to construct mathematical model for its description. Obtained results can be used for forecasting time series of Earth rotation parameters.

VLBI Intensive sessions are scheduled to provide operational Universal Time (UT1) determinations with low latency. UT1 estimates obtained from these observations heavily depend on the model of the celestial pole motion used during data processing. However, even the most accurate precession-nutation model, IAU 2000/2006, is not accurate enough to realize the full potential of VLBI observations. To achieve the highest possible accuracy in UT1 estimates, the CPO correction should be applied. Three CPO models are currently available for users.

In this paper, these models have been tested and the differences between UT1 estimates obtained with those models are investigated. It has been shown that neglecting CPO modelling during VLBI UT1 Intensive processing causes systematic errors in UT1 series of up to 20 µs. It has been also found that using different CPO models causes the differences in UT1 estimates reaching 10 µs [Malkin, 2011c]. The obtained results are applicable to the satellite data processing as well. Analogous result was obtained for the Russian domestic program of operational UT1 determinations on the Quasar VLBI-network of IAA RAS [Malkin, Skurikhina, 2013]. It was found that the systematic differences between the UT1 estimates computed with different CPO models (trend and seasonal terms) are at a level of 1–3 µs. On the other hand, the formal error of the UT1 estimates practically does not depend on the CPO model used.

In [Malkin, 2011b], several publicly available empiric models of the celestial pole offset (CPO) and free core nutation (FCN), included those developed by the author, are investigated and compared each other from different points of view, such as representation of the observation data, FCN parameters variation, prediction accuracy. Based on this study, some practical recommendations are proposed.

FCN amplitude and phase variations are associated with different processes in the Earth's fluid core and core-mantle coupling. The same processes are generally caused the variations in the geomagnetic field (GMF) particularly the geomagnetic jerks (GMJs), which are rapid changes in GMF secular variations. Therefore, the joint investigation of variations in FCN and GMF can elucidate the Earth's interior and dynamics. In this paper, we investigated the FCN amplitude and phase variations derived from VLBI observations. Comparison of the epochs of the changes in the FCN amplitude and phase with the epochs of the GMJs indicated that the observed extremes in the FCN amplitude and phase variations were closely related to the GMJ epochs [Malkin, 2013k]. In particular, the FCN amplitude begins to grow one to three years after the GMJs. Thus, processes that cause GMJs are assumed as sources of FCN excitation.

Free Inner Core Nutation (FICN) is another free rotational mode of the Earth. According to the literature, the FICN period is in the interval 930-1140 days. Detecting of this signal in the observational data is a very important scientific task allowing us to substantially improve our knowledge about the Earth's interior and rotation. Due to small expected amplitude of the FICN oscillation its detection can be successful only from the most accurate nutation series obtained from the VLBI observations. In [Malkin, 2013a, 2014d], some results are presented of further steps in this direction. The author investigated several VLBI nutation series by means of spectral and wavelet analysis. It has been shown that there are several periodic signals with close amplitude around the expected FICN period without a prevailing one, which can be associated with the FICN. So, it seems to be necessary to improve the theoretical estimates of the FICN period to make its search in the observational series more promising.

The accuracy of the current CPO data was analyzed in [Malkin2012f]. CPO time series are initially computed at the IVS Analysis Centers (ACs) routine products. They are then used in the IVS Coordinator Office to derive the IVS combined CPO series. In turn, IERS Combination Centers use original ACs' and/or IVS combined CPO series to derive the IERS combined product. All these transformations between the original series derived by the IVS ACs and the final IERS products are recommended and usually used by users introduce random and systematic differences between CPO series, which in turn lead to differences and inconsistencies between results of users' applications. This situation requires clear recommendations on using CPO series.

Corrections to the IAU 2000/2006 parameters of the theory of precession and nutation are calculated using five different series – two individual series and three combined series that have been used in the literature for this purpose [Malkin, 2014c]. A comparison of the sets of corrections obtained using the different datasets indicates significance systematic differences between them, which often substantially exceed the corresponding random errors. At the same time, existing studies have usually used data obtained from one or two series chosen by the authors without special justification. When refining the theory of precession and nutation, it is necessary to consider and compare various available series of VLBI data if one wishes to reduce the systematic errors in an improved model.

Prediction of the Earth rotation parameters (ERP) is not only an interesting scientific task, but also has many important practical applications, such as ground-based and satellite navigation systems, operational navigation, space missions control, etc. The users' requirements become much more precise during the last few years, which causes intensification of the scientific researches in the field of ERP prediction including improvement of old methods and development of new ones. As a rule each such study is accompanied by the accuracy assessment of the method under investigation. Different methods are often used for this purpose, not always compatible. Besides, methods of the accuracy assessment not always meet the users' requirements. A comparison of several methods of the ERP prediction accuracy assessment has been made, which allows us to obtain more objective data on the quality of the prediction method and its suitability for various applications. A comparison is made of three main methods of the ERP prediction accuracy assessment based on a differences analysis between the predicted and final values: root-mean-square error (RMS), mean absolute error (MAE), and maximum error (MaxE). For the test computations the predictions made at USNO and JPL in the framework of the EOPCPPP campaign were used. The results of this test have shown that the RMS and MAE statistics are practically equivalent for the prediction method comparison. On the other hand, MaxE statistics gives valuable information about the quality of prediction by different methods. The obtained results can be useful also in other fields where a time series prediction is used [Malkin, 2013f].

A new method developed at the Siberian Research Institute of Metrology (SNIIM) for highly accurate prediction of UT1 and Pole coordinates was studied in detail in [Malkin, Tissen, 2012a, 2012b; Malkin et al, 2012]. The method is based on construction of a general polyharmonic model of the variations of the Earth rotation parameters using all the data available for the last 80–100 years, and modified autoregression technique. In this presentation, a detailed comparison was made of real-time UT1 predictions computed making use of this method in 2006–2010 with simultaneous predictions computed at the International Earth Rotation and Reference Systems Service (IERS). Obtained results have shown that proposed method provides better accuracy at different prediction lengths.

Analysis of the Chandler wobble (ChW) shows that its amplitude is decreasing in $2010^{th}$ like in $1930^{th}$ [Zotov, Bizouard, 2014a], it seems it has 70-year amplitude change. At the same time Atlantic Multidecadal Oscillation – the main driver of 70-year changes in Global mean Earth temperature (T) and Sea Level (SL) had maxima in $1930^{th}$ and is in maxima phase now, defining the pause in global warming. Singular spectrum analysis of climatological (T, SL) and Earth rotation (ChW, LOD) time series brought the author of [Zotov, 2013] to the conclusion, that they can be interrelated. In [Zotov, Bizouard, 2014b] atmospheric influence on Chandler wobble was studied, Chandler wobble excitation and variations in global Earth temperature were compared. They show distinctive similarities. The attempts are made to find the possible mechanism, including one based on Earth gravity field monitoring from space [Zotov et al., 2015] and recent works of Yu.G. Markov.

Combined prediction of Earth rotation parameters is performed by L. Zotov in SAI MSU in collaboration with Shanghai observatory. The results are sent to EOPCPPP.

The study of particular spectrum features of Atmospheric Angular Momentum (13.66-day period in CRF), performed in [Bizouard et al, 2014, Sidorenkov et al., 2014] have proved their tidal origin. Thus, an influence of the moon tide on the winds and pressure distribution in the upper atmosphere has been clearly shown.

**References**

Bizouard C., L. Zotov, N. Sidorenkov (2014), Lunar influence on atmospheric angular momentum, Journal of Geophysical Research: Atmospheres, 2014, Wiley, DOI: 10.1002/2014JD022240

Chubey M.S., Kupriyanov V.V., L'vov V.N., Tsekmeyster S.D., Tolchelnikova S.A., Baholdin A.V., Tsukanova G.I., Markelov S.V. (2012) Orbital celestial stereo observatory: scientific and applied significance of the project. Чубей М.С., Куприянов В.В., Львов В.Н., Цекмейстер С.Д., Толчельникова С.А., Бахолдин А.В., Цуканова Г.И., Маркелов С.В. Орбитальная звездная стереоскопическая обсерватория: научное и прикладное значение проекта / Сб. тезисов: ВАК «Пулково-2012», 1–5 октября 2012 г., Санкт-Петербург, ГАО РАН, стр. 76–77.

Finkelstein A., Ipatov A., Smolentsev S., Salnikov A., Surkis I., Gayazov I., Skurikhina E., Kurdubov S., Rahimov I., Dyakov A., Sergeev R., Shpilevsky V. (2012) EOP determination from observations of Russian VLBI-network «QUASAR» // 7th IVS General Meeting "Launching the Next-Generation IVS Network". Madrid (Spain), March 4-9 2012. Abstract's Book, P. 37.

Finkelstein AM.., Ipatov A.V., Gayazov I.S., Skurikhina E.A., Kurdubov S.L., Surkis I.F., Smolentsev S.G., Salnikov A.I., Fedotov L.V., Ivanov D.V., Rahimov I.A., Dyakov A.A., Sergeev R.Yu. (2012) Earth rotation parameter determination from VLBI observation at Quazar-KVO network. Финкельштейн А.М., Ипатов А.В., Гаязов И.С., Скурихина Е.А., Курдубов С.Л,, Суркис И.Ф., Смоленцев С.Г., Сальников А.И., Федотов Л.В., Иванов Д.В., Рахимов И.А., Дьяков А.А., Сергеев Р.Ю. Определение параметров вращения Земли из наблюдений на РСДБ-сети «Квазар-КВО» // Труды ИПА РАН, вып. 23, 2012. С.55-60.

Gorshkov V.L. (2014) Study of pole tide triggering of seismicity. In: Proc. of X International Conference "Problem of Geocosmos", Editors: V.N. Troyan, N. Yu. Bobrov, A. A. Kosterov, A. A. Samsonov, N. A. Smirnova, and T. B. Yanovskaya. SPb., 2014, P. 163-167.

Gorshkov V.L., Miller N.O. (2011) Solar trace in dynamics of Earth's rotation. Горшков В.Л., Миллер Н.О. Солнечный след во вращательной динамике Земли. В сб.: Избранные проблемы астрономии. Материалы III Всероссийской астрономической конференции «Небо и Земля», Иркутск, 22-24 ноября 2011 г., изд. ИГУ, 2011. С. 239-247.

Gorshkov V.L., Miller N.O., Vorotkov M.V. Горшков В.Л., Миллер Н.О., Воротков М.В. (2011) Solar activity manifestation in time series structure of Earth's rotation parameters. Проявление солнечной активности в структуре рядов параметров вращения Земли. Труды Всероссийской ежегодной конференции по физике Солнца. Солнечная и солнечно-земная физика-2011, 2-8 октября 2011 г., С-Петербург, с. 323-326.

Gorshkov V.L., Miller N.O., Vorotkov M.V. (2012a) Manifestation of solar and geodynamic activity in the dynamics of the Earth's rotation. Geomagnetism and Aeronomy, 2012, V. 52, No. 7, 944-952.

Gorshkov V., M. Vorotkov. (2012) On the pole tide excitation of seismicity. Proc. of the 9th International Conference "Problems of geocosmos", St.Petersburg, Petrodvorets, October 8-12, 2012, Eds: V. Troyan, V. Semenov, M. Kubyshkina, pp. 142-145.

Gorshkov V., Vorotkov M., Malkin Z., Miller N., Chapanov Ya. (2012c) On manifestation of solar activity in sea level and Earth's rotation variations. Горшков В., Воротков М., Малкин З., Миллер Н., Чапанов Я. О проявлении солнечной активности в вариациях уровня моря и вращения Земли. В кн.: Тр. Всеросс. ежегодной конф. по физике Солнца "Солнечная и солнечно-земная физика - 2012", СПб: ГАО РАН, 2012, 507-510.

Haas, R., Hobiger T., Nothnagel A., Kingham K., Luzum B., Behrend D., Kurihara S., Uunila M., Malkin Z., Gipson J. (2012) Report on the IVS Task Force on UT1 Intensives. 7th IVS General Meeting: Launching the Next-Generation IVS Network, Madrid, Spain, March 4-9 2012, Abstract's Book, 121. http://elibrary.ru/item.asp?id=17966512

Kaufman M.B., Pasynok S.L. (2011) Rapid EOP calculations using VieVS software, Abstract book of Journeys 2011 "Earth rotation, reference systems and celestial mechanics: Synergies of geodesy and astronomy", 19-21 September 2011, BEV Vienna, Austria, p. 15.

Krilov V.I., Nepoklonov V.B., Yashkin S.N. (2014) Development of resonance perturbation from the Earth represented by a point mass on the base of limited space elliptic task of three bodies for asteroids dangerous approaching to the Earth. Крылов В.И., Непоклонов В.Б., Яшкин С.Н. Вывод резонансных возмущений от Земли, представленной точечной массой, на основе ограниченной, пространственной эллиптической задачи трёх тел для астероидов, опасно сближающихся с Землёй. // Сборник статей по итогам научно-технических конференций: Приложение к журналу «Известия вузов. Геодезия и аэрофотосъёмка», № 6. – вып.7, в двух частях. Часть первая. 2014. С. 8-10.

Liu J.-C., Capitaine N., Lambert S., Malkin Z., Zhu Z. (2012) Systematic effect of the Galactic aberration on the ICRS realization and the Earth orientation parameters. In: IAU XXVIII General Assembly, 2012, Abstract Book, 942-943. http://elibrary.ru/item.asp?id=17966512

Malkin Z. M. (2011a) The Influence of Galactic Aberration on Precession Parameters Determined from VLBI Observations. Astronomy Reports, 2011, Vol. 55, No. 9, 810-815. DOI: 10.1134/S1063772911090058

http://elibrary.ru/item.asp?id=16555930 Малкин З. М. Влияние галактической аберрации на параметры прецессии, определяемые из РСДБ-наблюдений. Астрон. журн., 2011, т. 88, N 9, 880-885.

Malkin Z.M. (2011b) Comparison of CPO and FCN empirical models. In: Proc. Journees 2010: New challenges for reference systems and numerical standards in astronomy, Paris, France, 20-22 Sep 2010, ed. N. Capitaine, Paris, p. 172-175. http://syrte.obspm.fr/journees2010/PDF/Malkin1.pdf

Malkin Z. (2011c) The impact of celestial pole offset modelling on VLBI UT1 intensive results. J. of Geodesy, v. 85, No. 9, p. 617-622. http://www.springerlink.com/content/xl67k7l7u814k1h3/

Malkin Z.M. (2011d) Comparison of CPO and FCN empirical models. In: Proc. Journees 2010 Systemes de Reference Spatio-temporels, Observatoire de Paris, 20-22 Sep 2010, ed. N. Capitaine, Paris, 2011, 172-175.

Malkin Z. (2012a) On the impact of Galactic aberration on parameters of precession-nutation model. In: Schuh H., Boehm S., Nilsson T., Capitaine N. (Eds.) Proc. Journees 2011: Earth rotation, reference systems and celestial mechanics: Synergies of geodesy and astronomy, Vienna, Austria, Sep 19-21, Vienna: Vienna University of Technology, 2012, 168-169.http://syrte.obspm.fr/jsr/journees2011/malkin2.pdf

Malkin Z. (2012b) Celestial pole offset: from initial analysis to end user. 7th IVS General Meeting: Launching the Next-Generation IVS Network, Madrid, Spain, March 4-9 2012, Abstract's Book, 66. http://www.oan.es/gm2012/gm2012AbstractsFinal.pdf

Malkin Z. (2012c) Celestial Pole Offsets: From Initial Analysis to End User. In: IVS 2012 General Meeting Proc., ed. D. Behrend, K.D. Baver, NASA/CP-2012-217504, 2012, 375-379. http://ivscc.gsfc.nasa.gov/publications/gm2012/malkin.pdf

Malkin Z. (2012d) Consistency assessment of celestial pole offset series. Geophysical Research Abstracts, 2012, v. 14, EGU2012-3911. http://elibrary.ru/item.asp?id=17966512

Malkin, Z. (2012e) Pulkovo IVS Analysis Center (PUL) 2011 Annual Report. In: IVS 2011 Annual Report, Eds. D. Behrend, K. D. Baver, NASA/TP-2012-217505, 2012, 256-258. ftp://ivscc.gsfc.nasa.gov/pub/annual-report/2011/pdf/acpul.pdf

Malkin Z.M. (2012f) On accuracy assessment of prediction of Earth rotation parameters. Малкин З.М. Об оценивании точности прогноза параметров вращения Земли. Тр. Всероссийской астрометрической конф. "Пулково-2012", Изв. ГАО, 2013, No. 220, 111-114. http://www.gao.spb.ru/russian/publ-s/izv_220/conf_2012_astr.pdf

Malkin Z. (2012g) The current best estimate of the Galactocentric distance of the Sun based on comparison of different statistical techniques. arXiv:1202.6128, 2012.

Malkin Z. (2012h) Celestial Pole Offsets: From Initial Analysis to End User. In: IVS 2012 General Meeting Proc., ed. D. Behrend, K.D. Baver, NASA/CP-2012-217504, 2012, 375-379.

Malkin Z. (2013a) On the observability of the inner core Earth's nutation. Малкин З.М. О наблюдаемости свободной нутации внутреннего ядра Земли. Тр. Всероссийской астрометрической конф. "Пулково-2012", Изв. ГАО, 2013, No. 220, 115-118.

Malkin Z. (2013b) On the Impact of the Seasonal Station Motions on the Intensive UT1 Results. In: Proc. 21st Meeting of the EVGA, Eds. N. Zubko, M. Poutanen, Rep. Finn. Geod. Inst., 2013, 2013:1, 89-93. ISBN: 978-951-711-296-3. http://evga.fgi.fi/sites/default/files/u3/Proceedings_EVGA2013.pdf

Malkin Z. (2013c) Impact of seasonal station motions on VLBI UT1 Intensives results. J. of Geodesy, v. 87, 2013, No. 6, 505-514. DOI: 10.1007/s00190-013-0624-5. http://link.springer.com/article/10.1007/s00190-013-0624-5

Malkin Z. M. (2013d) Catalogue of optical characteristics of astrometric radio sources OCARS. Малкин З.М. Каталог оптических характеристик астрометрических радиоисточников OCARS. Тр. Всероссийской астрометрической конф. "Пулково-2012", Изв. ГАО, 2013, No. 220, 507-510. http://www.gao.spb.ru/russian/publ-s/izv_220/conf_2012_astr.pdf

Malkin Z. M. (2013e) On assessment of the stochastic errors of source position catalogues. Малкин З.М. Об определении случайных ошибок каталогов координат радиоисточников. Тр. Всероссийской астрометрической конф. "Пулково-2012", Изв. ГАО, 2013, No. 220, 59-64. http://www.gao.spb.ru/russian/publ-s/izv_220/conf_2012_astr.pdf

Malkin Z. (2013f) On the impact of the seasonal station motions on the Intensive UT1 results. In: 21st Meeting of the European VLBI Group for Geodesy and Astrometry, Espoo, Finland, March 5-8, 2013, Book of abstracts, 30. http://evga.fgi.fi/sites/default/files/Abstract_book.pdf

Malkin Z. (2013g) Statistical analysis of the determinations of the Sun's Galactocentric distance. In: Advancing the Physics of Cosmic Distances, Proc. IAU Symp. 289, R. de Grijs (Ed.), 2013, 406-409. DOI: 10.1017/S1743921312021825

Malkin Z. (2013h) Some results of statistical analysis of the Sun galacsy centric distance determination. Малкин З.М. Некоторые результаты статистического анализа определений галактоцетрического расстояния Солнца. Тр. Всероссийской астрометрической конф. "Пулково-2012", Изв. ГАО, 2013, No. 220, 401-406.

Malkin Z.M. (2014a) On accuracy of the precession-nutation theory. Малкин З.М. О точности теории прецессии и нутации. Астрономический журнал. 2014. Т. 91. № 6. С. 490.

Malkin Z.M. (2014b) On weighting of astrometric VLBI observation. Малкин З.М. О взвешивании астрометрических РСДБ-наблюдений. Вестник Санкт-Петербургского университета. Серия 1: Математика. Механика. Астрономия. 2014. Т. 1. № 4. С. 640-649.

Malkin Z.M. (2014c) On the Accuracy of the Theory of Precession and Nutation. Astronomy Reports, 2014, v. 58, No. 6, 415-425. DOI: 10.1134/S1063772914060043

Malkin Z. (2014d) On detection of the free inner core nutation from VLBI data. In: Proc. Journees 2013 Systemes de Reference Spatio-temporels, Paris, France, 16-18 Sep 2013, Ed. N.Capitaine, Paris, 2014, 224-225.

Malkin Z. (2014e) On the Galactic aberration constant. In: Proc. Journees 2013 Systemes de Reference Spatio-temporels, Paris, France, 16-18 Sep 2013, Ed. N.Capitaine, Paris, 2014, 44-45.

Malkin Z.M., Miller N. O. (2011) Amplitude and phase variations of the Chandler wobble from 164-yr polar motion series. In: Proc. Journees 2010: New challenges for reference systems and numerical standards in astronomy, Paris, France, 20-22 Sep 2010, ed. N. Capitaine, Paris, 2011, p. 208-209. http://syrte.obspm.fr/journees2010/PDF/Malkin2.pdf

Malkin Z.M., Skurikhina E.A. (2013) Dependence of the rapid UT1 results obtained from the VLBI network "Quasar" on nutation model. Малкин З.М., Скурихина Е.А. Зависимость результатов оперативных определений UT1 на РСДБ-сети "Квазар" от используемой модели нутации. Тр. Всероссийской астрометрической конф. "Пулково-2012", Изв. ГАО, 2013, No. 220, 119-124. http://www.gao.spb.ru/russian/publ-s/izv_220/conf_2012_astr.pdf

Malkin Z. M., Tissen V. M. (2011) Accuracy assessment of ERP prediction method based on analysis of 100-year series. In: Schuh H., Boehm S., Nilsson T., Capitaine N. (Eds.) Proc. Journees 2011: Earth rotation, reference systems and celestial mechanics: Synergies of geodesy and astronomy, Vienna, Austria, Sep 19-21, Vienna: Vienna University of Technology, 170-171. http://syrte.obspm.fr/jsr/journees2011/malkin3.pdf

Malkin Z.M., Tissen V.M. (2012a) Accuracy research of Earth's rotation parameter forecast by SNIIM technique. Малкин З.М., Тиссен В.М. Исследование точности прогноза параметров вращения Земли методом СНИИМ. Вестн. СПбГУ, Сер. 1, 2012, Вып. 3, 143-152. http://elibrary.ru/item.asp?id=17966512

Malkin Z.M., Tissen V.M. (2012b) Research of the forecast accuracy of the Earth's rotation parameter prediction by SNIIM thechnique. Малкин З. М., Тиссен В. М. Исследование точности прогноза параметров вращения Земли методом СНИИМ. Вестн. СПбГУ, Сер. 1, 2012, Вып. 3, 143-152. http://elibrary.ru/item.asp?id=17966512

Malkin Z., Tissen V., Tolstikov A. (2012) Accuracy assessment of the UT1 prediction method based on 100-year series analysis. Тр. ИПА РАН, 2012, вып. 26, 34-38.

Miller N.O. (2011) Chandler wobble in Pulkovo latitude changes along 170 years. Миллер Н.О. Чандлеровское колебание в изменениях широты Пулкова за 170 лет. Астрон. вестник, т. 45, N. 4, 2011, с. 342-353. http://www.maikonline.com/maik/showArticle.do?auid=VAGQJBJLGR

Miller N.O. (2013) Fine structure and parameters of Chandler pole movement. Миллер Н.О. Тонкая структура и параметры чандлеровского движения полюса. Труды Всероссийской астрометрической конференции «Пулково-2012», Известия ГАО. 2013. № 220. С.125-130.

Miller N., Malkin Z. (2012a) Joint analysis of the Polar Motion and Celestial Pole Offset time series. 7th IVS General Meeting: Launching the Next-Generation IVS Network, Madrid, Spain, March 4-9 2012, Abstract's Book, 129. http://elibrary.ru/item.asp?id=17966512

Miller N., Malkin Z. (2012b) Joint Analysis of the Polar Motion and Celestial Pole Offset Time Series. In: IVS 2012 General Meeting Proc., ed. D. Behrend, K.D. Baver, NASA/CP-2012-217504, 2012, 385-389. http://ivscc.gsfc.nasa.gov/publications/gm2012/miller.pdf

Miller N., Malkin Z. (2012c) Analysis of polar motion variations from 170-year observation series. Тр. ИПА РАН, 2012, вып. 26, 44-53.

Miller N., Malkin Z. (2012d) Joint Analysis of the Polar Motion and Celestial Pole Offset Time Series. In: IVS 2012 General Meeting Proc., ed. D. Behrend, K.D. Baver, NASA/CP-2012-217504, 2012, 385-389.

Miller N.O., Prudnikova E.A. (2011) Early Pulkovo latitude observation. Миллер Н.О., Прудникова Е.А. Ранние пулковские наблюдения широты. Кинем. физ. неб. тел, т. 27, N 1, 2011, с. 40-52. ftp://ftp.mao.kiev.ua/pub/kfnt/27/1/kfnt-27-1-2011-03.pdf

Miller N.O., Vorotkov M.V. (2013) Analysis of residuals after comprehention of main components of the Earth's pole movement. Миллер Н.О., Воротков М.В. Анализ остатков после выделения основных компонент движения полюса земли. Труды Всероссийской астрометрической конференции «Пулково-2012», Известия ГАО. № 220. С.125-130.

Molodensky S.M. (2011a) Models of density distribution and mechanical Q-factor parameters using new data on nutation and proper Earth's oscillations I. Молоденский С.М. Модели распределений плотности и параметров механической добротности по новым данным о нутации и собственных колебаниях Земли I. Неоднозначность решения обратной задачи // Физика Земли.-2011.- № 4. С. 3-18. http://elibrary.ru/item.asp?id=21229011

Molodensky S.M. (2011b) Models of density distribution and mechanical Q-factor parameters using new data on nutation and proper Earth's oscillations II. Молоденский С.М. Модели распределений плотности и параметров механической добротности по новым данным о нутации и собственных колебаниях Земли II. Сравнение с астрометрическими данными // Физика Земли. 2011. № 7. http://elibrary.ru/item.asp?id=16525443

Molodensky S.M., Molodensky M.S. (2012) Allowable value area of parameters of the rigid inner core Q-factor received by nutation and proper Earth's oscillations. Молоденский С.М., Молоденская М.С. Область возможных значений параметров добротности внутреннего твердого ядра по данным о нутации и о собственных колебаниях Земли // Физика Земли. 2012. № 7-8, с. 10-19. http://elibrary.ru/item.asp?id=17795820

Molodensky S.M., Molodensky M.S. (2013a) On diapason of allowable values of mass and moment of inertia of the liquid core. I. Inverse problem of nutation and proper Earth's oscillations solving by decomposition of mechanical parameters using ortogonalised basis. Молоденский С.М., Молоденский М.С. Диапазоны допустимых значений массы и момента инерции жидкого ядра. I. Решение обратной задачи о нутации и собственных колебаний Земли методом разложений механических параметров по ортогонализованному базису.//Физика Земли. 2013. № 4. с. 3-12. http://elibrary.ru/item.asp?id=19086143

Molodensky S.M., Molodensky M.S. (2013b) On diapason of allowable values of mass and moment of inertia of the liquid core. II. Numerical calculation results. Молоденский С.М., Молоденский М.С. Диапазоны допустимых значений массы и момента инерции жидкого ядра. II. Результаты численных расчетов//Физика Земли. 2013, № 4, с. с13-17. http://elibrary.ru/item.asp?id=19086144

Schuh H., Huang C., Seitz F., Brzezinski A., Bizouard C., Chao B., Gross R., Kosek W., Salstein D., Titov O., Richter B., Malkin Z. (2012) Commission 19: Rotation of the Earth. Proc. IAU, v. 7, Transactions T28A, Reports on Astronomy, Ed. I. Corbett, 2012, 33-46. DOI: 10.1017/S1743921312002608 http://journals.cambridge.org/action/displayAbstract?fromPage=online&aid=8527089

Sidorenkov N., C. Bizouard, L. Zotov, D. Salstein. (2014) Atmospheric Angular Momentum , Priroda, 2014, Vol. 4, p. 22-28, RAS

Sokolova Ju., Malkin Z. (2012a) Impact of covariance information on the orientation parameters between radio source position catalogues. 7th IVS General Meeting: Launching the Next-Generation IVS Network, Madrid, Spain, March 4-9 2012, Abstract's Book, 130. http://elibrary.ru/item.asp?id=17966512

Sokolova Ju., Malkin Z. (2012b) Impact of Covariance Information on the Orientation Parameters Between Radio Source Position Catalogs. In: IVS 2012 General Meeting Proc., ed. D. Behrend, K.D. Baver, NASA/CP-2012-217504, 2012, 339-341. http://ivscc.gsfc.nasa.gov/publications/gm2012/sokolova.pdf

Tolchelnikova S.A., Chubey M.S. (2012) Study of inertial movement of the Solar system. Толчельникова С.А., Чубей М.С. К изучению инерциального движения Солнечной системы. // Геодезия и картография.-2012.- № 1.- с. 8–15. http://elibrary.ru/item.asp?id=21623965

Vityazev V.V., Miller N.O., Prudnikova E. Ja. (2012) The use of the Singular Spectrum Analysis for investigating the pole motion. Витязев В.В., Н.О. Миллер, Е.Я. Прудникова. Использование сингулярного спектрального анализа при исследовании движения полюса. Вестник СПбГУ, Серия 1, 2012, вып. 2, с. 139-147. http://elibrary.ru/item.asp?id=17789285

Yurkina M.I. (2013) Development of rotation theory of a rigid celestial body on the base of Hamilton–Jacobi equation. Юркина М.И. Развитие теории вращения твердого небесного тела на основе уравнения Гамильтона-Якоби /

Физическая геодезия. Научно-технический сборник ЦНИИГАиК. – М.: Научный мир, 2013. – с.44-54.

Yurkina M.I. (2013) To the history of determination the Earth's precession and nutation corrections / Физическая геодезия. Научно-технический сборник ЦНИИГАиК. – М.: Научный мир, 2013. – с.198-199.

Zotov L., Christian Bizouard (2014), Prediction of the Chandler wobble, Journees-2014, Pulkovo, 22-24 September 2014 http://syrte.obspm.fr/jsr/journees2014/pdf/

Zotov L., C. Bizouard (2014), Regional atmospheric influence on the Chandler wobble, Advances in Space Research, Elsevier, doi:10.1016/j.asr.2014.12.013

Zotov L.V. (2013), Sea Level and Global Earth Temperature Changes have common oscillations, Odessa Astronomical Publications, vol. 26/2 p 289-291.

Zotov L., Bizouard C. (2013) Study of the prograde and retrograde excitation at the Chandler frequency. Proceedings of Journees 2013, 17 September, 2013, Paris, France.

Zotov L., Bizouard C. (2014) Reconstruction of prograde and retrograde

# Positioning and Applications

Dokukin P.[1], Ustinov A.[2]

[1]People's Friendship University of Russia, Moscow, Russia
[2]JSC Institute Hydroproject, Moscow, Russia

During the last years, much attention has been paid to the monitoring of the structures, GNSS techniques and use of terrestrial laser scanning technology for the different applications.

The problems of use the GNSS monitoring technology to continuously control the hydro-power structures are studied and described in [Kaftan, Ustinov, 2012c, 2012e]. The feasibility of using global radio-navigation satellite systems (GNSS) to improve functional safety of high-liability water-development works — dams at hydroelectric power plants and, consequently, the safety of the population in the surrounding areas is examined on the basis of analysis of modern publications. Characteristics for determination of displacements and deformations with use of GNSS, and also in a complex with other types of measurements, are compared. It is demonstrated that combined monitoring of deformations of the ground surface of the region, and engineering and technical structures are required to ensure the functional safety of HPP, and reliable metrological assurance of measurements is also required to obtain actual characteristics of the accuracy and effectiveness of GNSS observations.

Nowadays more and more attention is paid to permanent observation for structural monitoring using global navigation satellite systems (GNSS). It is conditioned by permanent GNSS development and a possibility of structural control in a near real time mode. Now it is difficult to realize real time relative point position determination with the better than 1 cm accuracy. For that reason high temporal resolution static mode approach is more extended. Thus relative point position is determined hourly and is transferred to analysis center of monitoring system. Now such systems operate at hydropower plant dams.

It is possible to sense intraday periodicities in high frequency baseline vector determination, caused by periodical natural changes and systematic GNSS errors. Such intraday oscillations were registered as an example in station coordinates of regional permanent GNSS networks and are associated to earth tides.

The task of the present research is to study periodical changes in short baseline vectors of local GNSS monitoring networks.

Hourly baseline determination results of local deformation monitoring are analyzed for the purpose to reveal hidden periodicities. All monitoring duration was equal to two months. Baseline lengths were varies from 0.17 to 4.3 km.

Three different spectral analysis techniques were used for more available periodicity determination. The wavelet transformation (Morlet function), Fast Fourier transformation, and sequential dominate harmonic analysis were used in the research. All results received by every technique are rather close to each other. Diurnal and semidiurnal oscillations are dominated in the received spectra.

The causes of it can be both natural and artificial. It is necessary to make further study of a nature of the oscillations. It can be seen that diurnal and semidiurnal waves are more clearly reflected in horizontal vector components and vertical components have high and low frequency oscillations. Amplitudes of diurnal and semidiurnal oscillations are attained to 4 and 3 mm with the standard deviations about 0.1-0.4 mm correspondingly.

It is necessary to reveal the nature of the received oscillations. Further research will be devoted to the problem. It is necessary to note that the cause of the revealed oscillations can be such as real changes due to temperature deformation or artificial systematical errors of GNSS technique. [Kaftan, Ustinov, 2013c]

The general questions of use of the GNSS monitoring technology are studied and described in [Ustinov, 2014]. The authors describe the basic principles, advantages and limitations of the technology of satellite-geodetic deformation monitoring of hydraulic structures. The recommendations for optimal composition, configuration and functional structure of the systems of satellite monitoring are given. The main factors affecting the accuracy of GNSS observations, and methods of accounting for their influence are also listed. Is was noted that the use of satellite technology in combination with traditional methods of monitoring increases the reliability of determination of displacements and, as a consequence, the safety of hydro-power structures and the population of adjacent territories.

The organization of geodetic observations of deformations transport tunnels is represented in [Afonin et al., 2014]. The algorithm for calculating the required accuracy in the observation of deformations is described. The questions of the accuracy of the destination in the observation of deformations in the horizontal and vertical plane are considered/ together with the results and conclusions of the calculations.

The several problems of development of the software package for monitoring dangerous objects are studied in [Brin et al., 2014].

In [Krolichenko, 2012] the authoers tell about the methods of supervision over deformations of the bases and constructions. The calculation of stability of an engineering construction is made and on the basis of this calculation the forecast of behavior of an engineering construction that allows preventing a wrong choice of a platform under building of an engineering construction was done. In the article the method of tilt observations which gives the possibility to continuously receive deformations of a ground under the influence of external pressure and construction deposits is given.

The research [Zarzura, Mazurov, 2014a] deals with the several questions of bridge monitoring using GNSS. The geodynamic bridges safety during their operation is an urgent problem. This research introduces an integrated monitoring system for observing and evaluation of structural deformation behavior of bridges using modern geodetic positioning systems (GNSS). The aim is the selection and realization of the mathematical model of complex analysis of the results of measurements. Displacement of the bridges points depends not only on time, the impact of traffic capacity and wind, but a natural technical system of suspension bridge should also be taken into consideration. This makes it difficult to construct a predictive model.

In [Zarzura, Mazurov, 2014b] the authors describe the study of bridge dynamics from the results of geodetic monitoring using GNSS technologies in conditions of wind and transport traffic. Engineering structures such as bridges are an important and widely used element of regional and urban infrastructure for traffic and transportation. One of the elements of the system to ensure their safety and security is geodetic monitoring using GNSS technologies. Analysis of the dynamics of suspension (cable-stayed) bridges should be based on external influences. The most significant are as follows: the temperature change, the impact of wind and vehicle movement. Here is some analysis of these effects on the dynamics of suspension bridge on real experimental data.

Accuracy estimation of results of landslide process by geodetic monitoring using regression analysis is described in [Kuznetsov, 2011]. The regression data analysis of geodetic monitoring is widely used for determining laws governing landslide process and prognostication. Besides this, with the known nature of the development of deformations, an accuracy analysis of geodetic data according to the actual deviations of separate results from the line of trend is possible. The article describes an example of the estimation of the accuracy of the geodetic monitoring landslide process with the use of linear regression.

A tropospheric delay in GNSS measurements is studied in [Antonovich, 2012]. The theory of troposphere delay calculation under the GNSS measurements is given. The neutral atmosphere parameters of interest and methods of their determination as well as simulation methods are described.

[Gorshkov, Shcherbakova, 2012] investigated the noise and systematic errors of GPS observations inside the Pulkovo Observatory territory. Using the data of three permanent GPS-stations located inside the Pulkovo observatory, the behavior of noise and low-frequency components of their coordinates has been analyzed. There are low-frequency variations of station coordinates from parts of year to the dominant seasonal variations conditioned by atmospheric and hydrological loading. Dynamics of the free from the low-frequency components of the station coordinates and base lines between them were used to estimate the type of distribution of errors. Using the data of all stations the components of the weighted average velocities and its errors were calculated for different types of noise.

In [Karpik, Avrunev, Varlamov, 2014] the authors proposed a methodology for monitoring the accuracy of satellite positioning when creating geospatial software territorial education.

The research [Karpik et al., 2014a] concerns the creation of reference station network to provide a monitoring of oil transportation objects. The authors developed an automated monitoring system with plan-height position of the axis of the main pipeline, including the network reference frame GLONASS / GPS stations, server monitoring and data processing, communication lines and equipment monitoring RTK modes, post-processing and continuous monitoring.

Methodological principles of a system of precise satellite navigation of moving objects with use of terrestrial GLONASS infrastructure are described in [Karpik et al., 2014b]. An accurate real-time satellite navigation system for moving objects is drawing increasing attention of specialists and scientists. Methodological principles for the creation of a single technology platform, navigation and information system competitiveness traffic control municipal and regional levels using GLONASS ground infrastructure and mass consumer equipment to decimeter level accuracy are described in detail. The research points out the principal features and structural elements of navigation and information system for traffic management. It provides some technological and technical solutions for the development and improvement of automobile navigating system as a whole.

The latest research was devoted to the terrestrial laser scanning and its applications.

Today most laser scanner manufacturers make instrument calibration indoors. Such approach has its own disadvantage. A special field test determination of terrestrial laser scanner quality was carried out at the high precision geodetic reference baseline. These problems have been discussed in [Kaftan, Nikiforov, 2011, 2012a, 2012b]

In [Ustinov, Tveritin, 2013, 2014] the authors described the procedure for control survey of water-development works with use of terrestrial laser scanning. The problems that can be resolved with use of ground laser scanning are listed. A proposed procedure for a control survey with use of the method of laser scanning is described in detail. The advantages of laser scanning for updating control documentation are noted. The basic steps in studies related to the control survey are described. The procedures for determination and analysis of deviations in actual geometry of an entity obtained as a result of laser scanning from the design are cited.

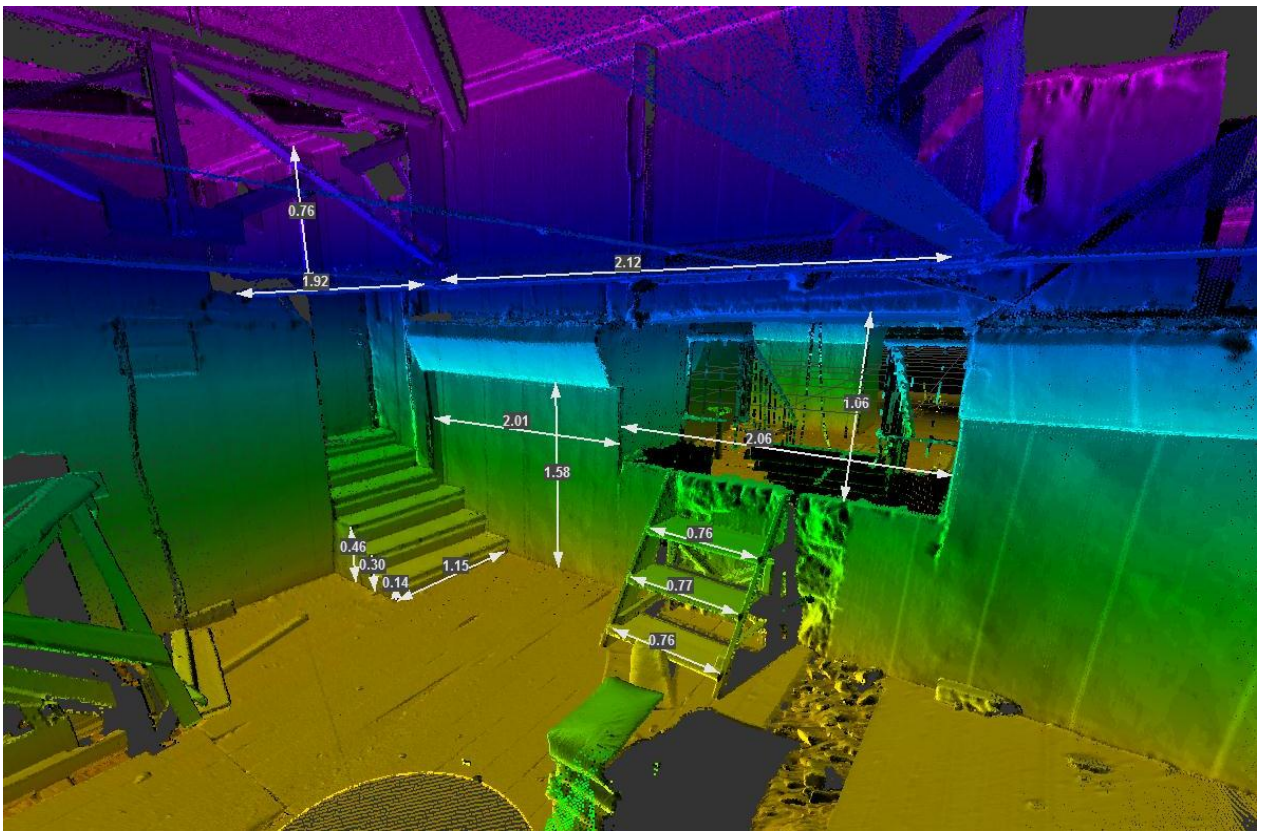

Fig.1 Measuring of point-cloud

During the reporting period several studies in the field of remote sensing and geoinformatics were carried out.

Research of mesoscale cyclone process impact on upper atmosphere and ionosphere of the Earth is accomplished and described in [Bondur, Pulinets, 2012]. The authors describe the mechanisms of emergence and intensification of such hazardous atmospheric vortex phenomena like tropical cyclones (TC), as well as the processes of their electromagnetic interaction with the Earth's ionosphere. The analysis of different models of TC, including the ionization, was carried out. The mechanisms of helical velocity field in the formation of TC, as well as the physical mechanism explaining the established statistical relationship of short-term variations of galactic cosmic rays (Forbush-decreases), the frequency of nucleation and increased TC were also analysed. It is shown that these impacts are due to a decrease of ionization during the Forbush-decreases on the level of the tropopause, and accordingly, by lowering the temperature at the level of the upper edge of the clouds at the expense of reduction of the latent heat associated with condensation of moisture on the newly formed ions. This process leads to the increase of temperature difference between the ocean surface and the upper part of a tropical cyclone, and, consequently, increases the vertical convection, which leads to the intensification of the cyclone. The author concludes that the study of such mesoscale eddies in the atmosphere, as TC, is necessary to consider not only the hydrodynamic features, but also thermodynamic and electromagnetic properties of these structures. The results are important for the organization of research and monitoring of TC, including space methods.

The paper [Savinikh, 2012b] considers the application of remote sensing in the management of transport: the application of remote sensing techniques in the application of intelligent transportation systems (ITS). The basic problems that arise in the management of common transport and the use of ITS were shown and disclosed. The paper deals with the major issues: ensuring the unity of the terminology of the field, ensuring the unity of time, ensuring the unity of the origin, maintenance of ITS in real time, coordinate the implementation of the medium. It shows a possible solution to these problems and the features of the solution of problems in Russia. It is proved that without the use of remote sensing techniques present the solution of problems of transport management cannot be found.

The paper [Savinikh, 2012b] considers the concepts of application of geoinformatics for engineering surveys (ES). The GIS approach as a new investigation method in relation to ES is disclosed. The integrated use of earth remote sensing methods including the engineering survey sphere is described. The navigation field is considered. The main tasks and features of satellite navigation use for ES as well as application of artificial intelligence methods in geoinformatics and implementation of them in engineering surveys are revealed. Geoinformation monitoring as a tool for control and analysis of the state of engineering objects is described. The author pays attention to geoinformation, logistics tasks of which have to be always solved for engineering surveys.

The results of the study of the reaction of the base line (located in Chile) on remote seismic events are described in [Dokukin, Poddubsky, 2011]. An evidence of the reaction of the baseline for the strongest earthquakes - in accordance with the Meshcheryakov scheme is discussed.

The paper [Dokukin, 2011] presents the main results of research carried out at a scientific and educational base "Gornoe" in 2004-2011. New satellite geodetic network for analyzing movements and deformations of the Earth's surface in the vicinity of the Pachelma aulacogene is created. Satellite measurements of the geodetic network are compared with a specially conducted geological research. As a result, there have been new empirical data on the movements of the Earth's surface, caused by seismic events. Authors also performed the experimental gravimetric measurements of a steep bank of the Osetr River. The results of gravimetric measurements allowed to clarify some parameters of a quasigeoid elevation map for the territory.

In the article [Dokukin, Poddubsky, Poddubskaya, 2012a] the results of research of motions and deformations of the Earth's surface using satellite observation in local network are considered. The local satellite geodetic network consists of four points of the Fundamental astronomic-geodetic network, which is located on a rather steady site of the Earth's crust within the East European platform (Moscow region). The research shows that such territories come under the influence of remote seismic events that is possible to register by means of repeated satellite geodetic measurements. However these measurements are in their turn the subject of influence by various factors (including meteoparameters), deforming the results. The present article makes an attempt to reveal the connection between the changes of parameters of deformations of a geodetic network and various external phenomena.

The results of the study of the parameters of movements and deformations of the Earth's surface covered by satellite geodetic network are described in [Dokukin, Poddubsky, Poddubskaya, 2012b]. The analysis of the repeated measurements of the local geodetic network in Moscow region has shown the real possibility of its use for studying the movements and deformations of the territory, and also for the search of geodetic reaction on the strongest earthquakes. The values of the parameters of spatial vectors of displacement of points, and also sizes of the parameters of deformations exceed the errors of measurements in most cases so it is possible to evaluate the reality of movements and deformations of the terrestrial surface and the physical nature of their origin.

The article [Dokukin, Poddubsky, Poddubskaya, 2012c] considers the results of research of exactness of GNSS observation, executed on a standard base in different combinations.

The paper [Dokukin, Melnikov, Bayramov, 2014] considers the results of the evaluation of the stability of reference stations. The conclusion about the stability of reference stations and the possibility of its use for cadastral survey in RTK mode was done.

The possibility of applying the methods of analysis of movements and deformations of the Earth's surface to assess the stability of geodetic networks located in tectonically quiet regions is considered in [Dokukin, 2014].

# References

Abramov O.I., Bondur V.G., Vasilchikov P.M., Pelevin V.V. (2012) Aviation fluorescent scanning lidar (fluorovisor) for monitoring of areas of production and transportation of hydrocarbons. Абрамов О.И., Бондур В.Г., Васильчиков П.М., Пелевин В.В. Авиационный флуоресцентный сканирующий лидар (флуоровизор) для мониторинга районов добычи и транспортировки углеводородов // в кн. «Аэрокосмический мониторинг объектов нефтегазового комплекса» / под ред. Бондура В.Г. М.: Научный мир, 2012, С. 478–486.

Afonin D.A. (2014) Presettlement accuracy estimation of geodetic measurements during organization of deformation monitoring of portal transport tunnel constructions. Афонин Д.А. Предрасчет точности геодезических измерений при организации мониторинга деформаций портальных частей транспортных тоннелей / Д. А. Афонин, Н.Н. Богомолова, М. Я. Брынь // Геодезия и картография. – 2014. – №1 – С.7 – 11. http://elibrary.ru/item.asp?id=21546026

Alekseenko A.G. (2014) Design of mine survey networks taking into account robustness factors. Алексенко А.Г., Зубов А.В. Проектирование маркшейдерско-геодезических сетей с учетом параметров надежности. Маркшейдерский вестник. 2014. №5. С. 31-33.

Alekseeva A.A., Bondur V.G., Dobrozakov A.D., Zhuravel N.E., Kurekin A.S., Pichugin A.P. (2012) Oil-gas territory research by radiolocation technique (Shebelin deposit example). Алексеева А.А., Бондур В.Г., Доброзаков А.Д., Журавель Н.Е., Курекин А.С., Пичугин А.П. Исследование нефтегазоносных территорий радиолокационным методом (на примере Шебелинского месторождения) // в кн. «Аэрокосмический мониторинг объектов нефтегазового комплекса» / под ред. Бондура В.Г. М.: Научный мир, 2012, С. 175–187.

Antonovich K. (2012) Tropospheric delay in GNSS measurements. Антонович К.М. Тропосферная задержка при ГНСС измерениях. Известия высших учебных заведений. Геодезия и аэрофотосъемка. 2012. № 2/1. С. 6-11. http://www.miigaik.ru/journal.miigaik.ru/2012/20120726112123-9081.pdf

Bondur V.G. (2012) Aerospace monitoring of oil-gas territories and objects. Бондур В.Г. Аэрокосмический мониторинг нефтегазоносных территорий и объектов нефтегазового комплекса. Реальности и перспективы // в кн. «Аэрокосмический мониторинг объектов нефтегазового комплекса» / под ред. Бондура В.Г. М.: Научный мир, 2012, С. 15–37.

Bondur V.G. (2013) Resent approaches to the treatment of hyperspectral aerospace images. Бондур В.Г. Современные подходы к обработке гиперспектральных аэрокосмических изображений // Материалы научно-технической конференции «Гиперспектральные приборы и технологии». 17–18 января 2013. г. Красногорск, 2013. С. 14–18.

Bondur V.G., Dobrozakov A.D., Kurekin A.S., Pichugin A.P. (2012) Bi-static radiolocation method for sea surface and hydrocarbon objects monitoring in areas of production and transportation of hydrocarbons. Бондур В.Г., Доброзаков А.Д., Курекин А.С., Пичугин А.П. Метод бистатической радиолокации для контроля состояния морской поверхности и объектов нефтегазового комплекса в районах добычи и транспортировки углеводородов // в кн. «Аэрокосмический мониторинг объектов нефтегазового комплекса» / под ред. Бондура В.Г. М.: Научный мир, 2012, с. С. 466–477.

Bondur V.G., Kuznetsova T.V. (2012) Research of natural oil-gas pollutions on sea surface by space images. Бондур В.Г., Кузнецова Т.В. Исследования естественных нефте- и газопроявлений на морской поверхности по космическим изображениям // в кн. «Аэрокосмический мониторинг объектов нефтегазового комплекса» / под ред. Бондура В.Г. М.: Научный мир, 2012, С. 272–287.

Bondur V.G., Makarov V.A. (2012) New method of remote sensing of geological medium with use of streams of elementary particles. Бондур В.Г., Макаров В.А. Новый активный метод дистанционного зондирования геологической среды с использованием потоков элементарных частиц // в кн. «Аэрокосмический мониторинг объектов нефтегазового комплекса» / под ред. Бондура В.Г. М.: Научный мир, 2012, С. 455–465.

Bondur V.G., Matveev I.A., Murinin A.B., Trekin A.N. (2012) Burn out territories recognition on multispectral images with adaption cloud mask. Бондур В.Г., Матвеев И.А., Мурынин А.Б., Трекин А.Н. Распознавание выгоревших территорий на мультиспектральных изображениях с адаптируемой маской облачности // Известия Южного федерального университета. Технические науки. 2012. Т. 131. № 6. С. 153-156.

Bondur V.G., Murinin A.B., Matveev I.A., Trekin A.N., Yudin I.A. (2013) Method of computational optimisation in the task of collating of raster and vector information in satellite data analysis. Бондур В.Г., Мурынин А.Б., Матвеев И.А., Трекин А.Н., Юдин И.А. Метод вычислительной оптимизации в задаче сопоставления растровой и векторной информации при анализе спутниковых данных // Современные проблемы дистанционного зондирования. 2013. Т. 10. №4. С. 98–106.

Bondur V.G., Murinin A.B., Richter A.A., Shahramanian M.A. (2012) Development of soil degradation level estimation by multispectral images. Бондур В.Г., Мурынин А.Б., Рихтер А.А., Шахраманьян М.А. Разработка алгоритма оценки степени деградации почвы по мультиспектральным изображениям. // Известия Южного федерального университета. Технические науки. 2012. Т. 131. № 6. С. 130-134.

Bondur V.G., Pulinets S.A. (2012) Mesoscale cyclone process impact on upper atmosphere and ionosphere of the Earth. Бондур В.Г., Пулинец С.А. Воздействие мезомасштабных вихревых процессов на верхнюю атмосферу и ионосферу Земли // Исследование Земли из космоса, –2012, №3, с. 3–11.

Bondur V.G., Reznev A.A. (2012) On using of supercomputers for a treatment of aerospace image streams. Бондур В.Г., Резнев А.А. О применении суперкомпьютеров для обработки потоков аэрокосмических изображений // Материалы 2-й Всероссийской научно-технической конференции «Суперкомпьютерные технологии», Дивноморское, Геленджик. – 2012. С.338-345.

Bondur V.G., Sabinin K.D., Grebeniuk Yu.V. (2013) Anomalous variability of inertial oscillations of ocean waves at Hawaiian shelf. Бондур В.Г., Сабинин К.Д., Гребенюк Ю.В. Аномальная изменчивость инерционных колебаний океанских волн на Гавайском шельфе // Доклады Академии наук. 2013. Т. 450. №1. С. 100–104. DOI:10.7868/S0869565213130173 http://www.aerocosmos.net/pdf/2013/anomalnaya_izmenchivost_1_2013.pdf

Bondur V.G., Tikunov V.S. (2013) Developing a Model of Transformation of Cities Based on the Principles of Eco-development and Using Space Monitoring Technologies Бондур В.Г., Тикунов В.С. Разработка модели трансформации городов на основе принципов экоразвития с использованием технологий космического мониторинга. // Сборник статей научно-технической конференции. Построение экологически чистых городов на основании инноваций. Евразийский экономический форум – 2013. С. 59–63. (Valery G. Bondur, Vladimir S. Tikunov. Developing a Model of Transformation of Cities Based on the Principles of Eco-development and Using Space Monitoring Technologies // S&T Sub-Forum's Documents. Innovation-Driven Urban Ecological Development. 2013 Euro-Asia Economic Forum. pp. 65–74).

Bondur V.G., Vorobiev V.E. (2012) Space image treatment methods in hydrocarbon brain objects monitoring. Бондур В.Г., Воробьев В.Е. Методы обработки аэрокосмических изображений, полученных при мониторинге объектов нефтегазовой отрасли // в кн. «Аэрокосмический мониторинг объектов нефтегазового комплекса» / под ред. Бондура В.Г. М.: Научный мир, 2012, С. 395–409.

Bondur V.G., Vorobiev V.E., Grebeniuk Yu.V., Sabinin K.D., Serebrianiy A.N. (2012) Researches of fields of currents and pollution of coastal water at Gelenjik shelf of the Black sea using space data. Бондур В.Г., Воробьев В.Е., Гребенюк Ю.В., Сабинин К.Д., Серебряный А.Н. Исследования полей течений и загрязнений прибрежных вод на Геленджикском шельфе Черного моря с использованием космических данных. // Исследование Земли из космоса, – №4, 2012, с.3-12+4 цв. вклейки.

Bondur V.G., Vorobiev V.E., Zhukov M.A., Zamshin V.V., Karachevtseva I.P., Cherepanova E.V. (2012) Ecological problems of arctic regions in relation with production and transportation of hydrocarbons and possibilities of solving it on the base of space monitoring results. Бондур В.Г., Воробьев В.Е., Жуков М.А., Замшин В.В., Карачевцева И.П., Черепанова Е.В. Экологические проблемы арктических регионов, связанные с добычей и транспортировкой углеводородов, и возможности их решения на основе результатов космического мониторинга // в кн. «Аэрокосмический мониторинг объектов нефтегазового комплекса» / под ред. Бондура В.Г. М.: Научный мир, 2012, С. 329–342.

Bondur V.G., Zamshin V.V. (2012) Space radiolocation monitoring of a sea surface in regions of production and transportation of hydrocarbons. Бондур В.Г., Замшин В.В. Космический радиолокационный мониторинг морских акваторий в районах добычи и транспортировки углеводородов // в кн. «Аэрокосмический мониторинг объектов нефтегазового комплекса» / под ред. Бондура В.Г. М.: Научный мир, 2012, С. 255–271.

Bondur V.G., Zverev A.T., Zima A.L. (2012) Space monitoring of seismic emergency of hydrocarbon areas (Kaliningrad earthquake of September 21 2004) Бондур В.Г., Зверев А.Т., Зима А.Л. Космический мониторинг сейсмоопасности нефтегазоносных районов (на примере Калининградского землетрясения 21 сентября 2004 г.) // в кн. «Аэрокосмический мониторинг объектов нефтегазового комплекса» / под ред. Бондура В.Г. М.: Научный мир, 2012, С. 362–371.

Brin M. (2014) A priory accuracy estimation of taheometer traverses wile execution of addition linear-angular measurements. Брынь М. Априорная оценка точности теодолитных ходов при выполнении дополнительных линейно-угловых измерений / М. Брынь, Н. Богомолова, В. Иванов, Ю. Щербак // Сучасні досягнення геодезичної науки та виробництва: Зб. наук. пр. – Львів, 2014 – вип. II (28). – С. 29–31.

Brin M. (2014) Software package for monitoring of dangerous objects. Брынь, М.Я. Программный комплекс для мониторинга деформаций особо опасных объектов / М.Я. Брынь, А.Д. Хомоненко, В.П. Бубнов, А.А. Никитчин, С.А. Сергеев, П.А. Новиков, А.И. Титов // Проблемы информационной безопасности. Компьютерные системы. – 2014. –№1 – С. 36–41.

Gorshkov V.L., Shcherbakova N.V. (2012) Investigation of the noise and systematic errors of GPS observations inside the Pulkovo Observatory territory. Горшков В.Л., Щербакова Н.В. (2012) Исследование случайных и систематических ошибок GPS-наблюдений на территории Пулковской обсерватории. Науки о Земле, 2012, N 4, 12-22. http://issuu.com/geo-science/docs/geo-science-04-2012

Dokukin P.A., Poddubsky A.A. (2011) Application experience of space geodesy methods for seismic events analyze. Докукин П.А., Поддубский А.А. Опыт применения методов космической геодезии для анализа сейсмических событий (на примере Чили) // Землеустройство, кадастр и мониторинг земель. – 2011. - №4. – с.81-87.

Dokukin P.A. (2011) Research work on scientific and educational base «Gornoe» of the State university of land use planning. Докукин П.А. Результаты научных исследований на научно-учебной базе «Горное» Государственного университета по землеустройству // Международный научно-технический и производственный журнал «Науки о Земле». 2011. - №2. – с. 14-27.

Dokukin P.A., Poddubsky A.A., Poddubskaya O.N. (2012) Research of deformations of the local satellite geodetic network // Международный научно-технический и производственный журнал «Науки о Земле». – 2012. - №1. – с.45-48.

Dokukin P.A., Poddubsky A.A., Poddubskaya O.N. (2012) Monitoring of geodynamic processes in Moscow region, based on the satellite observations of the geodetic network// Международный научно-технический и производственный журнал «Науки о Земле». – 2012. - №2. – с.51-57.

Dokukin P.A., Poddubsky A.A., Poddubskaya O.N. (2012) Satellite measurements analysis of the reference basis // Международный научно-технический и производственный журнал «Науки о Земле». – 2012. - №3. – с.29-35.

Dokukin P.A., Melnikov A.Yu., Bayramov A.N. (2014) The analysis of the base station "Rama" stability of the satellite geodetic network in Ramenskoe district of Moscow region. Докукин П.А., Мельников А.Ю., Байрамов А.Н. Анализ стабильности базовой станции «Rama» спутниковой геодезической сети Раменского района Московской области // Землеустройство, кадастр и мониторинг земель. – 2014. - № 10. – с. 65-70.

Dokukin P.A. (2014) Technique of stability analysis of geodetic networks. Докукин П.А. Методика анализа стабильности геодезических сетей // Теоретические и прикладные проблемы агропромышленного комплекса. 2014 - №2. - с.38-39.

Kaftan V.I., Nikiforov M.V. (2011) Preliminary laser scanner accuracy analysis at etalon baseline of TsNIIGAiK. Кафтан В.И., Никифоров М.В. Предварительный анализ точности измерений лазерного сканера на эталонном базисе ЦНИИГАиК / 7-я Международная научно-практическая конференция «Геопространственные технологии и сферы их применения». Материалы конференции.- М.: Информационное агентство «ГРОМ», 2011.- с.56.

Kaftan V.I., Nikiforov M.V. (2012a) Field calibration of terrestrial laser scanner at etalon baselines. Кафтан В.И., Никифоров М.В. Полевая калибровка наземных лазерных сканеров на эталонных базисах/ 8-я Международная научно-практическая конференция «Геопространственные технологии и сферы их применения». Материалы конференции.- М.: Информационное агентство «ГРОМ», 2012.- с. 98-100.

Kaftan V.I., Nikiforov M.V. (2012b) Laser scanner calibration at short etalon geodetic baseline. Кафтан В.И., Никифоров М.В. Калибровка лазерного сканера на коротком эталонном геодезическом базисе // Геодезия и картография.- 2012.- №5.- с.15-19. http://elibrary.ru/item.asp?id=21769819

Kaftan V.I., Ustinov A.V. (2012) Possibility and necessity of the application of global navigation satellite systems for the monitoring of deformations of hydropower structures. Кафтан В.И., Устинов А.В. Возможность и необходимость применения глобальных навигационных спутниковых систем для мониторинга деформаций гидротехнических сооружений/ Международная научно-практическая конференция «Актуальные вопросы геодезии и геоинформационных систем». Тезисы докладов. Казань, 2012.- с. 25-26.

Kaftan V.I., Ustinov A.V. (2012) Analysis of modern methods of geodetic monitoring of hydropower structures. Кафтан В.И., Устинов А.В. Анализ современных методов геодезического мониторинга гидротехнических сооружений // Седьмая научно-техническая конференция «Гидроэнергетика. Новые разработки и технологии». Тезисы докладов. Санкт-Петербург, 2012.- с. 95-96.

Kaftan V.I., Ustinov A.V. (2012) Use of global navigation satellite systems for the monitoring of deformations of hydropower structures. Кафтан В.И., Устинов А.В. Применение глобальных навигационных спутниковых систем для мониторинга деформаций гидротехнических сооружений // Гидротехническое строительство.- 2012.- №12.- с.11-19. http://elibrary.ru/item.asp?id=18274426

Kaftan V.I., Ustinov A.V. (2013) Periodicities in results of local structural monitoring with the use of satellite radionavigation systems. Кафтан В.И., Устинов А.В. Периодичности в результатах локального мониторинга сооружений с использованием спутниковых радионавигационных систем / Инновационные процессы в АПК [Текст]: сборник статей V Международной научно-практической конференции преподавателей, молодых ученых аспирантов и студентов. Москва, 17-19 апреля 2013 г.-М.: РУДН, 2013.-433-435.

Kaftan V.I., Ustinov A.V. (2013) Use of global navigation satellite systems for monitoring deformations of water-development works. Power Technology and Engineering, May 2013, Volume 47, Issue 1, pp. 30-37. http://link.springer.com/article/10.1007/s10749-013-0392-7

Kaftan V., Ustinov A. (2013) Diurnal and semidiurnal periodicities in results of local structural monitoring using global navigation satellite systems. International Association of Geodesy, Scientific Assembly 150th Anniversary of the IAG, Book of Abstracts, Book of Abstracts, September 1-6, 2013, Potsdam, p.430. http://www.iag2013.org/IAG_2013/Publication_files/abstracts_iag_2013_2808.pdf

Karpik A.P., Lipatnikov L.A. (2014) Problems and prospects of precise positioning with use of common apparatuses of GNSS user. Карпик А.П., Липатников Л.А. Проблемы и перспективы точного позиционирования с использованием массовой аппаратуры потребителя ГНСС. Интерэкспо Гео-Сибирь. 2014. Т. 1. № 2. С. 113-117.

Karpik A.P., Avrunev A.P., Varlamov A.A. (2014) Improvement of quality control technique for satellite positioning in the process of creation of a territorial geoinformatic space. Карпик А.П., Аврунев Е.И., Варламов А.А. Совершенствование методики контроля качества спутникового позиционирования при создании геоинформационного пространства территориального образования. Известия высших учебных заведений. Геодезия и аэрофотосъемка. 2014. № S4. С. 182-186. http://elibrary.ru/item.asp?id=22477713

Karpik A.P., Antonovich K.M., Tverdovsky O.V., Lagutina E.K., Reshetov A.P. (2014) Creation of reference station network to provide a monitoring of oil transportation objects. Карпик А.П., Антонович К.М., Твердовский О.В., Лагутина Е.К., Решетов А.П. Создание сети референцных станций для обеспечения мониторинга объектов транспорта нефти и нефтепродуктов. Интерэкспо Гео-Сибирь. 2014. С. 151-161.

Karpik A.P., Ganagina I.G., Goldobin D.N., Kosarev N.S. (2014) Methodological principles of a system of precise satellite navigation of moving objects with use of terrestrial GLONASS infrastructure. Карпик А.П., Ганагина И.Г., Голдобин Д.Н., Косарев Н.С. Методологические принципы системы точной спутниковой навигации подвижных объектов с использованием наземной инфраструктуры ГЛОНАСС. Известия высших учебных заведений. Геодезия и аэрофотосъемка. 2014. № 5. С. 69-74. http://elibrary.ru/item.asp?id=22834224

Krolichenko O.V. (2011) Methods of deformation observation of foundations and constructions. Кроличенко О.В. Методы наблюдения за деформациями оснований и сооружений // Международный научно-технический и производственный журнал «Науки о Земле» - 2011. - №2 - с.35-38. http://geo-science.ru/wp-content/uploads/35-38.pdf

Kuznetsov A.I. (2011) Accuracy estimation of results of landslide process geodetic monitoring using regression analysis. Кузнецов А.И. Оценка точности результатов геодезического мониторинга оползневых процессов с использованием регрессионного анализа // Геодезия и картография.- 2011.-№10.- с. 8-13. http://elibrary.ru/item.asp?id=21831825

Kuznetsov A.I., Moiseenko S.A., Volkov V.A. (2011) An experience of geodetic monitoring data usage for landslide sliding surface construction. Кузнецов А.И., Моисеенко С.А., Волков В.А.. Опыт использования данных геодезического мониторинга для построения поверхности скольжения оползня // Инженерные изыскания .- 2011.-№ 2.- с. 56-59.

Lazareva N.S. (2011) Calibration the non-metric small format cameras for thereof use for the decision of some problems photogrammetry. Лазарева Н.С. Калибровка неметрических малоформатных камер и их применение для решения некоторых задач фотограмметрии // Международный научно-технический и производственный журнал «Науки о Земле» - 2011. - №1 - с.80-91. http://geo-science.ru/wp-content/uploads/GeoScience-01-2011-p-80-91.pdf

Mayorov A.N. (2013) Research of the accuracy of the digital elevation model SRTM. Майоров А.Н. Исследование точности цифровой модели рельефа из SRTM / Физическая геодезия. Научно-технический сборник ЦНИИГАиК. – М.: Научный мир, 2013. – с.99-114.

Savinikh V.P. (2012a) Use of methods of remote sensing for management of transport. Савиных В.П. Использование методов дистанционного зондирования для управления транспортом // Международный научно-технический и производственный журнал «НАУКИ О ЗЕМЛЕ».- 2012.- №2.- с. 58-62. http://issuu.com/geo-science/docs/geo-science-02-2012

Savinikh V.P. (2012b) Concepts of application of geoinformatics in engineering researches. Савиных В.П. Концепции применения геоинформатики в инженерных изысканиях // Инженерные изыскания.-2012.- № 7.- с. 8-11. http://elibrary.ru/item.asp?id=17889270

Savinikh V.P. (2012c) Investigation of Nordic areas using remote sensing data. Савиных В.П. Исследование северных территорий по материалам ДДЗ // Математические методы и модели анализа и прогнозирования развития социально-экономических процессов черноморского побережья Болгарии, Материалы Международной научно-практической конференции, 2012.- с.64-67.

Savinikh V.P., Chibunichev A.G. (2012) Earth remote sensing data in researches of the Moscow State University of Geodesy and Cartography. Савиных В.П., Чибуничев А.Г. Данные дистанционного зондирования Земли в исследовательских проектах Московского государственного университета геодезии и картографии // Научно-производственный журнал «Земля Беларуси», 2012.- с. 13-15.

Savinikh V.P., Malinnikov V.A., Mayorov A.A., Cvetcov V.Ya. (2012) Geoinformatics: Development tendencies. Савиных В.П., Малинников В.А., Майоров А.А., Цветков В.Я. Геоинформатика: Тенденции развития / 8-я Международная научно-практическая конференция «Геопространственные технологии и сфера их применения» // Информационное агентство «Гром», 2012.- с. 11-15.

Simonian V.V., Kuznetsov A.I., Chernenko E.S., Piatnitskaya T.A. (2011) Instrumental determination of wall tilt of Boris-Gleb Monastery. Симонян В.В., Кузнецов А.И., Черненко Э.С., Пятницкая Т.А. Инструментальное определение кренов стен Борисоглебского монастыря // Вестник МГСУ № 1/2011, Т.2, стр. 239-243.

Устинов А.В., Тверитин А.Л. (2013) Технология исполнительной съемки гидротехнических сооружений с использованием наземного лазерного сканирования // Гидротехническое строительство. 2013. №12. С.2-5. http://elibrary.ru/item.asp?id=20928716

Ustinov A.V., Tveritin A.L. (2014) Procedure for Control Survey of Water-Development Works with Use of Ground Laser Scanning // Power Technology and Engineering. 2014. Volume 48. Issue 1. P. 13-16, DOI: 10.1007/s10749-014-0475-0  http://link.springer.com/article/10.1007/s10749-014-0475-0 http://elibrary.ru/item.asp?id=20442656

Ustinov A.V. (2014) Technology of satellite geodetic monitoring of hydropower structures. Устинов А.В. Технология спутникового геодезического мониторинга гидротехнических сооружений. Гидротехническое строительство. 2014. № 6. С. 39-43. http://elibrary.ru/item.asp?id=21645039

Varvarina E.A. (2011) Creation experience ortophotoplans with use of the data of air laser scanning on linear object. Варварина Е.А. Опыт создания ортофотопланов с использованием данных воздушного лазерного сканирования на линейный объект// Международный научно-технический и производственный журнал «Науки о Земле» - 2011. - №1 - с.47-49. http://geo-science.ru/wp-content/uploads/47-49.pdf

Varvarina E.A., Gavrilova L.A. (2011) Improvement technology of creation ortophotoplans of linear objects. Варварина Е.А., Гаврилова Л.А. Совершенствование технологии создания ортофотопланов линейных объектов // Международный научно-технический и производственный журнал «Науки о Земле» - 2011. - №1 - с.76-80. http://geo-science.ru/wp-content/uploads/GeoScience-01-2011-p-76-80.pdf

Zaitsev A.K. (2011) Design of the specified accuracy polygonometric rut. Зайцев А.К. Проектирование полигонометрического хода заданной точности // Международный научно-технический и производственный журнал «Науки о Земле» - 2011. - №2 - с.7-13. http://geo-science.ru/wp-content/uploads/7-13.pdf

Zarzura F.H., Mazurov B.T. (2014a) Bridge monitoring using GNSS. Зарзура Ф.Х., Мазуров Б.Т. Мониторинг мостов с использованием ГНСС. Интерэкспо Гео-Сибирь. 2014. Т. 1. № 1. С. 176-181.

Zarzura F.H., Mazurov B.T. (2014b) Bridge dynamics from results of geodetic monitoring using GNSS technologies in conditions of a wind and transport traffic. Зарзура Ф.Х., Мазуров Б.Т. Динамика мостов по результатам геодезического мониторинга с использованием ГНСС-технологий в условиях влияния ветра и транспортного движения. Интерэкспо Гео-Сибирь. 2014. Т. 1. № 1. С. 181-186.

# Common and Related Problems

**Kaftan V.[1], Malkin Z.[2]**

[1]Geophysical Center of the Russian Academy of Sciences, Moscow, Russia
[2]Pulkovo Observatory, Saint Petersburg, Russia

The possibility of using topological methods to describe the fractal geometry of geodetic lines is considered by [Malinnikov et al, 2014]. It is shown that the geodetic lines are fractal objects. The propagation of light can be described as a process occurring in a Finsler space. Meaningful implementation of the Finsler geometry in geodesy can help take a new look at its traditional tasks, but will also contribute to the construction of new approaches to problem areas of space geodesy and astrometry.

Methods of classical mechanics and astrometry are shown to be sufficient for deriving the exact mathematical formulae necessary for the solution of the problems of astrometry and connected to it sciences. From exact formulae it is not difficult to proceed to formulae of any approximation required by practice. Incompatibility of principles of the SRT and GRT with the principles of classical mechanics is shown and the question is raised of validity of the opinion submitted in academic magazine «Common sense», according to which the relativistic mechanics might be considered as more precise than mechanics of Newton [Tolchelnikova, 2014].

Astronomical method for clock synchronization is compared with the one based on light travel time which Einstein considers as universal, applicable not only in rest system, but for the inertial motion of bodies and systems of bodies as well. In spite of the process of differentiation of science in XX century, empirical basis securing the close cooperation of astronomy with geodesy and gravimetry was preserved up to "revolution in astrometry". Synthetic approach is urgent for efficient development of fundamental research, example of the one is demonstrated in the monograph «Gravimetry and Geodesy» [Tolchelnikova, 2013].

The possibility of determination of the Solar system motion's velocity by means of observations of Jupiter satellites was questioned by James Maxwell in 1879. The answer to Maxwell question is of interest to astronomers and physicists because the impossibility to determine the considered velocity follows from the fundamental principle of SRT. The authors [Tolchelnikova & Chubey, 2012] show the possibility to obtain the direction of projection (v) of the Solar system velocity to the plane of ecliptic and impossibility to obtain the value of v without the additional observations from Space.

Construction of a Lunar Base is discussed as an important long-term space program [Savinykh et al, 2014]. The data is given on the progress in studying the Moon, on the states-participants of lunar expeditions and scheduled projects-both domestic and foreign. Various aspects of this problem are considered: delivery of cargo to the Moon's surface, prospects of industrialization, the expediency of using lunar space stations located in the libration points of the Earth-Moon system as "a trans-shipment terminal". National and planetary problems to be solved while using the Lunar Base are listed. It is necessary to build a system of time-coordinate and navigation support on the Moon, which will be required in the lunar base constructing and functioning and at various stages of performing this work. The paper proves the importance of establishing a lunar base for resolving the Earth's problems and those of people further getting into space.

Problems of building the global positioning system for uninhabited planets and natural satellites are discussed in [Shirenin et al, 2014]

Analysis of the geodetic time series involves many statistical methods. Among them are computation of correlation between time series and the Allan variation of time series used to estimate their noise characteristics. However, the classical definitions of these statistics have been don't take into account the uncertainty of measurements under consideration. The latters can substantially bias the statistical estimates when the measurements have very different errors. To overcome this shortcoming, the weighted modifications was proposed of coefficient of correlation [Malkin, 2013e] and the Allan variance [Malkin, 2011b, 2013d], which proved to be an effective statistical tool for analysis of the real observational data.

Combining several independent measurements of the same physical quantity is one of the most important tasks in metrology. Small samples, biased input estimates, not always adequate reported uncertainties, and unknown error distribution make a rigorous solution very difficult, if not impossible. For this reason, many methods to compute a common mean and its uncertainty were proposed, each with own advantages and shortcomings. To overcome some problems of known approaches to compute the WA uncertainty, a new combined estimate has been proposed [Malkin, 2011a, 2013a, 2013b, 2013c]. It has been shown that the proposed method can help to obtain more robust and realistic estimate suitable for both consistent and discrepant measurements.

The studies were continued on application of the Singular Spectral Analysis (SSA) to analysis of the time series of astrometric and geodynamical data. This method allows separating different (quasi)periodical and trend-like signals with good frequency and time resolution even in presence of time-variable noise. Detailed analysis of the method was given in [Vityazev et al., 2012]. Using the SSA method several important results were obtained in analysis of the Chandler wobble [Miller, 2011, 2013, 2014; Miller, Malkin, 2012a; Miller, Vorotkov, 2013] and nutation series [Miller, Malkin, 2012b]. Advantages of singular spectral analysis for studying the long-period time series with involved structure have been demonstrated.

The authors [Gerasimenko & Kamorny, 2014] examine the adjustment questions of repeated geodetic measurements performed for the study of crustal deformation or engineering structures. It is shown that in the separate adjustment systematic errors will not affect the determination of the displacement vector only under the same weights in both periods of measurement, but the mean square error of unit weight will be distorted. This disadvantage is eliminated in the adjustment of differences of measurements. The reliability assessment, contrary to known results, is not deteriorated.

The algorithm of calculation of plan rectangular coordinates, declinations and scale of projection Gauss in 6º zone by geodetic coordinates is offered [Balandin et al, 2014]. This algorithm is used as alternative of the Gauss algorithm.

The solar activity is distinguished as a factor of influence to satellite measurements and the Earth's dynamic processes. Kinematic approach to solar activity modeling and forecasting is described in [Kaftan, 2012]. Proposed approximation model has provided a successful prediction of the $23^{rd}$ solar cycle (SC) and especially its final unusually long stage. The amplitude of the current $24^{th}$ SC was predicted as 130$\pm$20 sunspot number units Ri. Real maximum is equal to 102.3 Ri. Two median shape of the current cycle is predicted accurately. Long term behavior of solar activity is analyzed in [Komitov & Kaftan, 2013]. The authors suggest that the long term solar minimum can probably begins in the nearest decades.

The history of the Russian and international geodetic investigation is described in [Kaptüg, 2013; Mazurova, 2013, 2014; Teterin, 2013; Teterin, Sinianskaya, 2011].

Memories of famous scientists Euler, Molodensky, Krasovsky, Pellinen, Eremeev & Yurkina, and also biographies and anniversary congratulations are presented in articles [Brovar, 2013; Brovar & Gusev, 2013; Mazurov & Medvedev, 2014; Ogorodova et al, 2014; Prudnikova et al, 2013; Tolchelnikova, 2013; Yurkina, 2012; Yurkina et al, 2103].

Memory notes about famous Pulkovo latitude observers Lidia Kostina, Natalia Persiyaninova and Ivan Korbut are presented in [Malkin et al, 2013, Prudnikova et al., 2013].

Scientific event devoted to 150 anniversary of IAG is described in [Kaftan, 2013]. Brief review of scientific presentations and events is presented. The schedule of the assembly is described. Brief information of contrnt of some papers and lectures is delivered.

A historical book of the oldest Russian gravimetrist Nikolay Gusev is reviewed in [Baranov & Yuzefovich, 2014].

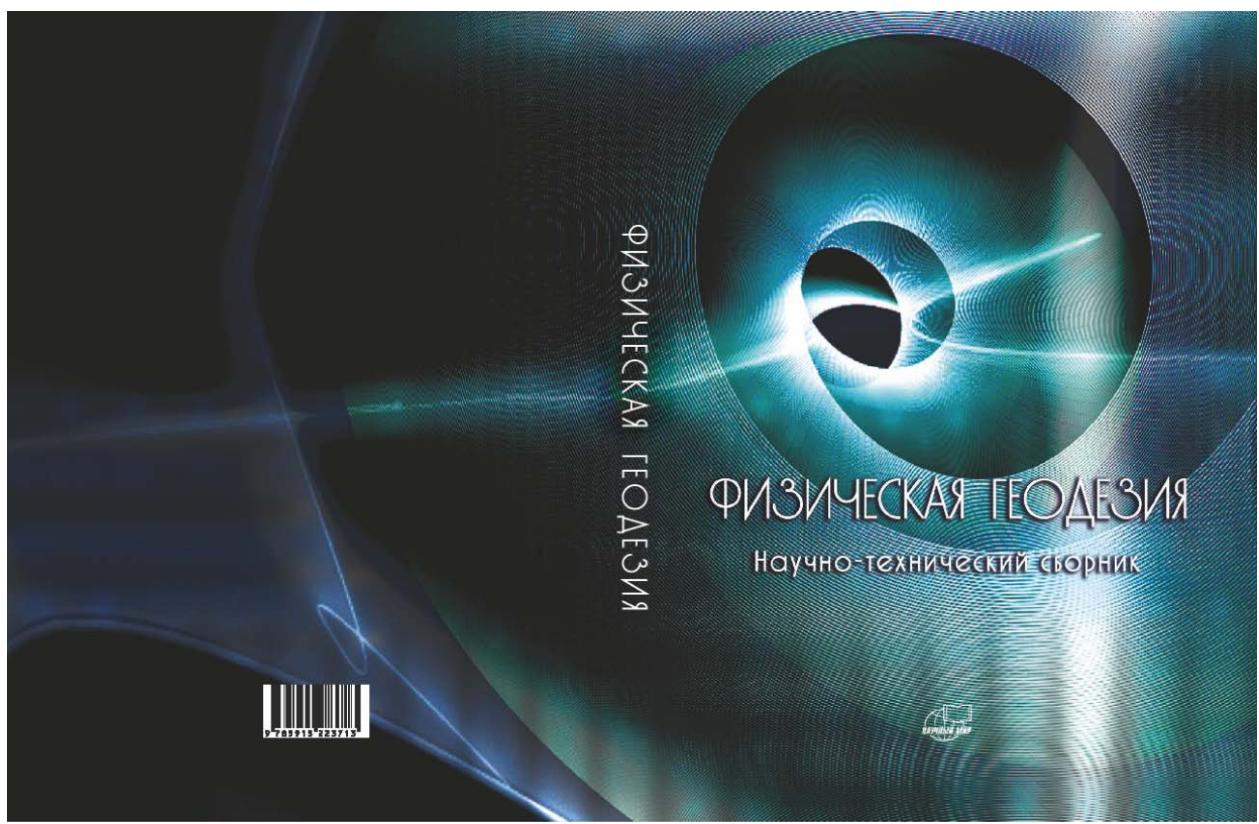

Fig. Physical geodesy. Scientific and technological paper collection of TsNIIGAiK.

**References**

Bakaliarov A.M., Bondur V.G., Karetnikov M.D., Lebedev V.I., Makarov V.A., Muradian G.V., Murinin A.B., Yakovlev G.V. (2012) Digital modeling use of shift emission for solving task of distant recognition of gamma quant source with given specter. Бакаляров А.М., Бондур В.Г., Каретников М.Д., Лебедев В.И., Макаров В.А., Мурадян Г.В., Мурынин А.Б., Яковлев Г.В. Использование численного моделирования переноса излучения для решения задачи дистанционного обнаружения источника гамма-квантов с заданным спектром // Известия Южного федерального университета. Технические науки. 2012. Т. 131. № 6. С. 73-77. http://elibrary.ru/item.asp?id=17800990

Balandin V.N., Bryn M.Ya., Menshikov I.V., Firsov Yu.G. (2014) Computation of the plane rectangular coordinates, declinations and scale of Gauss-Kruger projection in 60 zone using geographic coordinates. Баландин В.Н., Брынь М.Я., Меньшиков И.В., Фирсов Ю.Г. Вычисление плоских прямоугольных координат, сближения меридианов и масштаба проекции Гаусса в 6-градусной зоне по геодезическим координатам // Геодезия и картография. – 2014. –№2 – С. 11–13. http://elibrary.ru/item.asp?id=21831864

Baranov V.N., Yuzefovich A.P. (2014) Memoirs of veteran gravimetrist. Баранов В.Н., Юзефович А.П. Воспоминания старейшего гравиметриста. Геодезия и картография.- 2014. № 7. С. 61-62. http://elibrary.ru/item.asp?id=21831909

Bibliography of M.I.Yurkina publications. (2013) Библиография публикаций М.И.Юркиной / Физическая геодезия. Научно-технический сборник ЦНИИГАиК. – М.: Научный мир, 2013. – с.247-249.

Bondur V, Grebenyuk Yu., Ezhova E., Kandaurov A., Sergeev D., and Troitskaya Y. (2012a) Applying of PIV/PTV methods for physical modeling of the turbulent buoyant jets in a stratified fluid // InTech «PIV» Edited by Giovanna Cavazzini, ISBN 978-953-51-0625-8, Hard cover, 386 pages, Publisher: InTech, Published: May 23, 2012 under CC BY 3.0 license, in subject Mechanical Engineering, http://www.intechopen.com/books/the-particle-image-velocimetry-characteristics-limits-and-possible-applications

Bondur V.G., Krapivin V.F., Potapov I.I., Soldatov V.Yu. (2012b) Natural disasters and environment. Бондур В.Г., Крапивин В.Ф., Потапов И.И., Солдатов В.Ю. Природные катастрофы и окружающая среда // Проблемы окружающей среды и природных ресурсов. 2012. № 1. С. 3-160.

Bondur V.G., Zverev A.T., Gaponova E.V. (2014) Regularity of lineament precursor dynamics registered from the space in the process of earthquakes. Бондур В.Г., Зверев А.Т., Гапонова Е.В. Закономерность предвестниковой динамики линеаментов, регистрируемых из космоса, при землетрясениях. Известия высших учебных заведений. Геодезия и аэрофотосъемка. 2014. № 1. С. 89-94. http://www.miigaik.ru/journal.miigaik.ru/arhiv_zhurnalov/vypuski_za_2014_/20150120172913-6843.pdf

Brovar B.V., Gusev N.A. (2013) On the occasion of the eighty year birthday of Vadim Andreevich Taranov. Бровар Б.В., Гусев Н.А. К 80-летию Вадима Андреевича Таранова / Физическая геодезия. Научно-технический сборник ЦНИИГАиК. – М.: Научный мир, 2013. – с.247-249.

Brovar B.V., Kaftan V.I., Yurkina M.I. (2013) On the occasion of the eighty year birthday of Nikolay Alexandrovich Gusev. Бровар Б.В., Кафтан В.И., Юркина М.И. К 80-ти летию Николая Александровича Гусева / Физическая геодезия. Научно-технический сборник ЦНИИГАиК. – М.: Научный мир, 2013. – с.241-246.

Brovar B.V., Rubtsova Z.V., Tutova T.A., Scherbakova A.B. (2013) On the life and activity of M.I.Yurkina and V.F.Eremeev. Бровар Б.В., Рубцова З.В., Тутова Т.А., Щербакова А.Б. О жизни и деятельности М.И.Юркиной и В.Ф.Еремеева. / Физическая геодезия. Научно-технический сборник ЦНИИГАиК. – М.: Научный мир, 2013. – с.250-268.

Gerasimenko M.D. (2013) On the question of "Geometric Interpretation of the Adjustment Theory." Герасименко М.Д. К вопросу "О геометрической интерпретации сущности уравнивания" // Известия высших учебных заведений. Геодезия и аэрофотосъемка, 2013, № 3, с.25-27. http://www.miigaik.ru/journal.miigaik.ru/2013/20130830165548-6482.pdf

Gerasimenko M.D., Kamorny V.M. (2014) Adjustment of repeated measurements in case the systematic errors presence. Герасименко М. Д., Каморный В. М. Уравнивание повторных геодезических измерений при наличии систематических ошибок // Геодезия и картография. 2014. № 9. С. 7-8. http://elibrary.ru/item.asp?id=22253720

Hanchuk A.I., Sorokin A.A., Smagin S.I., Korolev S.P., Makogonov S.V., Tarasov A.G., Shestakov N. V. (2013) On the elaboration of information-telecommunication systems in the Far East Division RAS. Ханчук А.И., Сорокин А.А., Смагин С.И., Королёв С.П., Макогонов С.В., Тарасов А.Г., Шестаков Н.В. О развитии информационно-телекоммуникационных систем в Дальневосточном отделении РАН // Информационные технологии и вычислительные системы, 2013. - №4. С. 45-57. http://elibrary.ru/item.asp?id=21016916

Kaftan V.I. (2011a) Three whales of geodesy: geometry, gravimetry, astrometry. Кафтан В.И. Три кита геодезии: геометрия, гравиметрия, астрометрия // Кадастр недвижимости .- 2011.- №1.-(22).-с.33-37. http://www.roscadastre.ru/?id=625

Kaftan V.I. (2011b) Is it easier to pull down than to build? Remarks to the Conception of reforming of the state geodesy and cartography administration to 2020 and to the project of the state law "About changes of the Federal low about geodesy and cartography". Кафтан В.И. Ломать - не строить? Замечания к Концепции развития отрасли геодезии и картографии до 2020 года и к проекту закона «О внесении изменений в Федеральный закон «О геодезии и картографии»»// Кадастр недвижимости .- 2011.- №2.-(23).-с.91-94. http://elibrary.ru/item.asp?id=16324076

Kaftan V. (2012a) Kinematic Approach to the 24th Solar Cycle Prediction, Advances in Astronomy, Volume 2012, Article ID 854867, 7 pages, doi:10.1155/2012/854867 http://www.hindawi.com/journals/aa/2012/854867/

Kaftan V.I. (2012b) On the new project of the low "On geodesy, cartography and …" Кафтан В.И. О новом проекте закона «О геодезии, картографии и …»// Кадастр недвижимости .- 2012.- №3.-(28).-с.37-40.

Kaftan V.I. (2013a) Geodetic satellite measurements and its processing. Кафтан В.И. Геодезические спутниковые измерения и их обработка: Учебное пособие для бакалавров по направлению 120700 «Землеустройство и кадастр».- М.: МИИТ, 2013. –111 с.

Kaftan V.I. (2013b) At the jubilee Scientific Assembly of the International Association of Geodesy (IAG 150 Years) Кафтан В.И. На юбилейной Научной ассамблее Международной ассоциации геодезии (150 лет МАГ) // Международный научно-технический и производственный электронный журнал «Науки о Земле» (International scientific, technical and industrial electronic journal «Geo Science»).- 2013.- №2.- С. 5-24. http://issuu.com/geo-science/docs/_____________2-2013

Kaftan V.I., Tcvetkov V.J. (2013) On the form and content of the concept of a spatial data infrastructure. Кафтан В.И., Цветков В.Я. О форме и содержании понятия «инфраструктура пространственных данных» // Геодезия и картография. - 2013. - N 6. - С. 46-50. http://elibrary.ru/item.asp?id=21599997

Kaptüg V.B. (2013) The Moloskovitsy baseline as a memorial place of the first application of a new technique. Капцюг В.Б. Молосковицкий базис – памятник первого применения новой технологии. Мир измерений, 2013, № 5 (147), 51-54. http://elibrary.ru/item.asp?id=19078046

Kaufman M.B., Pasynok S.L. (2011) Tropospheric delays from GPS and VLBI data. Abstract book of Journeys 2011 "Earth rotation, reference systems and celestial mechanics: Synergies of geodesy and astronomy", 19-21 September 2011, BEV Vienna, Austria, p. 32.

Komitov B., Kaftan V. (2013) The sunspot cycle no. 24 in relation to long term solar activity variation, Journal of Advanced Research (2013) 4, 279-282. http://dx.doi.org/10.1016/j.jare.2013.02.001

Kuzin A.A. (2014) Vertical accuracy estimation of aerial laser scanning points for territory zoning on a level of landslide hazard. Кузин А.А. Оценка точности высот точек воздушного лазерного сканирования для зонирования территорий по степени оползневой опасности. Вестник ИрГТУ.-2014.-№5(88). С. 57-61. http://elibrary.ru/item.asp?id=21676820

Malinnikov V.A., Malinnikova E.V., Uchaev D.V. (2014) Fractal geometry of geodetic lines. Малинников В.А., Малинникова Е.В., Учаев Д.В. Фрактальная геометрия геодезических линий. Известия высших учебных заведений. Геодезия и аэрофотосъемка. 2014. № 1. С. 3-12. http://elibrary.ru/item.asp?id=21976776

Malkin Z. (2011a) On computation of a common mean. ArXiv:1110.6639. http://arxiv.org/abs/1110.6639

Malkin Z.M. (2011b) Study of Astronomical and Geodetic Series using the Allan Variance. Малкин З.М. Исследование астрономических и геодезических рядов с помощью вариации Аллана // Кинем. физ. неб. тел., т. 27, N 1, 2011, с. 59-70. ftp://ftp.mao.kiev.ua/pub/kfnt/27/1/kfnt-27-1-2011-05.pdf

Malkin Z. (2012a) Statistical analysis of the determination of the Galactic rotation constants. In: IAU XXVIII General Assembly, 2012, Abstract Book, 113-114.

Malkin Z. (2012b) The current best estimate of the Galactocentric distance of the Sun based on comparison of different statistical techniques. arXiv:1202.6128, 2012. http://arxiv.org/abs/1202.6128

Malkin Z. M. (2013a) On computation of the error of the weighted mean. Малкин З.М. О вычислении ошибки среднего взвешенного. Тр. Всероссийской астрометрической конф. "Пулково-2012", Изв. ГАО, 2013, No. 220, 511-516. http://www.gao.spb.ru/russian/publ-s/izv_220/conf_2012_astr.pdf

Malkin Z.M. (2013b) On the Calculation of Mean-Weighted Value in Astronomy. Малкин З.М. О вычислении средневзвешенных значений в астрономии. Астрон. журн., 2013, т. 90, N 11, 959-964. DOI: 10.7868/S0004629913110042

Malkin Z.M. (2013c) On the weigted mean computation. Малкин З.М. О вычислении ошибки среднего взвешенного. Тр. Всероссий кой астрометрической конф. "Пулково-2012", Изв. ГАО, 2013, No. 220, 511-516.

Malkin Z. (2013d) Using modified Allan variance for time series analysis. In: Reference Frames for Applications in Geosciences, Z. Altamimi, X. Collilieux (eds.), IAG Symposia, 2013, v. 138, 271-276. DOI: 10.1007/978-3-642-32998-2_39

Malkin Z. (2013e) A new approach to the assessment of stochastic errors of radio source position catalogues. Astron. Astrophys., 2013, v. 558, A29. DOI: 10.1051/0004-6361/201322334

Malkin Z.M., Prudnikiva E.Ya., Soboleva T.V., Miller N.O. (2013) L.D. Kostina and N.R.Persiyaninova Scientists of the Pulkovo Latitude Service. Малкин З.М., Прудникова Е.Я., Соболева Т.В., Миллер Н.О. Пулковские широтницы Л.Д. Костина и Н.Р. Персиянинова. Тр. Всероссийской астрометрической конф. "Пулково-2012", Изв. ГАО, 2013, No. 220, 581-587. http://www.gao.spb.ru/russian/publ-s/izv_220/conf_2012_astr.pdf

Mazurov B.T., Medvedev P.A. (2014) Leonhard Euler – contribution to astronomy, Earth mechanics, geodesy, cartography, geodynamics. Мазуров Б.Т., Медведев П.А. Леонард Эйлер – вклад для астрономии, небесной механики, геодезии, картографии, геодинамики. Интерэкспо Гео-Сибирь. 2014. Т. 1. № 1. С. 186-192.

Mazurova E. (2014) The Russian-Scandinavian Geodetic arc, Wissenschaftliches Kolloquium, Band 119, Jahrgang 2014, Leibniz-Sozietat der Wissenschaften zu, Berlin, Germany, 2014, pp.75-90.

Mazurova E., Ogienko S.A. (2013) Display of geodetic data in ArcGIS, Izvestiya Vuzov. Geodeziya i Aerofotos'yomka (News of Higher schools. Geodesy and air photography), № 5, 2013, pp. 35-42. http://www.miigaik.ru/journal.miigaik.ru/2012/20121108172714-3921.pdf

Mazurova Elena (2013) The Russian-Scandinavian Geodetic Arc, Wissenschaftliches Kolloquium, 15 November, 2013, Berlin, Germany.

Miller N.O. (2011) Chandler Wobble in Variations of the Pulkovo Latitude for 170 Years. Solar System Research, 2011, Vol. 45, No. 4, pp. 342-353.

Miller N.O. (2013) Fine structure and parameters of Chandle polar motion. Миллер Н.О. Тонкая структура и параметры чандлеровского движения полюса. Труды Всероссийской астрометрической конференции «Пулково-2012», Известия ГАО. 2013. № 220. С.125-130.

Miller N.O. Vorotkov M.V. (2013) Analisis of the residuals after main component determination in Earth's pole motion. Миллер Н.О., Воротков М.В. Анализ остатков после выделения основных компонент движения полюса земли. Труды Всероссийской астрометрической конференции «Пулково-2012», Известия ГАО. 2013. № 220. С. 131-136.

Miller N., Malkin Z. (2012a) Analysis of polar motion variations from 170-year observation series. Тр. ИПА РАН, 2012, вып. 26, 44-53.

Miller N., Malkin Z. (2012b) Joint Analysis of the Polar Motion and Celestial Pole Offset Time Series. In: IVS 2012 General Meeting Proc., ed. D. Behrend, K.D. Baver, NASA/CP-2012-217504, 2012, 385-389.

Ogorodova L.V., Kamynina N.S., Baranov V.N., Zaitsev A.K., Shilkin P.A. (2014) 135 years of the birth of F.N.Krasovsky. Огородова Л.В., Камынина Н.С., Баранов В.Н., Зайцев А.К., Шилкин П.А. К 135-летию со дня рождения Ф. Н. Красовского. Геодезия и картография. 2014. № 6. С. 59-64. http://elibrary.ru/item.asp?id=21831895

Physical geodesy. Scientific and technological paper collection of TsNIIGAiK. Физическая геодезия. Научно-технический сборник ЦНИИГАиК.- М.: Научный мир, 2013.- 288 с.

Prudnikiva E.Ya., Soboleva T.V., Malkin Z.M. (2013) In memory of Ivan Fedotovich Korbut. Прудникова Е.Я., Соболева Т.В., Малкин З.М. Памяти Ивана Федотовича Корбута. Тр. Всероссийской астрометрической конф. "Пулково-2012", Изв. ГАО, 2013, No. 220, 601-606. http://www.gao.spb.ru/russian/publ-s/izv_220/conf_2012_astr.pdf

Savinykh V., Kaftan V. (2013) Geodesy section of the National Geophysical Committee of the Russian Academy of Sciences as a component of geodetic infrastructure (Advisory). International Association of Geodesy, Scientific Assembly 150th Anniversary of the IAG, Book of Abstracts, Book of Abstracts, September 1-6, 2013, Potsdam, p.442. http://www.iag2013.org/IAG_2013/Publication_files/abstracts_iag_2013_2808.pdf

Savinykh V.P., Vasiliev V.P., Kapranov Yu.S., Krasnorylov I.I., Kufal G.E., Perminov S.V., Shevchenko V.V. (2014) Revisiting the Lunar base creation. Савиных В.П., Васильев В.П., Капранов Ю.С., Краснорылов И.И., Куфаль Г.Э., Перминов С.В., Шевченко В.В. К вопросу о создании Лунной базы. // Изв. вузов. Геодезия и аэрофотсъемка.-2014. М.:- № 2.- С. http://elibrary.ru/item.asp?id=21976807

Shirenin A. M., E.M. Mazurova, A.V. Bagrov. (2014) Building the global positioning system for uninhabited planets and natural satellites, Space Colonization Journal, Vol. 15, May, 23, 2014, pp. 1-14.

Soloviev A. A., V. I. Kaftan, R. I. Krasnoperov, R. V. Sidorov. (2013) Modern technological approaches for development of intermagnet observatories in Russia. Materials of the Partnership Conference "Geophysical observatories, multifunctional GIS and data mining", 30 September - 3 October 2013, Kaluga, Russia, DOI: 10.2205/2013BS012_Kaluga

Teterin G.N. (2011a) Dangerous illness of geodesy. Тетерин Г.Н. Опасное заболевание геодезии «Геодезия и картография», 2011, №9.-56-57. http://elibrary.ru/item.asp?id=21831860

Teterin G.N. (2011b) Geometrical and geophysical aspects in geodesy. Тетерин Г.Н. Геометрическое и геофизическое в геодезии. Вестник СГГА, 2011, №1. - С. 26-31. http://elibrary.ru/item.asp?id=17719636

Teterin G.N. (2011c) Theoretical and methodological foundation of modern geodesy. Тетерин Г.Н. Теоретические и методологические основы современной геодезии «Геодезия и картография», 2011, №1. - С. 55-59. http://elibrary.ru/item.asp?id=21816602

Teterin G.N. (2012a) The language of geodesy. Тетерин Г.Н. Язык геодезии «Геодезия и картография», 2012, №1. - С. 53-58. http://elibrary.ru/item.asp?id=21623972

Teterin G.N. (2012b) The symbol of obsolete ideology. Тетерин Г.Н. Символ устаревшей идеологии. Вестник СГГА.- 2012.- №1.-47-52. http://elibrary.ru/item.asp?id=17844053

Teterin G.N. (2013) From where the geodesy went to eat? Тетерин Г.Н. «Откуда есть пошла» геодезия? // Изыскательский вестник.- 2013.-№2 (17) .- с. 16-21. http://www.spbogik.ru/images/download/vestnik_17.pdf

Teterin G.N., Sinianskaya M.L. (2011) Angular and distance measures in ancient time. Тетерин Г.Н., Синянская М.Л. Угловые и линейные меры измерений в древнее время сб. матер. VII Междунар. научн. конгресса. «ГЕО-Сибирь-2011», - С. 79-83.

Teterin G.N., Sinianskaya M.L. (2012) Biography and chronology handbook (Geodesy, Cartography - Twenty sentury). Тетерин Г.Н., Синянская М.Л. Биографический и хронологический справочник (Геодезия, картография – двадцатый век) – 2012. – 592 с.

Tolchelnikova S.A. (2012) Note about distance determination techniques. Толчельникова С.А. Замечание о методах определения расстояний // Геодезия и картография.- 2012.- № 7.- с. 6–12. http://elibrary.ru/item.asp?id=21769828

Tolchelnikova S.A. (2013a) On the occasion of the 150 years from the birthday of A.N. Krilov. Толчельникова С.А. К 150-летию со дня рождения А.Н.Крылова. Геодезия и картография, 2013.- №12.- с.50-52. http://elibrary.ru/item.asp?id=21623813

Tolchelnikova S.A. (2013b) Scientific revolution in physics of XX century and classical heritage. Толчельникова С.А. Научная революция в физике XX века и классическое наследие. Материалы докладов Международной конференции, Восьмые Окуневские чтения, Санкт-Петербург, 2013 г, с.435-438.

Tolchelnikova S.A. (2013c) Speed of light and the problem of simultaneity. Толчельникова С.А. Скорость света и проблема определения одновременности. Москва, Геодезия и картография.- 2013.-№3.-с. 8-15. http://elibrary.ru/item.asp?id=21591569

Tolchelnikova S.A. (2014) Scientific revolution in physics of XX century and classic heritage. Толчельникова С.А. Научная революция в физике XX века и классическое наследие. Геодезия и картография. 2014. № 6. С. 10-19. http://elibrary.ru/item.asp?id=21831888

Tolchelnikova S.A., Chubey M.S. (2013) To the study of inertial motion of the Solar system (Astronomical method of special relativity theory verification). Толчельникова С.А., Чубей М.С. К изучению инерциального движения Солнечной системы (Астрономический способ проверки СТО) Геодезия и картография. 2012. № 1. С. 8-15. http://elibrary.ru/item.asp?id=21623965

Vityazev V.V., Miller N.O., Prudnikova E. Ja. (2012) The use of the Singular Spectrum Analysis for investigating the pole motion. Витязев В.В., Н.О. Миллер, Е.Я. Прудникова. Использование сингулярного спектрального анализа при исследовании движения полюса. Вестник СПбГУ, Серия 1, 2012, вып. 2, с. 139-147. http://elibrary.ru/item.asp?id=17789285

Yurkina M.I. (2012) On the occasion of the ninety year birthday of Leonhard Pavlovich Pellinen. Юркина М.И. К 90-летию Леонарда Павловича Пеллинена / Физическая геодезия. Научно-технический сборник ЦНИИГАиК. – М.: Научный мир, 2013. – с.235-240.

Yurkina M.I., Brovar B.V. (2014) On evolution of contents of geodesy and gravimetry and its tasks. Юркина М.И., Бровар Б.В. Об эволюции содержания геодезии и гравиметрии и их задач. Геодезия и картография. 2014. № 9. С. 47-56. http://elibrary.ru/item.asp?id=22253727

Yurkina M.I., Demianov G.V., Brovar B.V., Kaftan V.I. (2013) On the occasion of centenary of the birth of Michail Sergeevich Molodensky. Юркина М.И., Демьянов Г.В., Бровар Б.В., Кафтан В.И. К 100-летию Михаила Сергеевича Молоденского / Физическая геодезия. Научно-технический сборник ЦНИИГАиК. – М.: Научный мир, 2013. – с.226-235.

Zaitsev A.K. (2011) Degree of point. Зайцев А.К. Степень точки// Международный научно-технический и производственный журнал «Науки о Земле» - 2011. - №3-4 - с.17-21.

http://geo-science.ru/wp-content/uploads/3-17-21.pdf

Zharov V.E. (2011) VLBI in astrometry – the present and future achievements. Abstract book of the International astronomical congress "AstroKazan-2011", 2011.